\documentclass[article,12pt]{article} 
\usepackage{amscd,amsmath,amssymb,amsfonts,latexsym,mathrsfs,amsthm,mathtools}
\usepackage{graphicx} 
\graphicspath{{figures/}} %Setting the graphicspath
\usepackage{setspace} 
\usepackage{graphics,epsfig}
\usepackage{sectsty}
\usepackage[numbers]{natbib}
\usepackage[english]{babel}
\usepackage{amsmath}
\usepackage{amsfonts}
\usepackage{amssymb,amsthm}
\usepackage{bm}
\usepackage{mathrsfs}
\usepackage{subfigure}
\usepackage[usenames]{color}
\usepackage{rotating}
\usepackage{url}
\usepackage{bbm}
\usepackage{float}

\newtheorem{rem}{{\bf Remark}}

\usepackage{flushend}
\usepackage{verbatim}
\usepackage{tabularx}
\usepackage{multirow} 
\usepackage{arydshln}
\usepackage{hyperref}
\usepackage[toc,page]{appendix}

\newcommand{\x}{{\bf x}}
\newcommand{\z}{{\bf z}}

\newcommand{\y}{{\bf y}}

\newcommand{\post}{\bar{\pi}}

\newcommand{\s}{{\bf s}}

\newcommand{\norm}[1]{\left\lVert#1\right\rVert}

  \title{%Regression-based 
  Deep Importance Sampling based on Regression for\\  Model Inversion and Emulation}% with applications in remote sensing   }
\author{F. Llorente$^\star$, L. Martino$^{\star\star,\dagger}$, D. Delgado$^\star$, G. Camps-Valls$^\dagger$ \\
{\small$^\star$  Universidad Carlos III de Madrid,  Legan\'es (Spain).}\\
{\small$^\star$$^\star$  Universidad Rey Juan Carlos,  Fuenlabrada (Spain).} \\
{\small $^\dagger$ Universitat de València, Valencia (Spain).}
}
\date{}
 
\begin{document}

\maketitle

\begin{abstract}
	Understanding systems by forward and inverse modeling is a recurrent topic of research in many domains of science and engineering. 
	In this context, Monte Carlo methods have been widely used as powerful tools for numerical inference and optimization.
	They require the choice of a suitable proposal density that is crucial for their performance. For this reason, several adaptive importance sampling (AIS) schemes have been proposed in the literature. 
	We here present an AIS framework called Regression-based Adaptive Deep Importance Sampling (RADIS).  
	In RADIS, the key idea is the adaptive construction via regression of a non-parametric proposal density (i.e., {\it an emulator}), which mimics the posterior distribution and hence  minimizes the mismatch between proposal and target densities. RADIS is based on a deep architecture of two (or more) nested IS schemes, in order to draw samples from the constructed emulator.
	The algorithm is highly efficient since employs the posterior approximation as proposal density, which can be improved adding more support points. As a consequence, RADIS asymptotically converges to an exact sampler under mild conditions. Additionally, the emulator produced by RADIS can be in turn used as a cheap surrogate model for further studies. 
	We introduce two specific RADIS implementations that use Gaussian Processes (GPs) and Nearest Neighbors (NN) for constructing the emulator. 
	Several numerical experiments and comparisons show the benefits of the proposed schemes. A real-world application in remote sensing model inversion and emulation confirms the validity of the approach.
\end{abstract}

%%%%%%%%%%%%%%
\section{Introduction}
%%%%%%%%%%%%%%

Modeling and understanding systems is of paramount relevance in many domains of science and engineering. The problems involve both forward and inverse modeling, and very often one resorts to domain knowledge (either in the form of mechanistic models, hypotheses, constraints or just data) and observational data to learn parametrizations and do inferences.
Among the many approaches possible, Bayesian methods have become very popular during the last decades. Bayesian inference is very active in the communities of machine learning, statistics and signal processing \cite{OHagan1994,MARTINO_book,Robert04}.
With them, there has been a surge of interest in the Monte Carlo (MC) techniques that are often necessary for the implementation of the Bayesian analysis.  Several families of MC schemes have been proposed that excel in numerous applications, including the popular Markov Chain Monte Carlo (MCMC) algorithms, particle filtering techniques and adaptive importance sampling (AIS) methods \cite{Robert04,bugallo2017adaptive}.

\noindent{\bf Adaptive Importance Sampling (AIS).} The performance of the MC algorithms depends strongly on the proper choice of a proposal probability density function (pdf).  In adaptive schemes, the proposal pdf is updated considering the previous generated samples. 
%In importance sampling (IS), this choice is guided by the general fact that the variance of the estimator scales with the mismatch between the posterior and proposal distributions  (see e.g. \cite{akyildiz2019convergence}).
In recent years, a plethora of AIS algorithms have been proposed in the literature \cite{bugallo2017adaptive}. In most of these algorithms, the complete proposal can be expressed as a  finite parametric mixture of densities \cite{Cappe04,pmc-cappe08,ElviraPMC15,el2019variational,APIS15}.  Unlike these schemes, we consider a non-parametric proposal based on an interpolating construction.

\noindent
{\bf Emulators in Bayesian Inference.} Furthermore, many Bayesian inference problems involve the evaluation of computationally intensive models, due to the use of particularly complex systems, consisting of many coupled ordinary or partial differential equations in high-dimensional spaces, or a large amount of available data. 
To overcome this issue, a successful approach consists in replacing the true model by a surrogate model (a.k.a. {\it an emulator})~\cite{OHagan2006,busby2009hierarchical,schmit2018emulation,razavi2012review,Daniel2020}. 

The resulting emulator can be employed in different ways inside a Bayesian analysis. A first possibility is to apply MC sampling methods considering the surrogate model as an approximate posterior pdf within the MC schemes \cite{christen2005markov,ying2020moving,conrad2016accelerating}\cite[Chapter 9.4.3]{liu2008monte} or within different quadrature rules \cite{kennedy1998bayesian,rasmussen2003bayesian,llorente2020adaptive}, instead of the evaluation of a costly true posterior. For instance, this is also the case of the strategy known as {\em calibrate, emulate, sample}, currently in vogue  \cite{cleary2020calibrate}.
In order to improve the efficiency of MC algorithms, a second option is to use the emulator as a proposal density within an MC technique. Here, we focus on the last approach. 
\newline
{\bf Contribution.}
In this work, we design a deep AIS framework where a non-parametric interpolating proposal density is adapted online. The new approach is called Regression-based Adaptive Deep Importance Sampling (RADIS).
In RADIS, the key idea is the adaptive construction of a non-parametric proposal pdf (i.e., {\it an emulator}), which mimics the posterior distribution in order to minimize the mismatch between proposal and target pdfs. Differently from other adaptive schemes,  the adaptation in RADIS not only uses the information of the previous samples, but also all the evaluations of the posterior for directly constructing the emulator.
Thus, unlike in a parametric approach, in our setting this discrepancy can be arbitrarily decreased to zero by adding more nodes. Hence, RADIS is asymptotically an exact sampler.  
The proposed methodology is based on a {\it deep architecture}: two nested IS schemes are employed, with an inner and an outer IS layers. The inner IS stage is used to generate samples from the emulator. The outer IS layer provides the final posterior approximation by a cloud of weighted samples. Thus, RADIS finally provides two approximations of the posterior, one in form of a weighted particle measure, and also the emulator adapted online.\footnote{The emulation  can be applied to the entire posterior or part of it, like a physical model.} Parsimonious constructions of the emulator have been also discussed.

We discuss two specific implementation of RADIS. These specific implementations differ on the choice of the emulator construction. In the first one, a Gaussian Process (GP) model is applied to the log-posterior function obtaining the novel scheme denoted as GP-AIS. In the second one, a piece-wise constant approximation based on Nearest Neighbors (NNs) is applied, providing the novel algorithm denoted as NN-AIS. In both cases, the resulting proposal pdf can be seen as an {\it incremental mixture} of densities. 
A deep structure with more than two layers is  described, where a chain of emulators is adapted and then employed as proposal pdfs within different nested IS stages.
Robust and sequential implementations are also discussed.
Several numerical comparisons show the advantages of RADIS with respect to benchmark algorithms. A real-word application illustrates the capabilities for sequential parameter retrieval and emulation of a well-known radiative transfer model (RTM) used in remote sensing. In the next section, a brief overview of the related approaches is provided.

%%%%%%%%%%%%%%%%%%%%%%%%%%%%%%%%%%%%%%%%%%%%%%
%%%%%%%%%%%%%%%%%%%%%%%%%%%%%%%%%%%%%%%%%%%%%%
\section{Other related works} 
%%%%%%%%%%%%%%%%%%%%%%%%%%%%%%%%%%%%%%%%%%%%%%
%%%%%%%%%%%%%%%%%%%%%%%%%%%%%%%%%%%%%%%%%%%%%%

The non-parametric interpolating construction of the proposal and related strategies are appealing from different points of views. This is proved by attention devoted by the previous attempts in the literature  shown above, and by other related approaches that we describe next.
\newline
{\bf Interpolating proposal.} The idea of using interpolating densities is particularly attractive since %since the resulting proposal density can become closer and closer to the true posterior.  
we can arbitrarily  decrease the mismatch between proposal and posterior by adding more support points.
For this reason, the resulting algorithms provide very good performance \cite{Gilks95,Meyer08,IA2RMS15,FUSS,martino2018adaptive}.
The first use of an interpolating procedure for building a proposal density  can be ascribed to the rejection sampling and adaptive rejection sampling schemes \cite{Gilks92,Hoermann95,Gorur08rev}. The well-know Zigurrat algorithm and table methods are other examples of fast rejection samplers employing interpolating proposals \cite{Zigurrat,martino2018independent}. They are state-of-the-art methods as random sample generators of specific univariate distributions in terms of speed of generation. 
In some rejection samplers and MCMC algorithms, the proposal is formed by polynomial pieces (constant, linear, etc.) \cite{Gilks95,Meyer08,IA2RMS15,FUSS}, \cite[Chapters 4 and 7]{martino2018independent}. 
The use of interpolating proposal pdfs within an {IS} scheme is also considered in \cite{felip2019tree}. The conditions needed for applying an emulator as a proposal density are discussed in  \cite{martino2018adaptive}. More specifically,  we need to be able to: {\bf (a)} update the construction of the emulator, {\bf (b)} evaluate the emulator,  {\bf (c)} normalize the function defined by the emulator, and {\bf (d)}  draw samples from the emulator. It is not straightforward to find an interpolating (or regression) construction which satisfies all those conditions jointly, and especially for an arbitrary dimensionality of the problem. This is the reason why the previous attempts of using an interpolating proposal pdfs are restricted to the univariate case. Our deep architecture solves these issues. 
\newline
{\bf Partitioning and stratification.}
Note also that  
the use of a proposal pdf formed by components restricted to disjoint regions of the domain (like in the piecewise constant proposal based on NN) is related to the stratification idea.
Indeed, different schemes based on partitioning and/or stratification 
divide the entire domain in disjoint sub-regions and consider different partial proposals in each of them \cite[Chapter 4.6.3]{Robert04}, \cite{lepage1978new,friedman1981nested,press1990recursive,lu2018exploration}. The complete proposal pdf is then a mixture of the partial proposals.
Moreover, this process can be iterated so that the partition is refined over the iterations increasing the number of partial proposals. In this case, the complete proposal is an {\it incremental mixture} as RADIS  (see also below) \cite{lepage1978new,lu2018exploration}.
Recent works propose using trees in order to partition the space and subsequently build the proposal \cite{felip2019tree, foster2020model}. In the context of MCMC, \cite{hanson2011polya} builds an approximation of the target using Polya trees. 
\newline
{\bf Incremental mixtures.} The use of non-parametric but {\it non-interpolating} proposals have been suggested in other works. 
A non-parametric IS approach is considered in \cite{zhang1996nonparametric}, where the proposal is built by a kernel density estimation. In \cite{steele2006computing},  a proposal pdf defined as a mixture with increasing number of components is also suggested. When a weighting strategy based on the so-called {\it  temporal deterministic mixture} is applied \cite{LAIS17, elvira2019generalized}, incremental mixture proposals appear also in other IS schemes ( e.g., \cite{LAIS17,CORNUET12}). 
\newline
{\bf Other approaches.} Surrogate GP models has been also employed within IS schemes in the context of rare event estimation \cite{balesdent2013kriging,dubourg2013metamodel}. Finally, other  IS schemes can be encompassed  in a similar ``deep'' approach \cite{Neal01,elvira2020importance}. In the first one, MCMC steps are used to jump from different tempered versions of the posterior, and a global IS weighting as product of intermediate weights \cite{Neal01}. In the second scheme a two-stages weighting procedure is used, where the first layer considers a Gauss-Hermite quadrature and the second layer is a standard IS method  \cite{elvira2020importance}.

%%%%%%%%%%%%%%%%%%%%%%%%%%%%%%%%%%%%%%%%%%%%%%
%%%%%%%%%%%%%%%%%%%%%%%%%%%%%%%%%%%%%%%%%%%%%%
\section{Preliminaries and motivation}
%%%%%%%%%%%%%%%%%%%%%%%%%%%%%%%%%%%%%%%%%%%%%%
%%%%%%%%%%%%%%%%%%%%%%%%%%%%%%%%%%%%%%%%%%%%%%

%%%%%%%%%%%%%%%%%%%%%%%%%%%%%%%%%%%%%%%%%%%%%%
\subsection{Problem statement}
%%%%%%%%%%%%%%%%%%%%%%%%%%%%%%%%%%%%%%%%%%%%%%

{\bf Bayesian inference.} In many real world applications, the goal is to infer a variable of interest given a set of data \cite{OHagan94}.
Let us denote the parameter of interest (static or dynamic) by ${\bf x}\in \mathcal{X}\subseteq \mathbb{R}^{d_x}$, and let ${\bf y}\in \mathbb{R}^{d_y}$ be the observed data. In a Bayesian analysis, all the statistical information is contained in the posterior distribution, which is given by
\begin{equation}
	\bar{\pi}({\bf x})= p({\bf x}| {\bf y})= \frac{\ell({\bf y}|{\bf x}) g({\bf x})}{Z(\y)},
	\label{eq:posterior}
\end{equation}
where $\ell({\bf y}|{\bf x})$ is the likelihood function, $g({\bf x})$ is the prior pdf, and $Z(\y)$ is the  Bayesian model evidence (a.k.a. marginal likelihood). The marginal likelihood $Z(\y)$ is important for model selection purposes \cite{ourRev,Martino15PF}.
Generally, $Z(\y)$ is unknown, so we are able to evaluate the unnormalized target function,
$\pi({\bf x})=\ell({\bf y}|{\bf x}) g({\bf x})$. The analytical computation of the posterior density $\bar{\pi}({\bf x}) \propto \pi({\bf x})$ is often unfeasible, hence numerical approximations are needed. Our goal is to approximate integrals of the form 
\begin{align}\label{eq:IntegralOfInter}
	I =  \int_\mathcal{X}f(\x)\post(\x)d\x =\frac{1}{Z} \int_\mathcal{X}f(\x)\pi(\x)d\x,
\end{align}
where $f(\x)$ is some integrable function, and 
\begin{align}\label{eq:MargLike}
	Z = \int_\mathcal{X}\pi(\x)d\x.
\end{align}
In the literature, random sampling or deterministic quadratures are often used \cite{Robert04,martino2018independent,Niederreiter92}. 
In this work, we focus on the so-called {IS} approach.
\newline
{\bf Emulation.} There exist many situations where the evaluation of $\pi$ is expensive (e.g., as in big data framework or when the observation model is costly). 
Hence, we are also interested in obtaining an emulator of $\pi(\x)$ (or just a part of the posterior), denoted $\widehat{\pi}_t(\x)$, such that  (i) $\widehat{\pi}_t(\x)$ is cheap to evaluate, and (ii) $\widehat{\pi}_t(\x) \to \pi(\x)$ (in some sense, e.g., $L_2$ norm) as $t\to\infty$.

%%%%%%%%%%%%%%%%%%%%%%%%%%%%%%%%%%%%%%%%%%%%%%
\subsection{Importance sampling (IS) and aim of the work}
%%%%%%%%%%%%%%%%%%%%%%%%%%%%%%%%%%%%%%%%%%%%%%

Let us consider a normalized proposal density $\bar{q}({\bf x})$.\footnote{We assume that $\bar{q}({\bf x})>0$ for all ${\bf x}$ where $\bar{\pi}({\bf x})> 0$, and $\bar{q}({\bf x})$ has heavier tails than $\bar{\pi}({\bf x})$.} The importance sampling (IS) method consists of  
drawing $N$ independent samples, ${\bf x}_1,\ldots,{\bf x}_N$, from $\bar{q}({\bf x})$ (also called particles), and then assign to each sample the following unnormalized weights
\begin{equation} 
	w_n=w({\bf x}_n)= \frac{\pi({\bf x}_n)}{{\bar{q}({\bf x}_n)}}, \quad n=1,\ldots,N.
	\label{is_weights_static}
\end{equation} 
An unbiased estimator of  the marginal likelihood $Z$ is given by the arithmetic mean of these unnormalized weights \cite{Liu04b,Robert04}, i.e.,
$$
\widehat Z= \frac{1}{N} \sum_{n=1}^N w_n.
$$ 
Defining also the normalized weights 
${\bar w}_n=\frac{ w_n }{\sum_{i=1}^N w_i}$, 
with $n=1,\ldots,N$, the self-normalized IS estimator of $I$ in Eq. \eqref{eq:IntegralOfInter} is given by 
$$
{\widehat I}=\sum_{n=1}^N  {\bar w}_n f({\bf x}_n).
$$  
More generally, regardless of the specific function $f(\x)$, we obtain a particle approximation of $\bar{\pi}$, i.e.,
$\widetilde{\pi}(\x)=\sum_{n=1}^N {\bar w}_n \delta(\x-\x_n)$,
where $\delta(\x)$ is a delta function. 
It is important to remark that with this  particle approximation, we can approximate several quantities related to the posterior $\bar{\pi}({\bf x})$, such as any moments and/or credible intervals (not just a specific integral).
The quality of this particle approximation is related to the discrepancy between the proposal $\bar{q}(\x)$ and the posterior $\bar{\pi}({\bf x})$. Indeed, in an ideal MC scenario, we can draw from the posterior, i.e., $\bar{q}(\x)=\bar{\pi}({\bf x})$, so that ${\bar w}_n=\frac{1}{N}$, which corresponds with the maximum  effective sample size (ESS) \cite{Liu04b,MARTINO2017386}. With a generic proposal $\bar{q}(\x)\propto \bar{q}(\x)$, we can obtain a very small ESS and a bad particle approximation $\widetilde{\pi}(\x)$ (i.e., poor performance of the algorithm). 

{\rem The variance of the marginal likelihood estimator $\widehat{Z}= \frac{1}{N} \sum_{n=1}^N w(\x_n)$ is given by
	\begin{align}
		\texttt{var}[\widehat{Z}] = \frac{1}{N}\texttt{var}[w(\x)],
	\end{align}
	where $w(\x)=\frac{\pi({\bf x})}{{\bar{q}({\bf x})}}$ and $\x \sim \bar{q}(\x)$. Since $\widehat{Z}$ is also unbiased, then we also have 
	\begin{align}
		\mathbb{E}[|Z - \widehat{Z}|^2]=\frac{1}{N}\texttt{var}[w(\x)].
	\end{align}
	{	For more details, see \cite{Robert04}.}
}
{\rem The variance of the IS weight function $w(\x)$ is proportional to the  Pearson divergence  between $\bar{q}(\x)$ and $\post(\x)$, denoted as $\chi^2(\post\|\bar{q})$ (also called $\chi^2$ distance), i.e.,  
	\begin{align}
		\texttt{var}[w(\x)] \propto \chi^2(\post\|\bar{q})= \int_{\mathcal{X}} \frac{(\post(\x)-\bar{q}(\x))^2}{\bar{q}(\x)}d\x.
	\end{align}
	See {\cite{MARTINO2017386,akyildiz2019convergence}} and \ref{App:varISweights} for further details.
	Regarding the mean squared error of the estimator $\widehat{I}$, we have 
	\begin{align}
		\mathbb{E}[|I - \widehat{I}|^2] \leq \frac{C_f}{N}(\chi^2(\post\|\bar{q}) + 1).
	\end{align}
	%$$, where 
	%\begin{align}
	%    \chi^2(\post\|q) = \int \frac{(\post(\x) - q(\x))^2}{q(\x)}d\x,
	%\end{align}
	The relationships with the $L_2$ and $L_\infty$ distances is also given in  \ref{App:varISweights}.} 
%{
%{\rem Optimal proposal for $\widehat{I}$.... que opinas confunde o ayuda?
%}}	
\newline
\newline
To reduce the discrepancy between the proposal $\bar{q}(\x)$ and the posterior $\bar{\pi}({\bf x})$, we consider a non-parametric adaptive construction of the proposal $\bar{q}_t(\x)$ where $t$ denotes a discrete iteration index. In order to make that the discrepancy becomes smaller and smaller, an interpolating procedure $\bar{q}_t(\x)$ based on a set of support points $\mathcal{S}_{t-1}$ is employed. Namely, we generate a sequence of proposal pdfs $\bar{q}_t(\x)$, $\bar{q}_{t+1}(\x)$, $\bar{q}_{t+2}(\x)$,...  which become closer and closer to $\bar{\pi}({\bf x})$, as the number of support points grows.
Throughout the paper, we denote $\bar{q}_t(\x) \propto \widehat{\pi}_t(\x)$ the non-parametric regression function which approximates the unnormalized posterior $\pi(\x)$ at iteration $t$. The normalized proposal is denoted as $\bar{q}_t(\x) = \frac{1}{c_t}\widehat{\pi}_t(\x)$, where $c_t = \int_{\mathcal{X}} \widehat{\pi}_t(\x)d\x$. Although the approximation $\widehat{\pi}_t$ depends on the set of nodes $\mathcal{S}_{t-1}$, for simplicity we use the simpler notation $\widehat{\pi}_t(\x) =\widehat{\pi}_t(\x;\mathcal{S}_{t-1})$.

{\rem If the sequence of proposals is such $\norm{\post - \bar{q}_t}_2 \to 0$ as $t\to \infty$, then $\chi^2(\post\|\bar{q}_t) \to 0$. See \ref{App:varISweights} for more details.
}

\begin{figure*}[!ht]
	\centering
	%	\centerline{
	\includegraphics[width=0.8\textwidth]{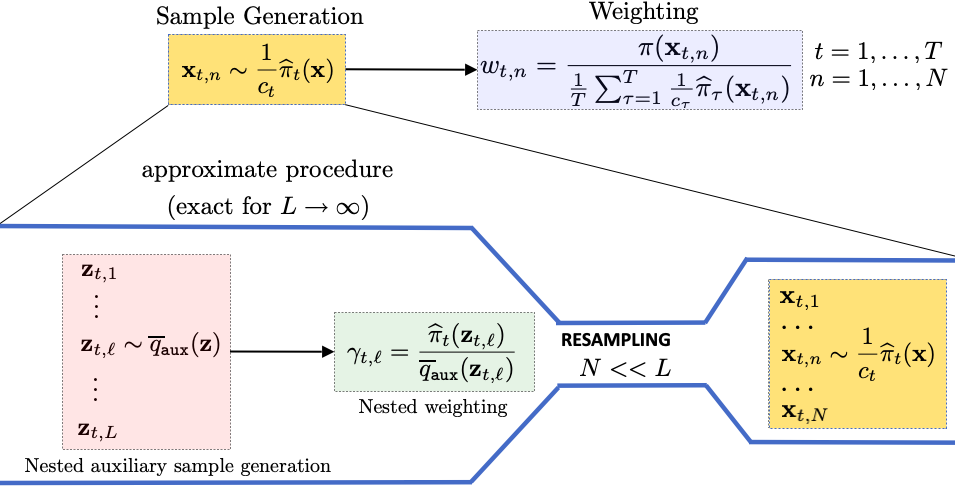}
	%	}
	% 	\vspace{-0.3cm}
	\caption{Approximate sampling from {$\frac{1}{c_t}\widehat{\pi}_t(\x) \propto \widehat{\pi}_t(\x)$} and final weighting scheme.}
	\label{fig_doubleIS}
\end{figure*}

\begin{figure*}[!ht]
	\centering
	%	\centerline{
	\includegraphics[width=0.4\textwidth]{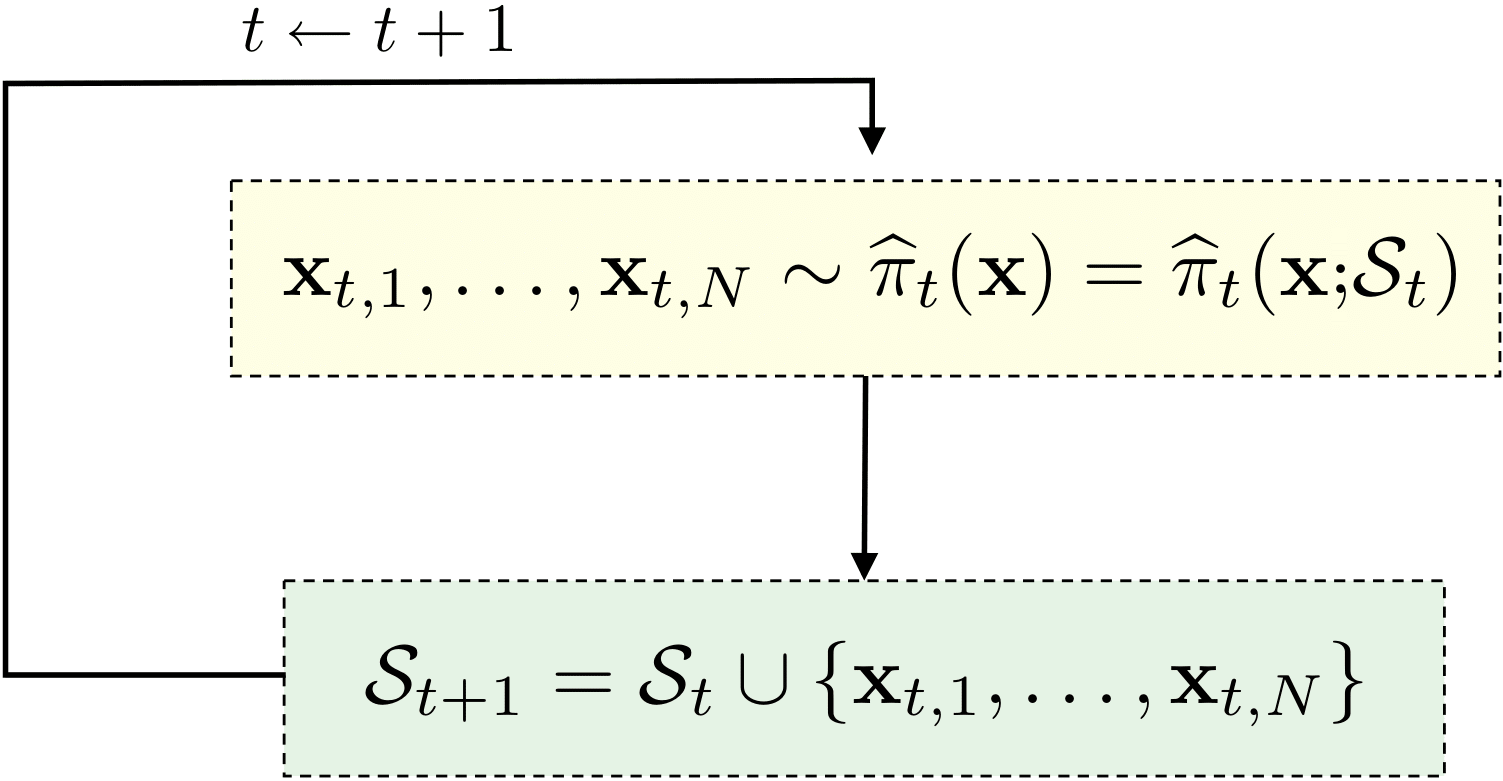}
	%	}
	% 	\vspace{-0.3cm}
	\caption{Graphical representation of the adaptation scheme.{ More parsimonious alternatives are introduced in Section \ref{sec_stickytests}.} }
	\label{fig_adaptation}
\end{figure*}

%%%%%%%%%%%%%%%%%%%%%%%%%%%%%%%%%%%%%%%%%%%%%% 
%%%%%%%%%%%%%%%%%%%%%%%%%%%%%%%%%%%%%%%%%%%%%% 
\section{Regression-based Adaptive Deep Importance Sampling}
%%%%%%%%%%%%%%%%%%%%%%%%%%%%%%%%%%%%%%%%%%%%%% 
%%%%%%%%%%%%%%%%%%%%%%%%%%%%%%%%%%%%%%%%%%%%%% 

In this section, we introduce the proposed scheme, called Regression-based Adaptive Deep Importance Sampling (RADIS). 
The resulting algorithm is an adaptive importance sampler with a non-parametric interpolating proposal pdf. 
We show how to implement the sampling and construction of the proposal density in Sect. \ref{sec:sampProp} and Sect. \ref{sec:buildProp} respectively. 
The novel scheme is summarized in Table \ref{table_DeepIS_algorithm}.
The proposal is adaptively built using a regression approach that considers the set of all previous nodes $\x_i$'s where $\pi$ is evaluated. In Section \ref{sec:buildProp}, we present two construction methodologies considered in this work. 
Samples from this proposal are drawn via an approximate procedure that can be interpreted as an additional ``inner'' IS. 
All the samples generated in the inner IS are then used in the ``outer'' IS.
Figure \ref{fig_doubleIS} outlines this procedure. The adaptation consists in sequentially adding the samples to the set of current nodes (see Figure \ref{fig_adaptation}). In the outer IS, we consider a temporal deterministic mixture approach to compute the weights. Note that the weighting step needs to be done only once at the end of the algorithm. 

%%%%%%%%%%%%%%%%%%%%%%%%%%%%%%%%%%%%%%%%%%%%%%
\subsection{RADIS: a two-layer Deep IS}\label{sec:sampProp}
%%%%%%%%%%%%%%%%%%%%%%%%%%%%%%%%%%%%%%%%%%%%%% 

RADIS is an adaptive IS scheme based on two IS stages.  
In the following, we describe the inner and outer stages as well as the possible construction and adaptation of the non-parametric proposal density. The extension with more than two nested layers is also discussed.

\subsubsection{Inner IS scheme}
The inner IS stage is repeated at every iteration. It generates samples approximately distributed from the current non-parametric proposal, denoted as $\widehat{\pi}_t$ (the unnormalized version). Furthermore, these samples are used to normalize $\widehat{\pi}_t$, i.e., in order to estimate $c_t = \int_\mathcal{X} \widehat{\pi}_t(\x)d\x$.  
\newline
%%%%%%%%%%%%%%%%%
{\bf Approximate sampling from the emulator.} It is not straightforward to sample from an interpolating proposal \cite{Gilks95,martino2018adaptive}. We propose using an approximate procedure based on IS.
Specifically, at each iteration, in order to sample from $\frac{1}{c_t}\widehat{\pi}_t(\x)$, we use sampling importance resampling (SIR)  with an auxiliary proposal $\bar{q}_{\text{aux}}$ \cite{Rubin88}.  First, a set of  $\{\z_{t,\ell}\}_{\ell=1}^L$ (with large $L$) are drawn from $\bar{q}_\text{aux}(\x)$. These auxiliary samples are weighted according to $\widehat{\pi}_t(\x)$
$$\gamma_{t,\ell} = 
%\gamma(\z_{t,\ell}) =
\frac{\widehat{\pi}_t(\z_{t,\ell})}{\bar{q}_\text{aux}(\z_{t,\ell})}\qquad \ell=1,\dots,L.$$
Finally, in other to obtain $\{\x_{t,n}\}_{n=1}^N$, we resample $N$ times within  $\{\z_{t,\ell}\}_{\ell=1}^L$ with probabilities $\{\bar{\gamma}_{t,\ell}\}_{\ell=1}^L$ where $\bar{\gamma}_{t,\ell}=\frac{\gamma_{t,\ell}}{\sum_{i=1}^L \gamma_{t,i}}$ for $\ell=1,...,L$, i.e.,
\begin{equation}
	\x_{t,n}\sim \sum_{\ell=1}^L \bar{\gamma}_{t,\ell} \delta(\x-\z_{t,\ell}), \quad \mbox{ for all } \quad n.    
\end{equation}
In this way, we obtain a set of samples $\{\x_{t,n}\}_{n=1}^N$ approximately distributed from $\widehat{\pi}_t$ \cite{Rubin88,smith1992bayesian}.

{\rem Under some mild conditions, as  $L \to \infty$, the SIR procedure is asymptotically exact. Namely, as $L \to \infty$ the density of the resampled particles becomes closer and closer to $q_t(\x)\propto \widehat{\pi}_t(\x)$. See, for instance, the following references
	\cite{Rubin88}, \cite[Sect. 6.2.4]{givens2012computational},  \cite[Sect. 3.2]{smith1992bayesian}. For further details, see \cite[page 6 ]{robert2020markov}, \cite[App. A]{martino2018group} and also \ref{APP:theo}.}

{\rem Note that the computation of the inner IS weights $\gamma_{t,\ell}$'s does not involve the evaluation of the posterior $\pi(\x)$, but only the evaluation of the emulator $\widehat{\pi}_t(\x)$. 
	Hence, assuming that the evaluation of the posterior is the main computational bottleneck, in this setting we can make $L$ arbitrarily large.
} 
\newline
\newline
Since we resample from a finite set, we can obtain duplicated samples, but it rarely happens when $L>> N$. An alternative to avoid these repetitions is to use a regularized resampling, i.e., 
\begin{equation}\label{RegRES}
	\x_{t,n}\sim \sum_{\ell=1}^L \bar{\gamma}_{t,\ell} K(\x-\z_{t,\ell}), \quad \mbox{ for all } \quad n,    
\end{equation}
where the deltas have been replaced by a kernel function $K(\x)$  \cite{Musso01}. The bandwidth of $K(\x)$ can tuned according to some kernel density estimation (KDE) criterion. For the computation of the outer IS weights (see below),  we need to approximate $c_t= \int_\mathcal{X}\widehat{\pi}_t(\x)d\x$ for $t=1,\dots, T$. They are estimated during the inner IS by the corresponding  estimator,
$\widehat{c}_t = \frac{1}{L}\sum_{\ell=1}^L\gamma_{t, \ell}$, for $t=1,\dots,T$. We have $\widehat{c}_t \to c_t$ when $L\to\infty$, by standard IS arguments \cite{Robert04}.

\subsubsection{Adaptation} \label{AdaptSect}

At each iteration, at the end of the inner IS stage, the algorithm performs the adaptation producing $\widehat{\pi}_{t+1}$. 
Specifically, the emulator $\widehat{\pi}_t(\x)$ is improved by incorporating the generated samples at each iteration as additional nodes (see Fig. \ref{fig_adaptation}).  Namely, the additional support points $\{\x_{t,n}\}_{n=1}^N$ to $\mathcal{S}_t$ are obtained by resampling $N$ times within $\{\z_{t,\ell}\}_{\ell=1}^L$ according to the probabilities $\bar{\gamma}_{t,\ell}=\frac{\gamma_{t,\ell}}{\sum_{i=1}^L \gamma_{t,i}}$ for $\ell=1,...,L$. Note that  the probability mass $\bar{\gamma}_{t,\ell}$ is directly proportional to $\widehat{\pi}_t(\z_{t,\ell})$. Therefore, the algorithm tends to add points where $\widehat{\pi}_t$ is higher. Indeed, as $L\to \infty$, the resampled particles are distributed as  $\widehat{\pi}_t$
\cite{Rubin88,smith1992bayesian,givens2012computational}.
%Since, as $t$ grows, $\widehat{\pi}$ approaches the posterior function $\pi$,  then the nodes are incorporated where actually needed, i.e., where the posterior is underestimated. 
If $L$ is not great enough, some $\x_{t,n}$ can be repeated. We do not include these repetitions as support points. Increasing $L$ or using  a regularized resampling as in Eq. \eqref{RegRES} avoids this issue \cite{Musso01}. Note that
the number of support points $J_t=|\mathcal{S}_t|$ increases as $t$ grows.
\newline
All the evaluations of the unnormalized posterior $\pi(\x)$ in the additional nodes are stored in the vector denoted as ${\bm \pi}_{t}$, in order to be used in the outer IS stage. Note also that all  evaluations of  $\pi$ are used to build the emulator.

%%%%%%%%%%%%%%%%%%%%%%%%%%%%%%%%%%%%%%
\subsubsection{Outer IS scheme}
%%%%%%%%%%%%%%%%%%%%%%%%%%%%%%%%%%%%%%%
At the end of the iterative part, we compute the final IS weights $w_{t,n}$, using all the posterior evaluations ${\pi}_{t,n}=\pi(\x_{t,n})$, which are stored in the inner layer. More specifically, we assign to each sample (drawn also in the inner stage) the weight 
\begin{align}\label{eq:OuterISweights}
	w_{t,n}%=\frac{\pi_{t,n}}{\Phi_{t}({\bf x}_{t,n})},%=\frac{\pi_{t,n}}{\frac{1}{t}\sum_{\tau=1}^t\widehat{\pi}_\tau(\x_{t,n})}.
	=\frac{\pi_{t,n}}{\frac{1}{T}\sum_{\tau=1}^T\frac{1}{\widehat{c}_\tau}\widehat{\pi}_\tau(\x_{t,n})}, \qquad \mbox{ for all }  \quad t=1,...,T, \quad n=1,...,N,
\end{align}
where we have employed a deterministic mixture weighting scheme \cite{veach1995optimally, elvira2019generalized}, i.e., the denominator consists of a temporal mixture (e.g., as also suggested in \cite{CORNUET12}). Note that the weights $w_{t,n}$ are not required in the iterative inner layer described above. Hence, they can be computed after the adaptation and sampling steps are finalized.
The output of the algorithm is then formed by all the sets of weighted particles $\{{\bf x}_{t,n},w_{t,n}\}_{n=1}^N$ for $t=1,...,T$, and  the final emulator $\widehat{\pi}_{T+1}(\x)=\widehat{\pi}_{T+1}(\x;\mathcal{S}_T)$.

\begin{table}[!ht]
	%	\centering
	%\small
	\caption{\textbf{Regression-based Adaptive Deep Importance Sampling (RADIS)}}
	\vspace{0.2cm}
	\begin{tabular}{|p{0.95\columnwidth}|}
		\hline
		%\footnotesize
		%\newline
		\vspace{0.1cm}
		{\bf - Initialization:} Choose the initial set $\mathcal{S}_0$ of nodes, and the values $T$, $L$, $N$ (with $L>> N$). Obtain the vector of initial evaluations $\bm{\pi}_0$.
		\newline
		{\bf - For $t=1,\ldots,T$:}
		\begin{enumerate}
			\item  {\bf Emulator construction:} Given the set $\mathcal{S}_{t-1}$  and the corresponding vector of posterior evaluations ${\bm \pi}_{t-1}$, build the proposal function $\widehat{\pi}_t(\x)=\widehat{\pi}_t(\x|\mathcal{S}_{t-1})$ with a non-parametric regression procedure (see Sect. \ref{sec:buildProp}).	
			\item {\bf Inner IS:}
			\begin{enumerate}
				\item {\it IS.} Sample $\{\z_{t, \ell}\}_{\ell=1}^L \sim q_{\text{aux}}(\x)$ and compute the following weights
				\begin{align}
					\gamma_{t, \ell} = \frac{\widehat{\pi}_t(\z_{t, \ell})}{q_{\text{aux}}(\z_{t, \ell})},
				\end{align}		
				for $\ell=1,\dots, L$. 
				\item {\it Resampling.} Resample $\{\x_{t,n}\}_{n=1}^N$ from $\{\z_{t, \ell}\}_{\ell=1}^L$ with probabilities $\{\bar{\gamma}_{t,\ell}\}_{\ell=1}^L$ where $\bar{\gamma}_{t,\ell}=\frac{\gamma_{t,\ell}}{\sum_{i=1}^L \gamma_{t,i}}$ for $\ell=1,...,L$.
				
				\item {\it  Normalizing constant.} Compute
				\begin{align}
					\widehat{c}_t = \frac{1}{L}\sum_{\ell=1}^L\gamma_{t, \ell}.
				\end{align}
				
				%	In the robust version, we will consider $\x_{t,n} \sim \beta q_{\texttt{par}}(\x)+(1-\beta) \widehat{\pi}_t(\x)$,  where $ q_{\texttt{par}}$ is a parametric density.								
			\end{enumerate} 
			\item {\bf Update:}\label{StepUpdate} Evaluate $\pi_{t,n}=\pi(\x_{t,n})$, for all $n=1,...,N$, and update the set of nodes						appending $\mathcal{S}_{t} =\mathcal{S}_{t-1} \cup \{\x_{t,1},...,\x_{t,N} \}$ and ${\bm \pi}_{t} = [{\bm \pi}_{t-1},\pi_{t,1}, ...,\pi_{t,N}]^\top$.	
			%\vspace{0.1cm}
			%	\item 
		\end{enumerate}
		{\bf-  Outer IS:}	 Assign to each sample the weight 
		$$
		w_{t,n}%=\frac{\pi_{t,n}}{\Phi_{t}({\bf x}_{t,n})},%=\frac{\pi_{t,n}}{\frac{1}{t}\sum_{\tau=1}^t\widehat{\pi}_\tau(\x_{t,n})}.
		=\frac{\pi_{t,n}}{\frac{1}{T}\sum_{\tau=1}^T\frac{1}{\widehat{c}_\tau}\widehat{\pi}_\tau(\x_{t,n})}, \qquad \mbox{ for all }  \quad t=1,...,T, \quad n=1,...,N.
		$$
		{\bf - Outputs:} Final emulator $\widehat{\pi}_{T+1}(\x)=\widehat{\pi}_{T+1}(\x|\mathcal{S}_T)$, and the set of weighted particles $\{{\bf x}_{t,n},w_{t,n}\}_{n=1}^N$ for $t=1,...,T$. 
		\\ 
		\hline 
	\end{tabular}
	\label{table_DeepIS_algorithm}
\end{table}

{\rem As $t\rightarrow \infty$ and $L\rightarrow \infty$, then $\widehat{c}_t\rightarrow c_t \rightarrow Z$, i.e., is an approximation of the marginal likelihood. Another estimator of the marginal likelihood $Z$ provided by RADIS is the arithmetic mean of all the outer weights, i.e., 
	$\widehat{Z}=\frac{1}{NT} \sum_{t=1}^T\sum_{n=1}^N w_{t,n}$.
}

\begin{figure*}[!ht]
	\centering
	%	\centerline{
	\includegraphics[width=1\textwidth]{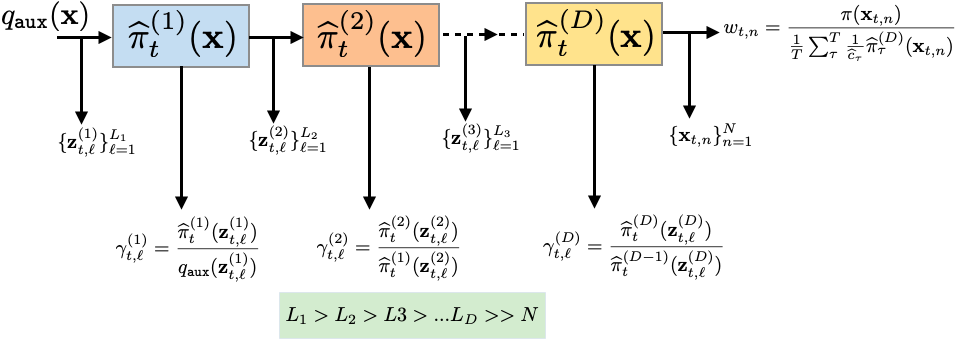}
	%	}
	% 	\vspace{-0.3cm}
	\caption{RADIS with $D+1$ layers in the deep architecture. Different emulator construction can be applied at each stage. In each $d$-th layer, the resampling is applied $L_{d+1}$ times for generating the next cloud of resampled particles $\{{\bf z}_{t,\ell}^{(d+1)}\}_{\ell=1}^{L_{d+1}}$ (with $d=1,...,D+1$). These samples  are used for the adaptation of $\widehat{\pi}_t^{(d)}$ and then are weighted again in the next stage. 
		Note that $L_{D+1}=N$ and $L_d>L_{d+1}$.  }
	\label{DeepIS_fig}
\end{figure*}

{\rem Additional layers can be included in the proposed deep architecture would consists in adapting a chain of several emulators. This is graphically represented in Figure \ref{DeepIS_fig}. One of the advantages of this deep approach with $D+1>2$ layers (where $D$ is the number of inner nested stages), is that different emulator constructions can be jointly applied. Each emulator serves as proposal of the next IS stage. In the additional layers, the evaluation of the posterior (true model) is not required.  In this scenario, RADIS also provides $D$ different emulators. 
	
	%See also the Section \ref{sec:robustSchemes}. 
}

%\clearpage

%%%%%%%%%%%%%%%%%%%%%%%%%%%%%%%%%%%%%%%%%%
\subsection{Construction of $\widehat{\pi}$ by regression} \label{sec:buildProp}
%%%%%%%%%%%%%%%%%%%%%%%%%%%%%%%%%%%%%%%%%%

We consider two different procedures to build the non-parametric proposal: a Gaussian process (GP) model and nearest neighbors (NN) scheme.  
In \ref{App:surrConv}, we show that these constructions converges to the true underlying function as the number of nodes ($J_t=|\mathcal{S}_t|$) grows.
\newline
\newline
\noindent {\bf GP construction.} Let us consider building the surrogate $\widehat{\pi}$ with Gaussian process (GP) regression in the log domain, i.e., over the $\log \pi(\x)$ 
\cite{rasmussen06,Gilks95}. 
GP regression provides with an approximation of a function from a set $\x_1,\dots,\x_{J_t}\in \mathcal{X} \subseteq \mathbb{R}^{d_x}$ (where $\mathcal{X}$ can be unbounded) and their corresponding function evaluation \cite{rasmussen2006gaussian,MartinoRead2020}.
To ensure the non-negativity of the approximation, we fit the GP to $\log \pi$ rather than directly on $\pi$ \cite{osborne2012active}. Let $\phi(\x) \equiv \log \pi(\x)$ and ${\bm \phi} = [\phi_1,\dots,\phi_{J_t}]^\top$ where $\phi_i=\log\pi(\x_i)$ for $i=1,\dots,J_t$. Given a symmetric and positive definite kernel $k(\x,\x')$ and some noise level $\sigma$, under the assumption that $\phi(\x)$ is a zero-mean GP with kernel $k$, the GP regression of $\phi( \x)$ is of the form
\begin{align}\label{GPint}
	\widehat{\phi}_t(\x) = \sum_{i=1}^{J_t}\beta_i k(\x,\x_i),
\end{align}
where the coefficients $\bm{\beta}=[\beta_1,\dots,\beta_{J_t}]^\top$ are given by
\begin{align}
	\bm{\beta} = \left( {\bf K} + \zeta {\bf I} \right)^{-1}\bm{\phi}
\end{align}
with $\left({\bf K}\right)_{i,j} = k(\x_i,\x_j)$ for $1 \leq i,j \leq {J_t}$ and ${\bf I}$ is the ${J_t} \times {J_t}$ identity matrix. 
% The approximation $\widehat{\phi}$ is also called best linear unbiased estimation of $\phi$ in the kriging literature {\cite{kriging}}.
Note that, for $\zeta=0$, $\widehat{\phi}$ corresponds to an interpolator of $\phi$. Note also that the cost of obtaining $\widehat{\phi}$ is $\mathcal{O}({J_t}^3)$ since it requires inverting a ${J_t} \times {J_t}$ matrix. 
As an example, a possible choice of kernel is the Gaussian $k(\x,\x') = \exp\{-\frac{1}{2\epsilon^2}\norm{\x-\x'}^2_2\}$, where the hyperparameter $\epsilon$ can be estimated, e.g., by maximizing the marginal likelihood \cite{rasmussen06}. Finally, the approximation of $\pi$ is given by
\begin{align}\label{GPint2}
	\widehat{\pi}_t(\x) = \exp\{\widehat{\phi}_t(\x)\}.
\end{align}
{ Instead of building on the emulator in the log-domain, a simpler alternative (to ensure non-negativity) consists in setting  $\phi(\x) \equiv  \pi(\x)$, ${\bm \phi} = [\phi_1,\dots,\phi_{J_t}]^\top$ where $\phi_i=\pi(\x_i)$ for $i=1,\dots,J_t$ Then, we set again $\bm{\beta} =[\beta_1,...,\beta_{J_t}] = \left( {\bf K} + \zeta {\bf I} \right)^{-1}\bm{\phi}$ and $\widehat{\phi}_t(\x) = \sum_{i=1}^{J_t}\beta_i k(\x,\x_i)$. The emulator is finally obtained as
	\begin{align}\label{GPint3}
		\widehat{\pi}_t(\x) = \max[\widehat{\phi}_t(\x),0].
\end{align}}
Note that these approximations can be directly applied for unbounded support $\mathcal{X}$. We call the scheme based on these constructions as Gaussian Process Adaptive Importance Sampling  (GP-AIS). 
\newline
\newline
{\bf NN construction.} Given $\x_1,\dots,\x_{J_t} \in \mathcal{X} \subset \mathbb{R}^{d_x}$ (where $\mathcal{X}$ is bounded) and evaluations $\pi(\x_1),\dots,\pi(\x_{J_t})$, the nearest neighbor (NN) interpolator at $\x$ consists of assigning the value of its nearest node. This is equivalent to consider the Voronoi partition $\mathcal{X} = \cup_{i=1}^{J_t} \mathcal{R}_i$, where 
\begin{align}\label{eq:DefVoronoiCells}
	\mathcal{R}_i = \{\x \in \mathcal{X}: \norm{\x - \x_i} < \norm{\x - \x_j}\enskip \text{for} \enskip j\neq i\},
\end{align}
is the $i$-th Voronoi cell.
The NN interpolator of $\pi$ is then given by
\begin{align}\label{NNint}
	\widehat{\pi}_t(\x) = \sum_{i=1}^{J_t} \pi(\x_i)\mathbb{I}_{\mathcal{R}_i}(\x), \qquad \x \in \mathcal{X}. 
\end{align}
where $\mathbb{I}_{\mathcal{R}_i}(\x)$ is the indicator function in $\mathcal{R}_i$. Note that $\widehat{\pi}$ above is an interpolating approximation of $\pi$. 
The NN search has a cost of $\mathcal{O}({J_t})$.  
We denote the scheme based on this construction as Nearest Neighbor Adaptive Importance Sampling (NN-AIS). The regression case consists in considering the $k$ nearest neighbours to $\x$, and taking the arithmetic mean of the values $\pi$ in those $k$ nearest nodes.
%In the next section, we also discuss how to use NN-AIS in an unbounded domain.  

{\rem Note that RADIS employs an  incremental mixture proposal density. Indeed, the emulator $\widehat{\pi}_t(\x)$ in Eqs. \eqref{GPint}-\eqref{GPint2} and \eqref{NNint} can be expressed as a mixture of pdfs where the number of components, $J_t$, increases as $t$ grows. For more details of the NN case, see \ref{AppQueMeGusta}.} 

{\rem Under mild conditions,
	the emulator $\widehat{\pi}_t \to \pi$ and $\widehat{c}_t \to Z$ as $t \to \infty$ (and $L\to \infty$), hence
	$\frac{1}{\widehat{c}_t}\widehat{\pi}_t \to \post$ (see \ref{App:surrConv}).
	Moreover, the SIR scheme to draw from  $\frac{1}{\widehat{c}_t}\widehat{\pi}_t$ is asymptotically exact when $L \to \infty$ (see \ref{App:SIRasymptotics}). Hence, RADIS is drawing samples from $\post$, i.e., it is asymptotically an exact sampler.
}
\newline
\newline
The GP construction provides smoother solutions that can be directly employed in unbounded domains. However, the GP requires the inversion of matrix (with a dimension that increases as the number of nodes grows) and the tuning of the hyperparameters of the kernel function. In contrast, the NN construction does not need any matrix inversion and, if we fix in advance the number $k$ neighbours (for instance in the interpolation case, we have $k=1$) no   hyperparameter tuning is required.

%%%%%%%%%%%%%%%%%%%%%%%%%%%%%%%%%%%%%%%%%%%%%% 
%%%%%%%%%%%%%%%%%%%%%%%%%%%%%%%%%%%%%%%%%%%%%% 
\section{Robust accelerating schemes }\label{sec:robustSchemes}
%%%%%%%%%%%%%%%%%%%%%%%%%%%%%%%%%%%%%%%%%%%%%%
%%%%%%%%%%%%%%%%%%%%%%%%%%%%%%%%%%%%%%%%%%%%%% 

In this section, we present some alternatives in order to {\bf (a)}   reduce the dependence from the initial nodes  and {\bf (b)}  increase the applicability of RADIS, {\bf (c)} speed up the convergence of the emulator covering quickly the state space and finally {\bf (d)} we discuss the computational cost of the proposed overall scheme.
%we introduce a parsimonious construction of the emulator in order to reduce the computational cost
The resulting methods are robust schemes, which can be also employed for extending the use of NN-AIS in unbounded supports. This is achieved combining the non-parametric proposal function $\widehat{\pi}_t(\x)$ with a parametric proposal density, $q_{\texttt{par}}(\x)$. Hence, the complete proposal, denoted as $\varphi_t(\x)$. will be a mixture of densities with a parametric and a non-parametric components.
% make it more robust and generally applicable. (i) We want to have a robust algorithm that is less reliant to a bad initialization, (ii) We propose combining the interpolating proposal with other AIS algorithms
\newline
\newline
{\bf Mixture with parametric proposal.}
%In the early iterations, the performance of the interpolator $\widehat{\pi}_t(\x)$  depends on the initialization. With bad initial nodes, the sampling will concentrate too much in a small region, biasing the estimation and also hurting the adaptation process.
%For this reason, 
%
The use of an additional parametric density $\bar{q}_\texttt{par}(\x)$ can (i) ensure that the complete proposal has fatter tails than target pdf, and (ii) foster the exploration of  important regions that could be initially ignored due to a possible bad initialization.
Thus,  we consider the following mixture as a proposal density in the inner IS layer,
\begin{align}\label{eq:stickyAndParam}
	\varphi_t(\x) = \alpha_t \bar{q}_\texttt{par}(\x) + (1-\alpha_t)\frac{1}{\hat{c}_t}\widehat{\pi}_t(\x),
\end{align}
where $\alpha_t\in [0,1]$ for all $t$, and $\alpha_t$ is a non-increasing function $t$. The idea is to set initially $\alpha_0=\frac{1}{2}$, and then decrease $\alpha_t\rightarrow \alpha_\infty$ as $t\to \infty$ (e.g., we can set $\alpha_\infty=0$).  {
	Note that $\varphi_t(\x)$ must be evaluated in the denominator of the outer layer weights $w_{t,n}$ in \eqref{eq:OuterISweights}, taking the place of $\frac{1}{\widehat{c}_t}\widehat{\pi}_t(\x)$ (see Table \ref{table_DeepIS_algorithm}).
}

%The proposal pdf $\varphi_t(\x)$ is used in the inner IS layer instead of simply the non-parametric density  $\frac{1}{c_t}\widehat{\pi}_t(\x)$.

{\rem Choosing $\bar{q}_\texttt{par}(\x)$ with fatter tails than $\post(x)$, then $\varphi_t(\x)$ has also fatter tails than $\post(x)$. 
	Hence, we avoid the infinite variance issue of the IS weights \cite{Robert04}.  }
\newline
\newline
See also \cite[Section 7.1]{ourRev} for a theoretical and numerical example of the infinite variance problem.
%We could set $\alpha_\infty=0$, but a more robust choice is to keep $\alpha_\infty>0$. 
As an example, if $\mathcal{X}$ is bounded, $\bar{q}_\texttt{par}(\x)$ could be a uniform density over $\mathcal{X}$. If $\mathcal{X}$ is unbounded, $\bar{q}_\texttt{par}(\x)$ can be, e.g., a Gaussian, a Student-t distribution or a mixture of pdfs (see below).

{\rem The fact that $\varphi_t(\x)$ has fatter tails than $\post(x)$ ensures to have a non-zero probability of adding new nodes in {any possible subset of the support} $\mathcal{X}$.}
\newline
\newline
%\noindent
This strategy also allows the use of the NN-AIS in an unbounded support. In \ref{AdaptiveSupportSect} we describe an extension of NN-AIS where the support of the NN approximation is also adapted.
%{\bf Safe proposal with unbounded $\mathcal{X}$ (i.e. $\mathcal{X} = \mathbb{R}^{d_x}$).} We can use a combination of $\bar{q}_\texttt{par}(\x)$ with unbounded support and $\widehat{\pi}_t$ with bounded support. Moreover, the bounded support of $\widehat{\pi}_t$ can be expanded along the simulation. See next section...
%\newline
%\newline
%The choice of $\bar{q}_\texttt{par}(\x)$ is important as well, so it makes sense that we should improve it along the iterations. Indeed, we can use an AIS algorithm to update $\bar{q}_\texttt{par}(\x)$, without increasing the cost in terms of target evaluations. Two-way benefit: interpolating proposal improves the exploitation of AIS algorithm, and the AIS algorithm helps to explore and not miss high-valued posterior regions.
\newline
\newline
{\bf Parametric mixture by other AIS schemes.} A more sophisticated option is to also update  $\bar{q}_\texttt{par}(\x)$ along the iterations. For instance,  $\bar{q}_\texttt{par}(\x)=\frac{1}{C}\sum_{c=1}^C q_c(\x|\bm{\mu}_{t,c}, {\bf \Sigma}_{t,c})$ can be itself a mixture, whose parameters are adapted following another AIS scheme, so that the complete proposal would be
\begin{align}\label{eq:stickyAndParam2}
	\varphi_t(\x) = \alpha_t \left(\frac{1}{C}\sum_{c=1}^C \bar{q}_c(\x|\bm{\mu}_{t,c}, {\bf \Sigma}_{t,c})\right) + (1-\alpha_t)\frac{1}{c_t}\widehat{\pi}_t(\x),
\end{align}
with  $\alpha_t\in [0,1]$ for all $t$.
As an example, the parametric mixture $\bar{q}_\texttt{par}(\x)$ can be obtained following a population Monte Carlo (PMC) method, or a layered adaptive importance sampling (LAIS) technique and/or adaptive multiple importance sampling (AMIS) scheme \cite{bugallo2017adaptive}. 
%In the second example considered in Sect. \ref{sec:num_exp} we show that LAIS can be made more robust by combining it with our scheme. 
The weight $\alpha_t$ is again a non-increasing function of the iteration $t$. 
\newline
{\bf Regression versus interpolation.} In the first iterations of RADIS, the use of $\zeta>0$ in the GP approximation and/or considering the $k$ nearest neighbours (instead only the closest one, $k=1$),  also decreases the dependence on the initial nodes. Namely, reducing the overfitting, at least in the first iteration of RADIS, also increases the robustness of the algorithm.  
\newline
{\bf More layers.} To leverage the benefits  of different emulator constructions in RADIS, one possible strategy is to employ additional layers in the deep architecture, as depicted in Fig. \ref{DeepIS_fig}. For instance, with one additional layer, we could use jointly the GP and the NN constructions. Another possibility is to consider several GP models with different kernel functions or several NN schemes with different $k$.   

{
	\subsection{Computational cost}
	%{{blue} discutir computational cost.... y connectar con la section siguiente... 
	%\newline
	%\newline
	In this section, we discuss computational details of our approach and hypothesize when our approach is convenient also in terms of computational time.  
	%\newline
	It is important to remark that RADIS is useful also for constructing a good emulator (not just for approximating integrals as other Monte Carlo schemes), 
	choosing the nodes in a proper way, similarly in an active learning scheme \cite{llorente2020adaptive,svendsen2020active}. Figures \ref{fig_banana_c} and \ref{fig_super_fer} in the numerical experiments provide a comparison with a random addition of nodes, showing the benefits of the adaptive construction employed in RADIS.  
	% and, for approximating integrals.
	
	RADIS requires $N$ evaluations of the posterior $\pi(\x)$ at each iteration, so that the total number of posterior evaluations is $E=N_0+NT$.
	%For instance, at each iteration in LAIS we run $N$ MCMC algorithms for one step each, hence requiring $N$ posterior evaluations per iteration. 
	Let denote as $C_{\text{eval-post}}$ the cost of evaluating $\pi(\x)$ once, so that the total cost of evaluating the posterior is $E C_{\text{eval-post}}$.  In addition to $E$ posterior evaluations, RADIS  carries out different other tasks, namely (i) evaluate $L$ times the current emulator per iteration, (ii) perform $N$ resampling steps per iteration over $L$ possible samples, and (iii) compute the denominator of thefinal IS weights at the end of the algorithm. Let $C_{\text{eval-emulator}}$, $C_{\text{resampling}}$ and $C_{\text{den-weights}}$ denote the {\it total} costs after $T$ iterations of RADIS, associated to tasks (i)-(iii). In term of computational time,  RADIS can be convenient with respect to other schemes,  when the inequality
	\begin{eqnarray}\label{EqSuperCost}
		C_{\text{eval-post}} > \frac{1}{E}\left(C_{\text{eval-emulator}}+C_{\text{resampling}}+C_{\text{den-weights}}\right),
	\end{eqnarray}
	is fulfilled. For an example, see the numerical experiment in Section \ref{sec_Astro} and the results in Table \ref{table_MSE_astro}.
	Recall that all the values  $C_{\text{eval-post}}$, $C_{\text{eval-emulator}}$, $C_{\text{resampling}}$, and  $C_{\text{den-weights}}$ also depend on the specific implementation and language of the code and the different processors/machines.
	\newline
	Generally, the term $C_{\text{eval-emulator}}$ dominates the other two since it is composed of evaluating $L$ times the emulator for $T$ iterations. Moreover, due to the non-parametric construction and the fact that we increase the set of active nodes in $N$, evaluating the interpolator becomes more costly with the iterations. More specifically, in the NN based approach, after $T$ iterations we have
	$C_{\text{eval-emulator}} \approx\sum_{t=1}^T\mathcal{O}(LNt)
	%=\mathcal{O}(LN{T(T-1)}/{2}) 
	= \mathcal{O}(LNT^2)$.  In the GP-AIS scheme, we have the additional cost of inverting the $J_t\times J_t$ matrix at each iteration (recall that  $J_t = N_0 + N(t-1)$). This cost at each iteration is $ \mathcal{O}(J_t^3) \approx \mathcal{O}(N^3t^3)$, for $t$ big enough. Then, in GP-AIS, $C_{\text{eval-emulator}} \approx \sum_{t=1}^T\mathcal{O}(N^3t^3)+ \mathcal{O}(LNT^2)= \mathcal{O}(N^3T^4)+\mathcal{O}(LNT^2)$. %\approx \mathcal{O}(N^3T^4)$. 
	\newline
	In the next section, we describe different procedures to decrease $C_{\text{eval-emulator}}$.
}

{
	\section{Construction of parsimonious emulators}\label{sec_stickytests}
	So far, we have considered updating the interpolant at each iteration $t$ by adding all the $N$ samples drawn at that iteration. 
	In order to control the computational cost of evaluating the emulator,
	we can design a strategy for accepting or rejecting some of the possible additional nodes. This can be done assigning acceptance probabilities, $p_\text{A}(\x_{t,n})\in[0,1]$, to each of the $N$ samples (in the same fashion of \cite{martino2018adaptive,PARS}). %In practice, we draw an uniform value $u$ in $[0,1]$ and accept the sample $\x_{t,n}$ (i.e. $\x_{t,n}$ is added to set $\mathcal{S}_t$) if $p_\text{A}(\x^{(m)}_t) < u$.
	Therefore, the update part of Step \ref{StepUpdate} in Table \ref{table_DeepIS_algorithm} would be replaced by the routine in Table \ref{table_Update_Sticky}.
	
	\begin{table}[!ht]
		%	\centering
		%\small
		{
			\caption{\textbf{Parsimonious update in Step \ref{StepUpdate}  of Table \ref{table_DeepIS_algorithm}. }} \label{table_Update_Sticky}
			\vspace{0.2cm}
			\begin{tabular}{|p{0.95\columnwidth}|}
				\hline
				%\footnotesize
				%\newline
				\vspace{0.1cm}
				- {\bf Initialization:} Choose an acceptance function $p_A(\x)$, set $\mathcal{S}_{t}=\mathcal{S}_{t-1}$, and consider the cloud of resampled particles $\{\x_{t,n}\}_{n=1}^N$, from the previous step of Table \ref{table_DeepIS_algorithm}.   
				\newline
				{\bf - For $n=1,\ldots,N$:}
				\begin{enumerate}
					\item  Draw $u\sim \mathcal{U}([0,1])$.
					\item If $u\leq p_A(\x_{t,n})$, then set $\mathcal{S}_{t} =\mathcal{S}_{t} \cup \{\x_{t,n} \}$. Otherwise, If $u> p_A(\x_{t,n})$, discard $\x_{t,n}$. 
				\end{enumerate}
				-{\bf Output:} Return $\mathcal{S}_{t}$ and $J_t=|\mathcal{S}_{t}|$.
				\\ 
				\hline 
			\end{tabular}
		}
		
	\end{table}
	%\newline
	%\newline

	\noindent
	{\bf Proper acceptance functions.} We say that an acceptance probability, $p_\text{A}(\x): \mathcal{X}\rightarrow [0,1]$, is {\it proper} if satisfies 
	\begin{equation}\label{SuperCondition}
		\mbox{ {\bf C1:} }  \quad  p_\text{A}(\x) \rightarrow 0,  \quad \mbox{ if } \quad |\pi(\x)-\widehat{\pi}_t(\x)| \rightarrow 0,
	\end{equation}
	for any $\x\in \mathcal{X}$, and 
	\begin{equation}\label{SuperCondition2}
		\mbox{ {\bf C2:} }  \quad p_\text{A}(\x)=0 \mbox{ if and only if }  |\pi(\x)-\widehat{\pi}_t(\x)|=0.
	\end{equation}
	Hence, for any node contained already in $\mathcal{S}_{t-1}$, i.e., ${\bf z}\in \mathcal{S}_{t-1}$, we have $p_\text{A}({\bf z})=0$. For this reason,  as we show below, the acceptance function often depends on the current emulator $\widehat{\pi}_t(\x)$, i.e.,  we should write $p_A(\x)=p_A(\x|\widehat{\pi}_t)$. Hence, a more precise and parsimonious construction would consider a sequential updating of the emulator (since $p_A(\x)$ also should change during the acceptance tests), as shown in Table \ref{table_Update_Sticky2}.
	\begin{table}[!ht]
		%	\centering
		%\small
		{
			\caption{\textbf{Alternative parsimonious update considering a sequential updating of the emulator. }} \label{table_Update_Sticky2}
			\vspace{0.2cm}
			\begin{tabular}{|p{0.95\columnwidth}|}
				\hline
				%\footnotesize
				%\newline
				\vspace{0.1cm}
				- {\bf Initialization:}  Set $\widehat{\pi}_t^{(0)}(\x)=\widehat{\pi}_t(\x)$, choose an acceptance function $p_A^{(0)}(\x)=p_A(\x|\widehat{\pi}_t^{(0)})$, set $k=0$  and $\mathcal{S}_{t}=\mathcal{S}_{t-1}$, and consider the cloud of resampled particles $\{\x_{t,n}\}_{n=1}^N$, from the previous step of Table \ref{table_DeepIS_algorithm}.  Note that, more generally, $p_A^{(k)}(\x)=p_A(\x|\widehat{\pi}_t^{(k)})$ where $k\geq 0$ is an index.  
				\newline
				{\bf - For $n=1,\ldots,N$:}
				\begin{enumerate}
					\item  Draw $u\sim \mathcal{U}([0,1])$.
					\item If $u\leq p_A^{(k)}(\x_{t,n})$, then set $\mathcal{S}_{t} =\mathcal{S}_{t} \cup \{\x_{t,n} \}$, and update the emulator construction $\widehat{\pi}_t^{(k+1)}(\x)$ considering the new set $\mathcal{S}_{t}$. Set also $k \leftarrow k+1$.  %(Otherwise, If $u> p_A(\x_{t,n}|\widehat{\pi}_t^{(k)})$, discard $\x_{t,n}$.) 
				\end{enumerate}
				-{\bf Output:} Return $\mathcal{S}_{t}$, $J_t=|\mathcal{S}_{t}|$  and $\widehat{\pi}_{t+1}(\x)=\widehat{\pi}_{t+1}(\x;\mathcal{S}_{t})=\widehat{\pi}_t^{(k)}(\x)$.
				\\ 
				\hline 
			\end{tabular}
		}
	\end{table}
	
	{\rem Note that the procedures in Table \ref{table_Update_Sticky} and \ref{table_Update_Sticky2} do not require additional evaluations of the target $\pi$, since all the values $\pi(\x_{t,n})$, for all $n$, are already obtained.}
	\newline
	\newline
	The difference between the schemes in Tables  \ref{table_Update_Sticky} and \ref{table_Update_Sticky2}, in term of performance and computational cost, becomes more relevant as $N$ grows. Note that the order of the tests in Table \ref{table_Update_Sticky2} could be also relevant and some strategies for ordering $\{\x_{t,n}\}_{n=1}^N$ (in a suitable way) could be designed.
	Below, we introduce some examples of proper acceptance functions and also some reasonable improper ones.
	%%%%%%%%%%%%%%%%%%%%%%%%%%%%%%%%%%%%%%%%
	%\subsubsection{Sequential updating (or not) during the acceptance tests}
	%%%%%%%%%%%%%%%%%%%%%%%%%%%%%%%%%%%%%%%%
	%---
	%lo podems hacer con dos pequena tablitas...quizas...
	
	%describe the problem....to respect the {\it properness}, we should update the emulator sequentially, during the test......but this increases a bit the cost...alternatevely, we do nothing :).....
	
	%\newline
	%\newline
	%%%%%%%%%%%%%%%%%%%%%%%%%%%%%%%%%%%
	\subsection{Examples of proper acceptance functions}
	%%%%%%%%%%%%%%%%%%%%%%%%%%%%%%%%%%%
	One possibility of proper acceptance function is
	\begin{align}\label{eq:AccProbStickyLuca}
		\textbf{A1:}\quad p_{\text{A}}(\x) = 1 - \frac{\min\{\pi(\x),\widehat{\pi}_t(\x)\}}{\max\{\pi(\x),\widehat{\pi}_t(\x)\}}=\frac{|\pi(\x)-\widehat{\pi}_t(\x)|}{\max\{\pi(\x),\widehat{\pi}_t(\x)\}},
	\end{align}
	where we have used $|\pi(\x)-\widehat{\pi}_t(\x)|=\max\{\pi(\x),\widehat{\pi}_t(\x)\}-\min\{\pi(\x),\widehat{\pi}_t(\x)\}$.
	Another possibility is to consider both the discrepancy between $\pi$ and $\widehat{\pi}_t$, and the distance to the closest node $\s^* \in \mathcal{S}_{t-1}$ to $\x$, i.e.,
	\begin{align}\label{eq:AccProbNew}
		\textbf{A2:}\quad p_\text{A}(\x) = \left(1 - e^{-\alpha |\pi(\x) - \widehat{\pi}_t(\x)|}\right)\left(1 - e^{-\beta \norm{\x - \s^*}}\right), \quad \alpha, \beta \geq 0.
	\end{align}
	%where $\s_*$ denotes the closest node to $\x$. 
	If either $\alpha=0$ or $\beta=0$ (or both), then  $p_\text{A}(\x) = 0$. As $\alpha \to \infty$ and $\beta \to \infty$ grow, then $p_\text{A}(\x) \to 1$.  When $\alpha = \infty$ and $\beta$ is finite, then $p_\text{A}(\x) =1 - e^{-\beta \norm{\x - \s^*}}$ and the acceptance probability is bigger when the point $\x$ is far from its closest node, i.e., we have a space-filling strategy. When $\alpha $ is finite and $\beta\to \infty$, then $p_\text{A}(\x) = 1 - e^{-\alpha |\pi(\x) - \widehat{\pi}_t(\x)|}$,  and the acceptance probability is bigger if there is a large discrepancy between $\pi$ and the interpolant at $\x$. 
	%The constants $\alpha, \beta > 0$ allows us to compare two limiting cases: (i) $\alpha \to \infty$ and $\beta=1$, hence the first term is 1 and the acceptance probability is bigger when the point $\x$ is far from its closest node, i.e., a space-filling strategy, and (ii) $\alpha= 1$ and $\beta \to \infty$, hence the second term is 1 and the acceptance probability is bigger if there is a large discrepancy between $\pi$ and the interpolant at $\x$. 
	Thus, unlike in \eqref{eq:AccProbStickyLuca},  in \eqref{eq:AccProbNew} we should tune the values $\alpha$, and $\beta$ according to the computational budget we have, or according to the trade-off between computational effort and performance. 
	\newline
	Note that, in the acceptance functions above, we have $p_\text{A}(\x)\in[0,1]$ for all $\x$, and the condition \eqref{SuperCondition} is fulfilled. Moreover, these acceptance functions depend only on $\x$, $\pi(\x)$ and  $\widehat{\pi}_t(\x)$. The decision is done considering  the quality of the approximation of $\widehat{\pi}_t(\x)$ and, in Eq. \eqref{eq:AccProbNew},  the relative position of $\x$ with respect to the nodes in $\mathcal{S}_{t-1}$. They do not depend on the rest of $N-1$ possible nodes within $\{\x_{t,n}\}_{n=1}^N$ to be tested. Nevertheless, if we use the sequential updating scheme of Table \ref{table_Update_Sticky2}, the acceptance probability will change depending on the order in which we test the candidate nodes.
	\newline
	An example of proper acceptance function depending on the population of candidate nodes is described next. Let us define  $R(\x)=|\pi(\x)-\widehat{\pi}_t(\x)|$. 
	%Note that the discrepancies $|\pi(\x)-\widehat{\pi}_t(\x)|$ where $\widehat{\pi}_t(\x)$ is small, are amplified (i.e., take more importance, having a greater $R(\x)$). 
	Considering $\x\in\{\x_{t,1},...,\x_{t,N}\}$ (i.e., one point within the set of possible nodes to be included) and defining $R_{\texttt{max}}=\max\limits_{\x\in \{\x_{t,1},...,\x_{t,N}\}} R(\x)$,  we can set
	\begin{equation}\label{eq_luca2}
		\textbf{A3:}\quad p_\text{A}(\x) =\frac{R(\x)}{R_{\texttt{max}}}, \quad\mbox{ with } \quad \x \in \{\x_{t,n}\}_{n=1}^N.
	\end{equation}
	Again $p_\text{A}(\x)\in[0,1]$ and the condition \eqref{SuperCondition} is satisfied.  Note that a normalization of $R(\x)$ using $\sum_{n=1}^N R(\x_{t,n})$ instead of $R_{\texttt{max}}$ would produce very small acceptance probabilities as $N$ grows (note that $R(\x) \geq 0$ for all $\x$). This is a non beneficial effect in our opinion, since the decrease of $p_\text{A}(\x)$ is not due to a good quality of the approximation $\widehat{\pi}_t$, but is generated by the increase of the possible alternative denominator $\sum_{n=1}^N R(\x_{t,n})$. %However, 
	%a probability of t
	Resampling schemes could be also employed but provide improper acceptance functions, as we discuss below.

	%%%%%%%%%%%%%%%%%%%%%%%%%%%%%%%%%%%%%%%%%%%
	\subsection{Examples of improper acceptance functions}\label{sec_improper_acc}
	%%%%%%%%%%%%%%%%%%%%%%%%%%%%%%%%%%%%%%%%%%%
	
	Let us define the auxiliary weights $\rho(\x)=\frac{F(\x)}{\widehat{\pi}_t(\x)}$ where $F(\x)$ is function  that can chosen in different ways, $F(\x)=\pi(\x)$, $F(\x)=|\pi(\x)- \widehat{\pi}_t(\x)|$ or $F(\x)=|\pi(\x)- \widehat{\pi}_t(\x)|\widehat{\pi}_t(\x)$, for instance.
	The nodes to be included are then selected resampling $N$ times within the set $\{\x_{t,n}\}_{n=1}^N$ according to the following probability mass,
	$$
	\bar{\rho}(\x_{t,i})=\frac{\rho(\x_{t,i})}{\sum_{n=1}^N \rho(\x_{t,n})}, \quad i=1,...,N,
	$$
	and taking only the {\it unique} values (i.e., without repetitions). Table \ref{table_Update_Sticky3} summarizes this idea.
	
	\begin{table}[!h]
		{
			%	\centering
			%\small
			\caption{\textbf{Parsimonious update in Step \ref{StepUpdate}  of Table \ref{table_DeepIS_algorithm} based on resampling. }} \label{table_Update_Sticky3}
			\vspace{0.2cm}
			\begin{tabular}{|p{0.95\columnwidth}|}
				\hline
				%\footnotesize
				%\newline
				\vspace{0.1cm}
				- {\bf Initialization:} Choose a numerator function  $F(\x)$ (e.g., $F(\x)=\pi(\x)$ or $F(\x)=|\pi(\x)- \widehat{\pi}_t(\x)|$) for the weight $\rho(\x)=\frac{F(\x)}{\widehat{\pi}_t(\x)}$. Set $\mathcal{S}_{t}=\mathcal{S}_{t-1}$, and consider the cloud of resampled particles $\{\x_{t,n}\}_{n=1}^N$, from the previous step of Table \ref{table_DeepIS_algorithm}. Then:   
				%\newline
				%{\bf - For $n=1,\ldots,N$:}
				\begin{enumerate}
					\item  Resample $N$ times within $\{\x_{t,n}\}_{n=1}^N$ according to the probability mass defined as
					$$
					\bar{\rho}_{t,i}=\bar{\rho}(\x_{t,i})=\frac{\rho(\x_{t,i})}{\sum_{n=1}^N \rho(\x_{t,n})}, \quad i=1,...,N,
					$$
					obtaining the new set $\{\widetilde{\x}_{t,n}\}_{n=1}^N$.
					\item Take the unique values in $\{\widetilde{\x}_{t,n}\}_{n=1}^N$ (i.e., removing the repetitions) obtaining $\{{\bf v}_{t,k}\}_{k=1}^K$ (where $K$ is the number of unique values in $\{\widetilde{\x}_{t,n}\}_{n=1}^N$).
					\item Set $\mathcal{S}_{t} =\mathcal{S}_{t} \cup \{{\bf v}_{t,1},...,{\bf v}_{t,K}\}$.
					%	\item If $u\leq p_A(\x_{t,n})$, then set $\mathcal{S}_{t} =\mathcal{S}_{t} \cup \{\x_{t,n} \}$. Otherwise, If $u> p_A(\x_{t,n})$, discard $\x_{t,n}$. 
				\end{enumerate}
				-{\bf Output:} Return $\mathcal{S}_{t}$ and $J_t=|\mathcal{S}_{t}|$.
				\\ 
				\hline 
			\end{tabular}
		}
	\end{table}

	The acceptance probability is, in this case,
	\begin{align}\label{Super_Fernando}
		p_A(\x_{t,i})= 1 - (1-\bar{\rho}(\x_{t,i}))^N.
	\end{align}
	Thus, the procedure in Table \ref{table_Update_Sticky3} is equivalent (in term of number of added nodes) to apply the procedure in Table \ref{table_Update_Sticky} and $p_A(\x)$ in \eqref{Super_Fernando} above.
	Observe also that, with these schemes, even in the ideal case $\widehat{\pi}_t(\x)= \pi(\x)$ for all $\x$, we always add at least one node to the new sets $\mathcal{S}_t$ (i.e., $K\geq 1$). This is due to the {\it improperness} of the acceptance functions. Then, these resampling-based schemes could possibly yield less parsimonious emulators. %{{blue}Moreover, the emulator is not updated....}
	Nevertheless, they are easy to implement and their implementation is computationally faster than the rest of approaches, described previously.  % can provide also good performance, as we show by simulations in Section \ref{}. 
	Starting from the samples ${\bf z}_{t,\ell}\sim q_{\text{aux}}(\x)$ in RADIS, the added points $\{{\bf v}_{t,k}\}_{k=1}^K$ in Table \ref{table_Update_Sticky3}  are then obtained as results of two resampling procedures and finally considering the unique values: 
	$$
	\{{\bf z}_{t,\ell}\}_{\ell=1}^L \xrightarrow[]{{\bar\gamma}_{t,\ell}}  \{\x_{t,n}\}_{n=1}^N   \xrightarrow[]{{\bar\rho}_{t,\ell}}  \{{\bf v}_{t,k}\}_{k=1}^K.
	$$
	In the vanilla version of RADIS, the nodes are obtained applying just the first resampling at each iteration.
	Another example of {\it improper} acceptance function that is not based on a resampling procedure (and does not take into account all the population $\{\x_{t,n}\}_{n=1}^N$, jointly)  is
	\begin{gather}
		p_\text{A}(\x) =\left\{
		\begin{split}
			&1 \quad \mbox{ if } \quad   |\pi(\x) - \widehat{\pi}_t(\x)| >\epsilon,\\
			&0\quad \mbox{ if }   \quad  |\pi(\x) - \widehat{\pi}_t(\x)| \leq \epsilon,\\ 
		\end{split}
		\right. \quad \mbox{ for } \quad  \epsilon\geq 0.
	\end{gather}
	Note that for a finite positive value of $\epsilon >0$, after some iterations, possibly we will have $p_\text{A}(\x)=0$, i.e.,  the adaptation of the emulator is stopped. This is the reason of its improperness, since it does not fulfill C2. If $\epsilon=0$, then we always have $p_\text{A}(\x) =1$, adding all the nodes. If $\epsilon=\infty$,   we have always $p_\text{A}(\x) =0$, and we never update the emulator. With a suitable choice of $\epsilon$ (tuned according to computational budget available), this acceptance function can be also a good option. A numerical comparison among these acceptance probabilities is given in Section \ref{sec:num_exp}.

}

%%%%%%%%%%%%%%%%%%%%%%%%%%%%%%%%%%%%%%%%%%%%%%   
%%%%%%%%%%%%%%%%%%%%%%%%%%%%%%%%%%%%%%%%%%%%%% 
%%%%%%%%%%%%%%%%%%%%%%%%%%%%%%%%%%%%%%%%%%%%%% 
\section{RADIS for model emulation and sequential inversion }
%%%%%%%%%%%%%%%%%%%%%%%%%%%%%%%%%%%%%%%%%%%%%% 
%%%%%%%%%%%%%%%%%%%%%%%%%%%%%%%%%%%%%%%%%%%%%% 
%%%%%%%%%%%%%%%%%%%%%%%%%%%%%%%%%%%%%%%%%%%%%%  
In this section, we describe the application of RADIS to solve Bayesian inverse problems.
We have already considered  the case of obtaining a surrogate function for the (unnormalized) density $\pi$ (or $\log \pi$).
We here focus on inverse inference problems where  our aim is also to obtain an emulator of the costly forward model.
More specifically, let us consider a generic Bayesian inversion problem
\begin{equation}
	\y = {\bf h}(\x) + {\bf v}.
\end{equation}
where ${\bf h}(\x): \mathbb{R}^{d_x}\rightarrow \mathbb{R}^{d_y}$ represents a non-linear mapping defining a physical or mechanistic model (e.g. a complex energy transfer model, a climate model subcomponent integrating subgrid physical processes, or a set of differential equations describing a chemical diffusion process) and ${\bf v}$ has a multivariate Gaussian pdf (e.g., with zero mean and a diagonal covariance matrix with $\sigma^2$ in the diagonal). 
Considering a prior $g(\x)$ over $\x$, the posterior   is 
$$ \post(\x) \propto \pi(\x)=\exp\left(-\frac{1}{2\sigma^2}\|\y - {\bf h}(\x)\|^2\right)g(\x),$$
which can be costly to evaluate if ${\bf h}(\x)$ is a complex model.
In this setting, it is often required to build an emulator of the physical model ${\bf h}(\x)$ instead of 
a surrogate function for the pdf $\pi$ \cite{OHagan12,Kennedy2001,CampsValls19nsr,Daniel2020}. However, we can build $\widehat{{\bf h}}(\x)$  using the same  procedures in Sect. \ref{sec:buildProp}, and then obtain
$$ \widehat{\pi}(\x) = \exp\left(-\frac{1}{2\sigma^2}\|\y - {\bf\widehat{ h}}(\x)\|^2\right)g(\x),
$$
which can be employed as proposal in our scheme. Hence, in this case, we obtain two emulators: $\widehat{{\bf h}}(\x)$ of the physical model, and $\widehat{\pi}(\x) $ of the posterior.

\begin{figure*}[!t]
	\centering
	%	\centerline{
	\subfigure[]{ \includegraphics[width=0.4\textwidth]{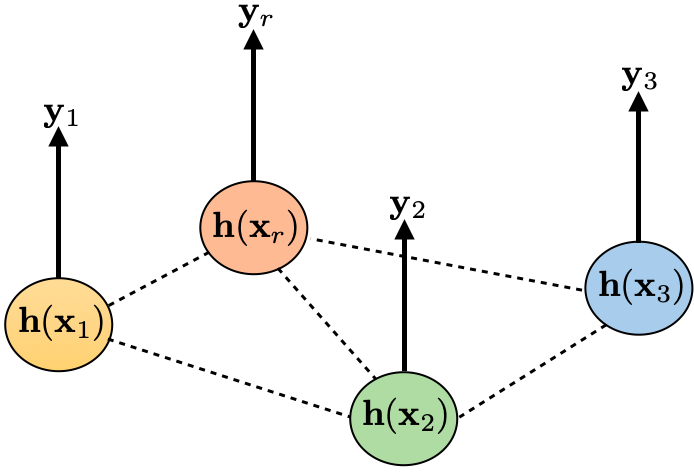}}
	\hspace{0.7cm}
	\subfigure[]{ \includegraphics[width=0.4\textwidth]{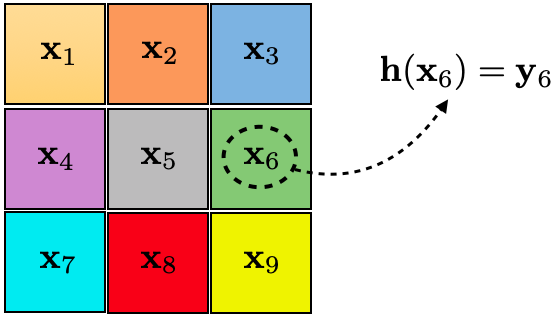}}
	%	}
	% 	\vspace{-0.3cm}
	\caption{{\bf (a)} Different inversion problems related to each other involving the same underlying physical model ${\bf h}(\cdot)$. Their relationships are represented by (dashed lines) edges between the nodes. {\bf (b)} Example of network in an image, where each pixel represents a node of the network. This is the scenario in remote sensing image processing, where $\x_i$ represents the physical state parameters to infer from a set of acquired (or simulated) spectra $\y_i$ (in this figure, we consider noise-free observations).  }
	\label{fig_Section5}
\end{figure*}

In many real-world applications, we have a sequence of inverse problems
\begin{equation}\label{ObsModelHere}
	\y_r = {\bf h}(\x_r) + {\bf v}_r,\quad r=1,\dots,R,
\end{equation}
where $R$ denotes the number of observation nodes in the network, but the physical model ${\bf h}$ is the same for all nodes. See an illustrative example in Fig. \ref{fig_Section5}(a).
The underlying graph represents different features and may have different statistical meanings. Moreover, it can contain prior information directly given in the specific problem. As an example, consider the case of an image where each pixel is represented as a node in the network, see Fig. \ref{fig_Section5}(b), and the goal is to retrieve a set of parameters $\x$ from the observed or simulated pixels $\y$. This is the standard scenario in remote sensing applications, where the observations $\y$ are very high dimensional (depending on the sensory system and satellite platform ranging from a few spectral channels to even thousands) and the set of parameters $\x$ describe the physical characteristics of each particular observation (e.g. leaf or canopy structure, observation characteristics, vegetation health and status, etc). In other settings the graph must be also inferred, i.e., the connections should be learned as well. A simple strategy is to consider the strength of the link is proportional to $\exp\left(-\|\y_r-\y_j\|\right)$, for instance. Other more sophisticated procedures can be also employed \cite{Rabbat19}. Given Eq. \eqref{ObsModelHere}, a piece of the likelihood function is 
$$
p(\y_r|\x_r) \propto  \exp\left(-\frac{1}{2\sigma^2}\|\y_r - {\bf h}(\x_r)\|^2\right), \quad r=1,\dots,R.
$$
Note that the observation model ${\bf h}(\cdot)$ is shared in all the $R$ nodes. The complete likelihood function is
$p(\y_{1:r}|\x_{1:r})=p(\y_{1},...,\y_R|\x_{1},...,\x_R) =\prod_{i=1}^R  p(\y_r|\x_r)$.
%This a typical case in parameter retrieval in remote sensing- paragraph - or two sentences}
%PARA GUS...hablar de RTM... citar o ayudarse con Fig. \ref{fig_Section5}(b)
%}
A complete Bayesian analysis can be considered in this scenario, implementing also RADIS within a particle filter for an efficient inference. However, it is out of the scope of this work and we leave it as a future research line.

%%%%%%%%%%%%%%%%%%%%%%%%%%%%%%%%%%%%%%%%%%%
\begin{figure*}[!ht]
	\centering
	\centerline{
		% 	\subfigure[{QUITAR} ]{\includegraphics[width=0.3\textwidth]{T_0}}
		%	\hspace{-0.70cm}
		\subfigure[]{\includegraphics[width=0.4\textwidth]{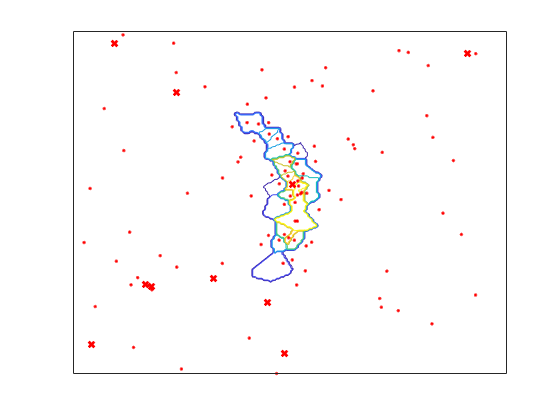}}
		\hspace{-0.7cm}
		\subfigure[]{\includegraphics[width=0.4\textwidth]{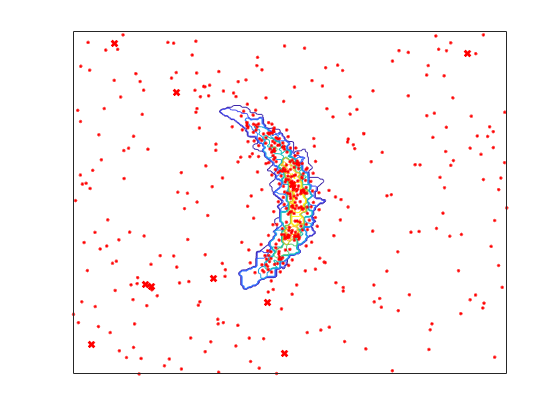}}
		\hspace{-0.7cm}
		\subfigure[]{\includegraphics[width=0.4\textwidth]{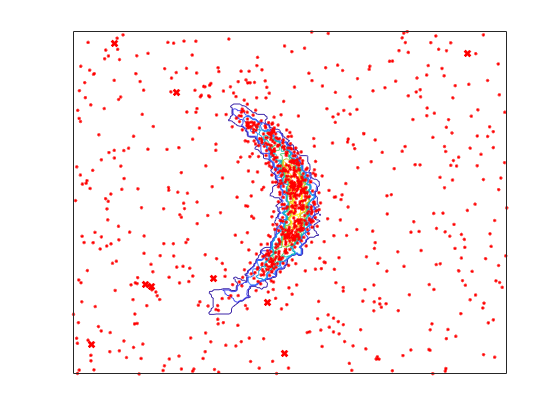}}
	}
	\vspace{-0.3cm}
	\caption{\footnotesize Evolution of $\widehat{\pi}_t$ from NN-AIS$+$U through iterations {\bf (a)} $t=10$, {\bf (b)} $t=50$, and {\bf (c)} $t=100$.
	}
	\label{fig_banana_emulators}
	% El script esta en C:\Users\Fernando\Dropbox\WITH_FERNANDO_LLORENTE\MATLAB\Adapt_Sticky_IS\VISION DOBLE IS\PINTA_BANANAS.m
\end{figure*}
%%%%%%%%%%%%%%%%%%%%%%%%%%%%%%%%%%%%%%%%%%%

%%%%%%%%%%%%%%%%%%%%%%%%%%
%%%%%%%%%%%%%%%%%%%%%%%%%%
%%%%%%%%%%%%%%%%%%%%%%%%%%
\section{Numerical experiments}\label{sec:num_exp}
%%%%%%%%%%%%%%%%%%%%%%%%%%
%%%%%%%%%%%%%%%%%%%%%%%%%%\\
%%%%%%%%%%%%%%%%%%%%%%%%%%\\
In this section, we provide several numerical tests in order to show the performance of the proposed scheme and compare them with benchmark approaches in the literature. 
The first example corresponds to a nonlinear banana shaped density in dimension $d_x=2$, where we compare NN-AIS against standard IS algorithms. 
The second test is a multimodal scenario with dimension $d_x=10$, where we test the combination of an AIS algorithm with NN-AIS against other AIS.
{ An application to an astronomical model is also given, where we provide a comparison in terms of computation time.}
Finally, we consider an application to remote sensing, specifically, we test our scheme in multiple bayesian inversions of PROSAIL.
% 

%%%%%%%%%%%%%%%%%%%%%%%%%%
\subsection{Toy example 1: banana-shaped density}\label{sec:banana}
%%%%%%%%%%%%%%%%%%%%%%%%%%
We consider a banana shaped target pdf,
\begin{align}\label{eq:BananaTarget}
	\post(\x) \propto \exp \left( -\frac{(\eta_1-Bx_1-x_2^2)^2}{2\eta_0^2}  - \sum_{i=1}^{d_x} \frac{x_i^2}{2\eta_{i}^2}\right),
\end{align}
with $B=4$, $\eta_0=4$ and $\eta_i= 3.5$ for $i=1,...,d_x$, where $\mathcal{X} = [-10,10]\times[-10,10]$, i.e., bounded domain. We consider $d_x = 2$ and compute in advance $Z$ and the mean of the target (i.e., the groundtruth) by using a costly grid, so that we can check the performance of the different techniques.

%%%%%%%%%%%%%%%%%%%%%%%%%%%%%%%%%%%%%%%%%%%
\begin{figure*}[!t]
	\centering
	%	\centerline{
	\subfigure[RMSE for Z]{\includegraphics[width=0.4\textwidth]{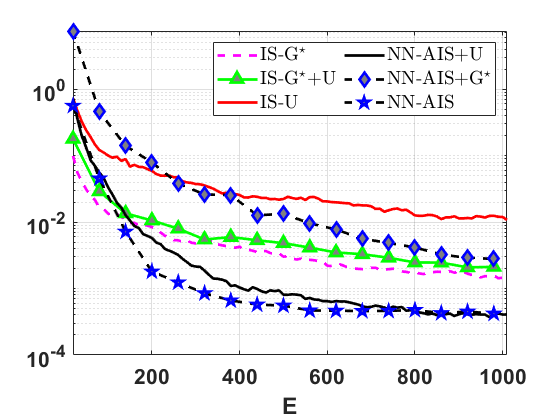}}
	\hspace{-0.3cm}
	\subfigure[RMSE for $\bm{\mu}$]{\includegraphics[width=0.4\textwidth]{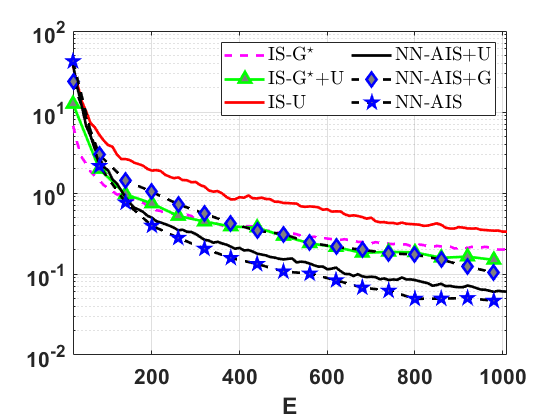}}
	%	\hspace{-0.4cm}
	\subfigure[RMSE for $\bm{\mu}$]{\includegraphics[width=0.4\textwidth]{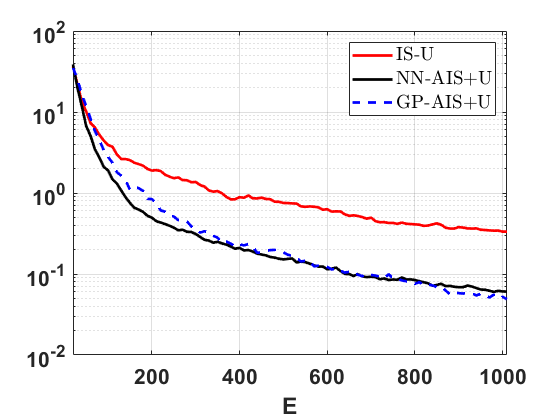}}
	\subfigure[$L_2$ distance between $\widehat{\pi}_t$ and $\pi$ versus $t$]{\includegraphics[width=0.4\textwidth]{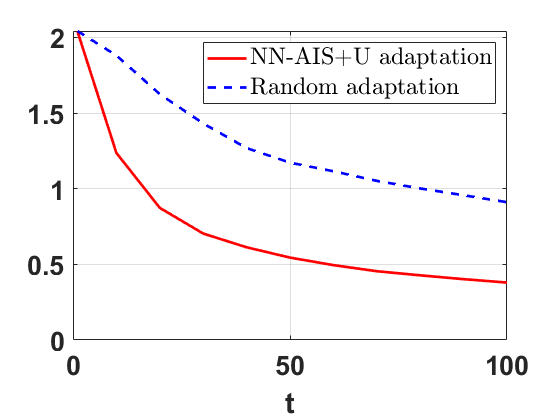} \label{fig_banana_c}}
	%	}
	\vspace{-0.3cm}
	\caption{\footnotesize 
		{\bf (a)} RMSE in log-scale for $Z$ as function of evaluations $E$. {\bf (b)} RMSE in log-scale for $\bm{\mu}$ as function of $E$. 
		{{\bf (c)}  RMSE of GP-AIS+U in log-scale for $\bm{\mu}$ as function of $E$.}
		{\bf (d)} $L_2$ distance between $\pi$ and  $\widehat{\pi}_t$ when the nodes are adaptively obtained by NN-AIS+U (in solid line), and when the nodes are random and uniformly chosen in the domain (in dashed line), as a function of $t$.}
	\label{fig_banana}
	% El script esta en C:\Users\Fernando\Dropbox\WITH_FERNANDO_LLORENTE\MATLAB\Adapt_Sticky_IS\REUNE_results.m
	% Para la otra figura, en C:\Users\Fernando\Dropbox\WITH_FERNANDO_LLORENTE\MATLAB\Adapt_Sticky_IS\para_L2_dist
\end{figure*}
%%%%%%%%%%%%%%%%%%%%%%%%%%%%%%%%%%%%%%%%%%%
\subsubsection{Estimating $Z$ and $\mu$}
\begin{figure*}[!t]
	\centering
	%	\centerline{
	\subfigure[]{\includegraphics[width=0.4\textwidth]{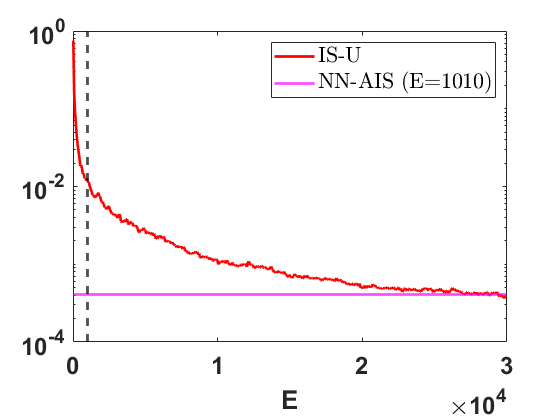}}
	\subfigure[]{\includegraphics[width=0.4\textwidth]{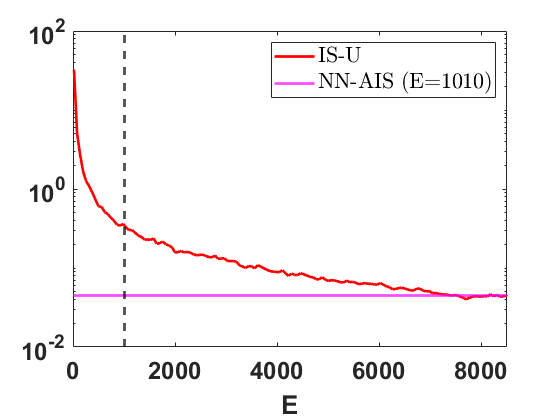}}
	%	}
	\vspace{-0.3cm}
	\caption{\footnotesize We show the number of additional evaluations required by IS-U to achieve the same RMSE than NN-AIS with $E=1010$ in {\bf (a)} the estimation of $Z$, and  {\bf (b)} the estimation of $\bm{\mu}$. The red line represents the RMSE of IS-U as a function of $E$, while the horizontal line is the RMSE achieved by NN-AIS with $E=1010$. The vertical dash line is at $E=1010$.}
	\label{fig_mismoErrorUnif}
	% El script esta en C:\Users\Fernando\Dropbox\WITH_FERNANDO_LLORENTE\MATLAB\Adapt_Sticky_IS\OLD_scripts_and_SIMUS
\end{figure*}

We aim to estimate $Z = 7.9976$ and $\bm{\mu}=[-0.4841, 0]$ with NN-AIS and compare it, in terms of relative mean squared error (RMSE), with different IS algorithms considering the same number of target evaluations. The results are averaged over 500 independent simulations. The goal is to investigate the performance of NN-AIS as compared to other parametric IS algorithms that consider a  proposal, well designed in advance. 
We set $T=100$ and $N=10$, and use $10$ starting nodes (random chosen in the domain) to build $\widehat{\pi}_1(\x|\mathcal{S}_0)$. 
%We consider building $\widehat{\pi}_t$ by nearest neighbors (NN). 
With the selected values of $T$ and $N$ the total budget of target evaluations is $E = 10 + NT = 1010$. 
\newline
\newline
{\bf Methods.} 
We consider three variants of NN-AIS to illustrate three different scenarios: in the first one (denoted as {\bf NN-AIS}) initial nodes uniform in $[-10,10]\times[-10,10]$, i.e. good initialization, without $\bar{q}_\texttt{par}(\x)$; ({\bf NN-AIS$+$U}) same initialization with $\bar{q}_\texttt{par}(\x) = \frac{1}{|\mathcal{X}|}$, i.e. good initialization and with a  good choice of $\bar{q}_\texttt{par}(\x)$; ({\bf NN-AIS$+$G}) initial nodes are uniform in $[5,10]\times[5,10]$ with Gaussian $\bar{q}_\texttt{par}(\x)=\mathcal{N}(\x|[2,2]^\top,3^2{\bf I}_2)$, i.e., a bad initialization with a bad choice of the parametric proposal $\bar{q}_\texttt{par}(\x)$. In all cases, we consider a fixed value of $\alpha_t=\frac{1}{2}$. 	
\newline	
Furthermore, we compare the NN-AIS schemes with three alternative IS methods: ({\bf IS-U}) with uniform proposal in $\mathcal{X}$, which is very good choice of proposal in this problem;  ({\bf IS-G$^\star$}) with Gaussian proposal matching the moments of $\post(\x)$, i.e.,  the optimal Gaussian proposal; ({\bf IS-G$^\star$+U}) with a proposal which is an equally weighted mixture of the two previous cases. {In addition, we also test our algorithm using GPs, denoted GP-AIS+U.}
\newline
\newline
{\bf Discussion.} 
As shown in Figures \ref{fig_banana}(a)-(b),  NN-AIS and NN-AIS+U outperform the rest. NN-AIS performs a bit better than NN-AIS+U: the use of a parametric proposal is safer but entails a loss of performance, trading off exploitation for exploration.
%\newline
In Figure \ref{fig_banana}(a),  NN-AIS+G shows worse performance in estimating $Z$ in the early iterations as a consequence of the bad initialization and bad parametric proposal. However, it quickly improves and start performing as good as IS-G$^\star$ and IS-G$^\star$+U.
%\newline	
In Figure \ref{fig_banana}(b), regarding the estimation of $\bm{\mu}$, our methods perform better than alternative IS algorithms. {Figure \ref{fig_banana}(c) shows that GP-AIS+U provides similar performance than NN-AIS+U.}
%\newline 
Overall, this simple experiment shows the range of performance of our method: it is best if we use only our method, provided that we have a good initialization; adding a good parametric proposal is safer if we do not trust our initialization, showing just a small loss of performance w.r.t. the first scenario. In the case both the initialization and parametric proposal are wrongly chosen, our method is able to achieve good results and recover quickly from a bad initialization. 
\newline
\newline
{\bf Additional comparison.} we have run IS-U for $E>1010$ until it reached the same error in estimation achieved by NN-AIS. The results are depicted in Figures \ref{fig_mismoErrorUnif}. Specifically, in Figure \ref{fig_mismoErrorUnif}(a) we see that around 29000 more evaluations are needed to obtain the same error in estimating $Z$, and Figure \ref{fig_mismoErrorUnif}(b) shows that around 7000 more evaluations to obtain the same error in estimating $\bm{\mu}$.

\subsubsection{Convergence of $\widehat{\pi}_t$ to $\pi$}
The convergence of $\widehat{\pi}_t$ to $\pi$ depends on the fact that nodes should fill the space enough (see \ref{APP:theo}). However, some filling strategies yield a faster convergence than others. In our simulations, we aim to show that the construction provided by NN-AIS+U converges faster than another construction using nodes random and uniformly chosen in the domain $\mathcal{X}$.
Figures \ref{fig_banana_emulators} and \ref{fig_banana}(d) show that the approximation $\widehat{\pi}_t$ obtained by NN-AIS$+$U is indeed converging to $\pi$ as $t$ increases.  In Figure \ref{fig_banana}(c), we show the $L_2$ distance between $\pi$ and  $\widehat{\pi}_t$ with random nodes (in dashed line), and  by NN-AIS+U (in solid line), along with the number of iterations $t$. 
As shown in Figure \ref{fig_banana}(d), the $\widehat{\pi}_t$ gets more rapidly closer to $\pi$ in $L_2$ when the nodes are sampled from NN-AIS+U rather than only adding random points, uniformly over the domain.

{
	\subsubsection{Comparing NN-AIS+U with different values of $L$}
	
	In our proposed approach, we need to evaluate $L$ times the approximation $\widehat{\pi}_t$ at each iteration. 
	The computation cost of the algorithm thus scales with $L$, which needs to be big enough (and bigger than $N$) so that the resampling step and the estimation of $c_t$ are accurate.
	Here, we investigate the performance of NN-AIS+U for several values $L \in \{5000,10000,25000,50000\}$. 
	As expected,
	Figure \ref{fig_varias_L} shows that the performance of the algorithm deteriorates as we lower the value of $L$.  However, note that all NN-AIS scheme with the considered $L$ perform better compared to standard IS with uniform proposal.

	\begin{figure*}[!h]
		{
			\centering
			%	\centerline{
			\subfigure[]{\includegraphics[width=0.4\textwidth]{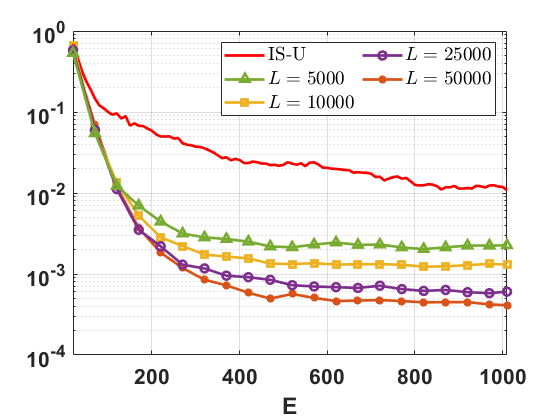}}
			\subfigure[]{\includegraphics[width=0.4\textwidth]{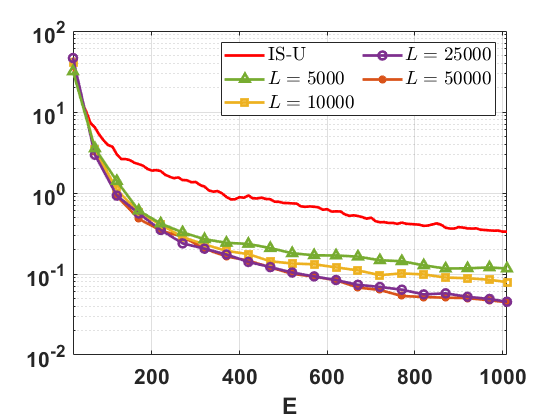}}
			%	}
			\vspace{-0.3cm}
			\caption{\footnotesize  Performance of NN-AIS+U with different choices of $L \in \{1000,5000,10000,25000,50000\}$ in {\bf (a)} the estimation of $Z$, and  {\bf (b)} the estimation of $\bm{\mu}$. The red curve represents the RMSE of IS-U as a function of $E$.}
			\label{fig_varias_L}
			% El script esta en C:\Users\Fernando\Dropbox\WITH_FERNANDO_LLORENTE\MATLAB\Adapt_Sticky_IS\Saca_results_diferentes_L
		}
	\end{figure*}
	
	\subsubsection{Results of the parsimonious constructions}

	In the vanilla version of RADIS, the approximation $\widehat{\pi}_t$ is refined by adding the $N$ samples drawn at iteration $t$ to the set of active nodes. Since we consider non-parametric approximations, this implies that $\widehat{\pi}_t$ becomes more complex, i.e. more costly to evaluate, as $t$ grows. In Sect. \ref{sec_stickytests}, we showed means of controlling the complexity of $\widehat{\pi}_t$ by the computation of acceptance probabilities: instead of adding all the samples, the $n$-th sample is added with certain probability. Here, we test the application of several acceptance probabilities to NN-AIS+U and compare the performance with respect to NN-AIS+U that accepts all nodes. We also examine the complexity, in terms of number of nodes, of the final emulator. Specifically, we consider the acceptance functions \textbf{A1} in Eq. \eqref{eq:AccProbStickyLuca}, \textbf{A2} in Eq. \eqref{eq:AccProbNew} and \textbf{A3} in Eq. \eqref{eq_luca2}. We also test three variants of the improper acceptance function in Sect. \ref{sec_improper_acc}, namely $F(\x)=\pi(\x)$, $F(\x)=|\pi(\x)-\widehat{\pi}_t(\x)|$ and $F(\x)=|\pi(\x)-\widehat{\pi}_t(\x)|\widehat{\pi}_t(\x)$. The results are given in Figures \ref{fig_Z_varias_acc}, Figure \ref{fig_numero_nodos} and Figure \ref{fig_super_fer}.  Note that NN-AIS+U (ALL) represents the vanilla version NN-AIS+U in Table \ref{table_DeepIS_algorithm}, adding all the nodes at the Step \ref{StepUpdate}.  
	%\newline
	%\newline
	\begin{figure*}[!h]
		{
			\centering
			\centerline{
				\subfigure[]{\includegraphics[width=0.35\textwidth]{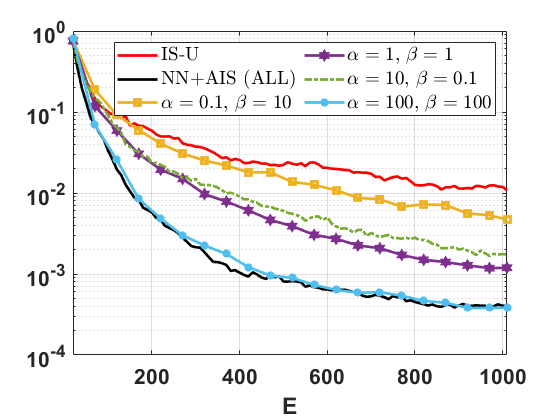}}
				\subfigure[]{\includegraphics[width=0.35\textwidth]{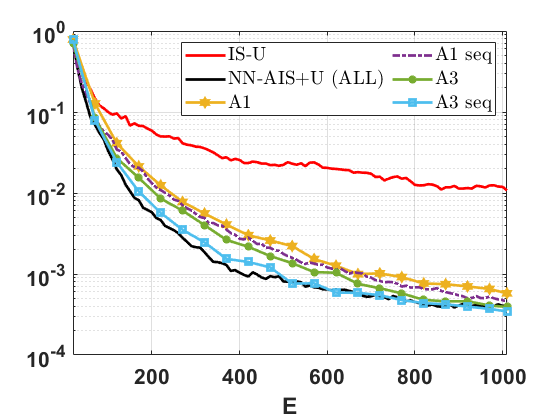}}
				\subfigure[]{\includegraphics[width=0.35\textwidth]{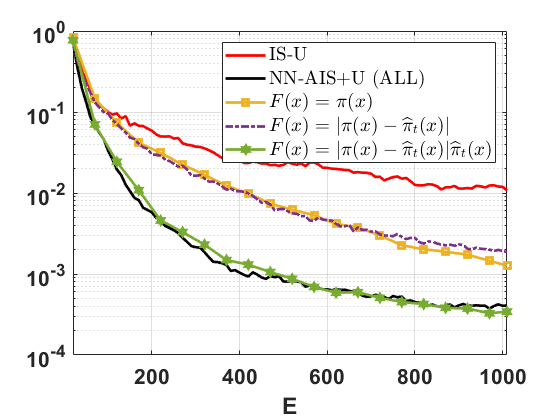}}
			}
			%		\subfigure[]{\includegraphics[width=0.4\textwidth]{mu_varias_alfa_beta}}
			%	}
			%		\subfigure[]{\includegraphics[width=0.3\textwidth]{nodes_vs_t_alfabeta}}
			%		\subfigure[]{\includegraphics[width=0.3\textwidth]{nodes_vs_t_alfabeta_LOG}}
			\vspace{-0.3cm}
			\caption{\footnotesize  Performance of NN-AIS+U with acceptance function from Eq. \eqref{eq:AccProbNew} for different choices of $\alpha$ and $\beta$, in {\bf (a)} the estimation of $Z$. 
				%		The red solid curve represents the RMSE of IS-U as a function of $E$, and the black solid curve represents NN-AIS+U accepting all the samples. 
				Number of nodes versus $t$ 
				%		of NN-AIS+U with acceptance function from Eq. \eqref{eq:AccProbNew} for different choices of $\alpha$ and $\beta$, 
				in {\bf (b)} in linear scale, and  {\bf (c)} in logarithm scale. The black solid curve represents the number of nodes of NN-AIS+U that accepts all.}
			\label{fig_Z_varias_acc}
			% El script esta en C:\Users\Fernando\Dropbox\WITH_FERNANDO_LLORENTE\MATLAB\Adapt_Sticky_IS\Saca_results_diferentes_L
		}
	\end{figure*}
	
	Figure \ref{fig_Z_varias_acc}(a) shows the application of the acceptance probability \textbf{A2} for different choices of $\alpha$ and $\beta$ using the updating scheme in Table \ref{table_Update_Sticky2}. Recall that, when $\alpha$ or $\beta$ are 0, the acceptance probability is 0.
	When $\alpha\gg 1$ and $\beta > 0$, the nodes are added in a space-filling fashion. On the contrary, when $\beta\gg 1$  and $\alpha>0$, the nodes are added by accounting for the discrepancy between $\pi$ and $\widehat{\pi}_t$. We note that the former strategy works better than the latter, as shown in Figure \ref{fig_Z_varias_acc}(a). Moreover, the performance is better when $\alpha=\beta=1$, that is, both strategies at the same time. As $\alpha$ and $\beta$ grow, we recover the performance of the NN-AIS+U accepting all samples. 
	Figure \ref{fig_numero_nodos}(a) shows the number of nodes of the final constructed emulators. We see that the choice $\alpha=\beta=100$ produces an approximation $\widehat{\pi}_t$ that has only half of the nodes of the algorithm accepting all the samples, but achieves the same level of precision in the estimation.  
	We also tested the acceptance functions based on resampling in Eq. \eqref{Super_Fernando}. %with the choices $F(\x) = \pi(\x)$, $F(\x) = |\pi(\x)-\widehat{\pi}_t(\x)|$ and $F(\x) = |\pi(\x)-\widehat{\pi}_t(\x)|\widehat{\pi}_t(\x)$. 
	The results are given in Figures \ref{fig_Z_varias_acc}(c) and \ref{fig_numero_nodos}(c).  We also tested the acceptance functions \textbf{A1} and \textbf{A3}, each one with the two possible updating schemes from Tables \ref{table_Update_Sticky} (non-sequential) and \ref{table_Update_Sticky2} (sequential). As shown in Figure \ref{fig_Z_varias_acc}(b), the acceptance function \textbf{A3} provides better results than \textbf{A1}. For both, the use of a sequential updating scheme improve the results. Figure \ref{fig_numero_nodos}(b) shows the number of final nodes of the emulator.
	We can observe that several parsimonious schemes provide very good performance, close to the vanilla NN-AIS+U (with a much smaller number of added nodes).
	\newline
	\newline
	Finally, in Figure \ref{fig_super_fer} we compare the best parsimonious schemes with the vanilla NN-AIS+U method, showing their RMSE as function of the total number of added nodes at each iterations.  Furthermore, as the dashed line in Figure \ref{fig_banana}(d), we have compared with an NN-AIS+U scheme where $N$ nodes are added at each iteration but chosen randomly in the space (instead of adding the nodes obtained in the inner resampling in Step \ref{StepUpdate} of Table \ref{table_DeepIS_algorithm}).  
	The corresponding curve is shown with a dashed line.
	The end point of each curve is highlighted with greater black circle. The reason is that this last point is completely comparable among the different curve since, at this point, we have the same number of target evaluations $E$.  Therefore, observing these last points, we can see that all the  parsimonious schemes achieve the same or smaller error than the vanilla NN-AIS+U, with a smaller number of added nodes.

	\begin{figure*}[!h]
		{
			\centering
			%	
			
			%		\subfigure[]{\includegraphics[width=0.4\textwidth]{mu_varias_alfa_beta}}
			%	}
			%		\subfigure[]{\includegraphics[width=0.3\textwidth]{nodes_vs_t_luca1luca2}}
			%		\subfigure[]{\includegraphics[width=0.3\textwidth]{nodes_vs_t_luca1luca2_LOG}}
			\centerline{
				\subfigure[]{\includegraphics[width=0.35\textwidth]{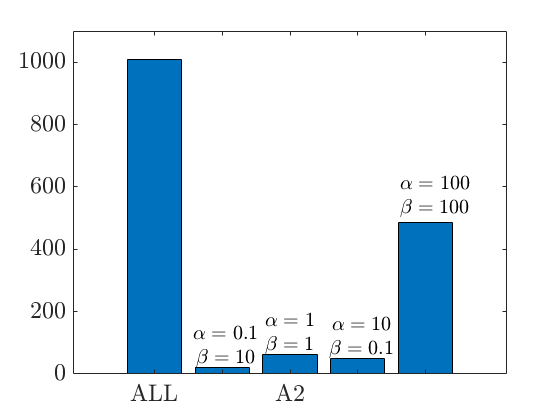}}
				\subfigure[]{\includegraphics[width=0.35\textwidth]{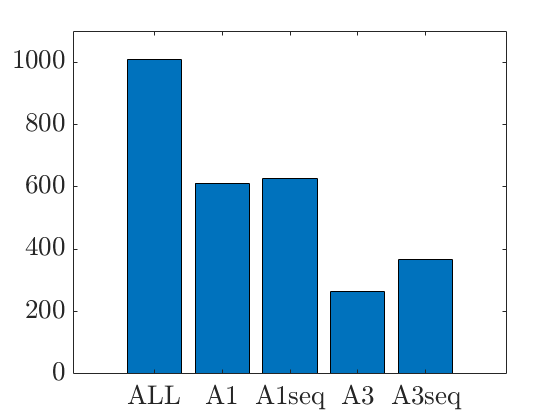}}
				\subfigure[]{\includegraphics[width=0.35\textwidth]{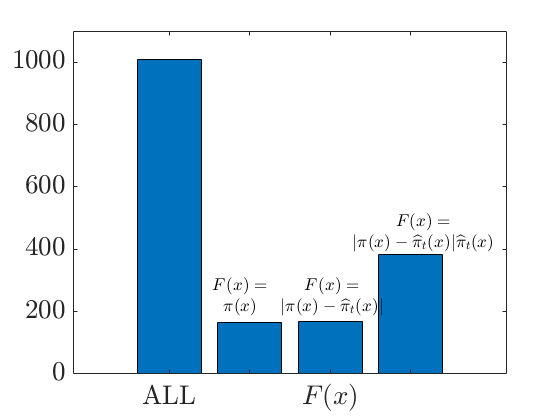}}
			}
			\vspace{-0.3cm}
			\caption{\footnotesize  Final number of added nodes for the construction of the emulator for NN-AIS+U with several acceptance functions of different parsimonious schemes. 
				%		from Eq. \eqref{eq:AccProbStickyLuca} and \eqref{eq_luca2} , in {\bf (a)} the estimation of $Z$. 
				%		The red solid curve represents the RMSE of IS-U as a function of $E$, and the black solid curve represents NN-AIS+U accepting all the samples. 
				%			Number of nodes versus $t$ 
				%		of NN-AIS+U with acceptance function from Eq. \eqref{eq:AccProbNew} for different choices of $\alpha$ and $\beta$, 
				%			in {\bf (b)} in linear scale, and  {\bf (c)} in logarithm scale. The black solid curve represents the number of nodes of NN-AIS+U that accepts all.
			}
			\label{fig_numero_nodos}
		}
	\end{figure*}

	%We note that both updating schemes perform similar {{blue}luca1} and {{blue}luca1}.
	%\newline
	%\newline
	%\begin{figure*}[!h]
	%	{
	%		\centering
	%		%	\centerline{
	%
	%		%		\subfigure[]{\includegraphics[width=0.4\textwidth]{mu_varias_alfa_beta}}
	%		%	}
	%		\subfigure[]{\includegraphics[width=0.3\textwidth]{nodes_vs_t_resampling}}
	%		\subfigure[]{\includegraphics[width=0.3\textwidth]{nodes_vs_t_resampling_LOG}}
	%		\vspace{-0.3cm}
	%		\caption{\footnotesize  Performance of NN-AIS+U with acceptance functions based on resampling, in {\bf (a)} the estimation of $Z$. 
	%			%		The red solid curve represents the RMSE of IS-U as a function of $E$, and the black solid curve represents NN-AIS+U accepting all the samples. 
	%			Number of nodes versus $t$ 
	%			%		of NN-AIS+U with acceptance function from Eq. \eqref{eq:AccProbNew} for different choices of $\alpha$ and $\beta$, 
	%			in {\bf (b)} in linear scale, and  {\bf (c)} in logarithm scale. The black solid curve represents the number of nodes of NN-AIS+U that accepts all.}
	%		\label{fig_resampling}
	%		% El script esta en C:\Users\Fernando\Dropbox\WITH_FERNANDO_LLORENTE\MATLAB\Adapt_Sticky_IS\Saca_results_diferentes_L
	%	}
	%\end{figure*}

	%\subsubsection{Summary of the results in Example 1}
	
	%despues de anadir todo, aqui yo voy a escribir un resumen ...lo tengo en la cabeza, no te preocupes!
	\begin{figure*}[!h]
		{
			\centering
			%	\centerline{
			\includegraphics[width=0.5\textwidth]{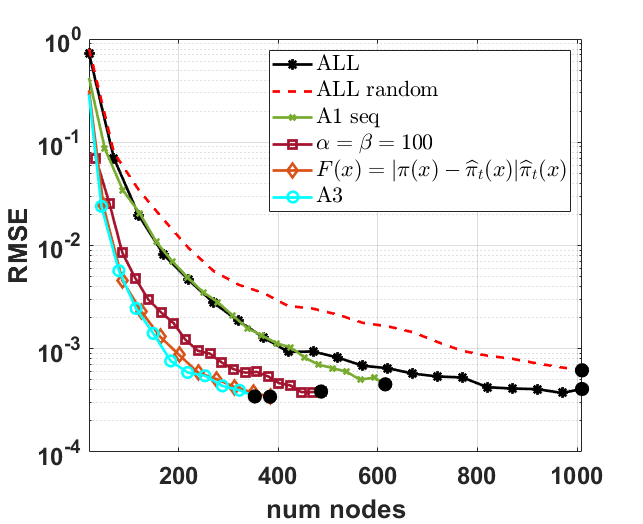}
			%	}
			%	
			\vspace{-0.3cm}
			\caption{\footnotesize  RMSE for NN-AIS+U with different acceptance functions (choosing the best schemes in the previous tests), versus the total number of added nodes at each iteration. We have also incorporated a curve (depicted with dashed line) of an NN-AIS+U scheme where $N$ nodes are added at each iteration but chosen randomly in the space (instead of adding the nodes obtained in the inner resampling in Step \ref{StepUpdate} of Table \ref{table_DeepIS_algorithm}).  The end point in each curve is highlighted with greater black circle. The reason is that this last point is completely comparable among the different curve since, at this point, we have the same number of target evaluations $E$.  Observing these end points, we see that all the  parsimonious schemes shown in the figure provide the same or smaller error than the vanilla NN-AIS+U, with a smaller number of added nodes.  
			}
			\label{fig_super_fer}
		}
	\end{figure*}
}

%%%%%%%%%%%%%%%%%%%%%%%%%%
\subsection{Toy example 2: multimodal density}\label{sec:multimodal}
%%%%%%%%%%%%%%%%%%%%%%%%%%
In this experiment, we consider a multimodal Gaussian target in $d_x=10$,
\begin{align*}
	\bar{\pi}(\x) = \frac{1}{3}\mathcal{N}(\x|\bm{\mu}_1, {\bf \Sigma}_1) + \frac{1}{3}\mathcal{N}(\x|\bm{\mu}_2, {\bf \Sigma}_2) + \frac{1}{3}\mathcal{N}(\x|\bm{\mu}_3, {\bf \Sigma}_3),
\end{align*}
with $\bm{\mu}_1 = [5,0,\dots,0]$, $\bm{\mu}_2=[-7,0,\dots,0]$, $\bm{\mu}_3=[1,\dots,1]$ and ${\bf \Sigma}_1={\bf \Sigma}_2={\bf \Sigma}_3 = 4^2{\bf I}_{10}$. 
We want to test the performance of the different methods in estimating the normalizing constant $Z=1$. Specifically, our aim is to test the combination of our NN-AIS scheme with an AIS algorithm against other AIS algorithms. 
The budged of target evaluations is $E = 1000$.
\newline
\newline
{\bf Methods.} 
We consider three sophisticated AIS schemes, namely {\it population Monte Carlo} (PMC)\cite{Cappe04}, {\it layered adaptive IS} (LAIS)\cite{LAIS17} and {\it adaptive multiple IS} (AMIS)\cite{CORNUET12}. These are AIS algorithms where the proposal (or proposals) gets updated at each iteration using information from previous samples. Specifically, PMC performs multinomial resampling to locate the proposals in the next iteration; AMIS matches the mean of the single proposal with the current estimation of the posterior mean using all previous samples; LAIS evolves the location parameters of the proposals with a MCMC algorithm. 
The goal is to compare the performance of PMC, LAIS and AMIS with a combination of our NN-AIS scheme and LAIS.  
\newline
We set Gaussian pdfs as the proposal pdfs for all methods. We also need to set the number of these proposals in PMC and LAIS, as well as the dispersion of the Gaussian densities. 
For PMC, we test different number of proposals $N_\text{PMC} \in \{10,100,200,500\}$, whose means are initialized at random in $[-15,15]^{10}$. At each iteration of PMC, one sample is drawn from each of the $N_\text{PMC}$ proposals, hence the algorithm is run for $T_\text{PMC} = \frac{1000}{N_\text{PMC}}$ iterations for a fair comparison. As a second alternative, we consider the deterministic mixture weighting approach for PMC, which is shown to have better overall performance, denoted DM-PMC \cite{owen2000safe, veach1995optimally}. 
\newline
For LAIS, we also test different number of proposals $N_\text{LAIS} \in \{10,100,200,500\}$. We consider the {\it one-chain} application of LAIS (OC-LAIS), that requires to run one MCMC algorithm targeting $\post(\x)$ to obtain the $N_\text{LAIS}$ location parameters, hence it requires $N_\text{LAIS}$ evaluations of the target. Then, at each iteration of LAIS, one sample is drawn from the mixture of proposals, hence we run the algorithm for $T_\text{LAIS} = 1000 - N_\text{LAIS}$ iterations for a fair comparison.  For simplicity, we also consider Gaussian random-walk Metropolis to obtain the $N_\text{LAIS}$ means.
\newline
Finally, we consider AMIS with several combinations of number of iterations $T_\text{AMIS}$ and number of samples per iteration $M$. At each iteration, $M$ samples are drawn from a single Gaussian proposal, hence the total number of evaluations is $E = MT_\text{AMIS}$. In this case, we test $E \in \{1000, 2000, 3000, 5000\}$, so the comparison is not fair (penalizing our approach) except for $E=1000$.
\newline
Regarding our method, we use a mixture of  $N_\text{LAIS}\in \{100, 200, 500\}$ proposal pdfs obtained by LAIS as $\bar{q}_\texttt{par}(\x)$ as in Eq. \eqref{eq:stickyAndParam2} (we also use the means of these proposals as initial nodes). %We then proceed sampling, $\frac{N}{2}$ from $\bar{q}_\texttt{par}(\x)$ and $\frac{N}{2}$ from the interpolating proposal. %%
We vary $N$, and run our combined scheme for $T = \frac{E-N_\text{LAIS}}{N}$, keeping the number of target evaluations  $E=1000$. For PMC, LAIS and AMIS, as well as for the random walk proposal within the Metropolis algorithm, the covariance of the Gaussian proposals was set to $\xi^2{\bf I}_{10}$ and we test $\xi=1,...,6$.
All the methods are compared through the mean absolute error (MAE) in estimating $Z$, and the results are averaged over 500 independent simulations.
\newline
%{\bf Discussion.}
The results are shown in Table \ref{table_AIS}, Table \ref{table_stickyLAIS} and Table \ref{table_amis}.
%We can see that LAIS and NN-AIS$+$LAIS clearly outperform the rest in terms of MAE.
We can see that NN-AIS$+$LAIS provides more robust results than only using LAIS. Namely, 
NN-AIS$+$LAIS obtains the same or a lower MAE than LAIS, depending on choice of the different parameters. 
%(i) when $\xi$ is small ($\xi=1$ and $\xi=2$), NN-AIS$+$LAIS is able to achieve lower MAE than LAIS; this is specially marked for $N_\text{LAIS}=500$ and $N_\text{LAIS}=200$; (ii) NN-AIS$+$LAIS is also more robust when $N_\text{LAIS}$ is small ($N_\text{LAIS}=100$), where it beats LAIS for every $\xi$.
%\newline
%For small $N_\text{LAIS}$ and/or small $\xi$, it is beneficial to run more iterations of our algorithm, i.e., smaller $N$. On the other hand, when the proposal produced by LAIS is sufficiently good, e.g. when $N_\text{LAIS}=500$ and $\xi>1$, the results show that just a couple of iterations of our method ($T=2$ for $N=250$, or $T=5$ for $N=100$) is better. 
Overall, the proposed scheme outperforms all the other benchmark AIS methods such as PMC, DM-PMC, LAIS and AMIS easily, even considering more target evaluations (penalizing our scheme) as shown in Table \ref{table_amis}.

\begin{table*}[!h]
	\centering
	%\small
	\caption{ \textbf{MAE for $Z$ with $E=1000$} (best and worst MAE of each method are boldfaced)}
	\vspace{0.2cm}
	%	{\footnotesize
	\begin{tabular}{|c c|c|c|c|c|c|c|}
		\hline	
		% & & & & & & &  \\
		\multicolumn{2}{|c|}{{\bf Methods}}  & $\xi = 1$ & $\xi = 2$ & $\xi = 3$ & $\xi = 4$ &  $\xi = 5$ & $\xi = 6$  \\		  
		%		\hline
		%		\hline
		%		\multicolumn{2}{|c|}{GK-AQ}     	& {\bf 0.4782} & 0.1741 & {\bf 0.0780}  & 0.1362 &  0.1497 &  0.2322\\
		\hline
		\hline
		\multirow{ 4}{*}{PMC} & $N_\text{PMC}=10$& 0.9993 & 0.9526	& 0.8603 & 0.6743 & 0.6024 & 0.6155 \\
		%\hline 
		& $N_\text{PMC}=100$& 0.9998 & 0.9896 & 0.8853 & 0.6761 & 0.5192 & {\bf 0.4544}\\ 
		& $N_\text{PMC}=200$& {\bf 1.0002} & 0.9893 & 0.8816 & 0.7099 & 0.6389  &  0.5384\\
		%\hline
		&$N_\text{PMC}=500$& 0.9995 & 0.9916 & 0.9741 & 0.8700 & 0.7421 & 0.6544 \\
		\hline
		\hline
		\multirow{ 4}{*}{DM-PMC}  & $N_\text{PMC}=10$&  0.9991 & 0.9478 &	0.8505 & 0.6009 & 0.5352 & 0.5814 \\
		%\hline 
		&$N_\text{PMC}=100$& 0.9997 & 0.8719 & 0.4490 & 0.2425 &  {\bf 0.1901} & 0.2193 \\ %\hline
		&$N_\text{PMC}=200$&  0.9999	& 0.9321 & 0.5708 & 0.3257 & 0.2374 & 0.2524 \\
		%\hline
		& $N_\text{PMC}=500$& {\bf 1.0000} & 0.9888 & 0.7969 & 0.5009 & 0.3684 & 0.3800 \\
		\hline
		\hline
		%		\multirow{ 4}{*}{Ideal LAIS} &$N=10$& {\bf 0.9992}	& 0.8114 & 0.2579 & 0.0863 & 0.0819 & 0.1091 \\
		%		%\hline 
		%		&$N=100$& 0.9918 & 0.3638 & 0.0547 &  0.0407 & 0.0598  & 0.1053 \\ %\hline
		%		&$N=200$&  0.9846 & 0.2486 & 0.0352 & 0.0411 & 0.0680 & 0.1093 \\
		%		%\hline
		%		& $N=500$& 0.9687 & 0.1852 & {\bf 0.0335} & 0.0473 & 0.0891 & 0.1353 \\
		%		\hline
		%		\hline
		\multirow{ 4}{*}{OC-LAIS} &$N_\text{LAIS}=10$& {\bf 1.0000}	& 1.0000 & 0.9992 & 0.9883 & 0.9468 & 0.9079 \\
		%\hline 
		&$N_\text{LAIS}=100$& 0.9999 & 0.8731 & 0.4434 & 0.2785 & 0.2392 & 0.2870  \\ %\hline
		&$N_\text{LAIS}=200$& 0.9982 & 0.7028 & 0.2418 & 0.1243 & 0.1406 & 0.2070 \\
		%\hline
		& $N_\text{LAIS}=500$& 0.9937 & 0.4949 & 0.1221 & {\bf 0.0857} & 0.1195 & 0.1786 \\
		\hline
	\end{tabular}
	%	}	
	\label{table_AIS}
\end{table*}

\begin{table*}[!h]
	\centering
	%\small
	\caption{ \textbf{MAE for $Z$ with $E=1000$} (best of each combination of $N_\text{LAIS}$ and $\xi$ are boldfaced)}
	\vspace{0.2cm}
	%	{\footnotesize
	\begin{tabular}{|c c|c|c|c|c|c|c|}
		\hline	
		% & & & & & & &  \\
		\multicolumn{2}{|c|}{{\bf Methods}}  & $\xi = 1$ & $\xi = 2$ & $\xi = 3$ & $\xi = 4$ &  $\xi = 5$ & $\xi = 6$  \\
		\hline
		\hline
		%\hline
		%		& $N=500$& 0.9937 & 0.4949 & 0.1221 & {\bf 0.0857} & 0.1195 & 0.1786 \\
		%		\hline
		%		\hline
		\multirow{ 3}{*}{NN-AIS$+$LAIS ($N_\text{LAIS}=100$)} &$N=50$& {\bf 0.9778}	& {\bf 0.3886} & {\bf 0.1334} & 0.1487 & {\bf 0.1624} & {\bf 0.1968} \\
		%\hline 
		&$N=100$& 0.9900 & 0.4152 & 0.1408 & 0.1519 & 0.1853 & 0.2502  \\ %\hline
		&$N=300$& 0.9907 & 0.4817 & 0.1761 & {\bf 0.1466} & 0.1869 & 0.2427 \\
		%\hline
		%		& $N=500$& 0.9937 & 0.4949 & 0.1221 & {\bf 0.0857} & 0.1195 & 0.1786 \\
		\hline
		\multirow{ 3}{*}{NN-AIS$+$LAIS ($N_\text{LAIS}=200$)} &$N=100$& {\bf 0.7662}	& {\bf 0.1607} & 0.1332 & {\bf 0.1179} & {\bf 0.1300} & 0.2000 \\
		%\hline 
		&$N=200$& 0.8195 & 0.2176 & {\bf 0.1001} & 0.1250 & 0.1418 & {\bf 0.1854}  \\ %\hline
		&$N=400$& 0.8417 & 0.2954 & 0.1512 & 0.1218 & 0.1522 & 0.2060 \\
		%\hline
		%		& $N=500$& 0.9937 & 0.4949 & 0.1221 & {\bf 0.0857} & 0.1195 & 0.1786 \\
		\hline
		\multirow{ 3}{*}{NN-AIS$+$LAIS ($N_\text{LAIS}=500$)} &$N=50$& {\bf 0.2428}	& 0.1801 & 0.1614 & 0.1313 &  0.1190 & 0.1642 \\
		%\hline 
		&$N=100$& 0.2905 & 0.1406 & {\bf 0.1144}  & 0.1046 & {\bf 0.1152} & 0.1851  \\ %\hline
		&$N=250$& 0.4139 & {\bf 0.1270} & 0.1226 & {\bf 0.0989} & 0.1262 & {\bf 0.1783} \\
		\hline
	\end{tabular}
	%	}	
	\label{table_stickyLAIS}
\end{table*}

\begin{table*}[!h]
	\centering
	%\small
	\caption{ \textbf{MAE for $Z$ of AMIS with $E\in \{1000, 2000, 3000, 5000\}$.} Note that the comparison is unfair (penalizing our approach) except for $E=1000$.}
	\vspace{0.2cm}
	%	{\footnotesize
	\begin{tabular}{|c c|c|c|c|c|c|c|}
		\hline	
		% & & & & & & &  \\
		\multicolumn{2}{|c|}{{\bf Methods}}  & $\xi = 1$ & $\xi = 2$ & $\xi = 3$ & $\xi = 4$ &  $\xi = 5$ & $\xi = 6$  \\		  
		%		\hline
		%		\hline
		%		\multicolumn{2}{|c|}{GK-AQ (E=1000)}     	& {\bf 0.4782} & 0.1741 & {\bf 0.0780}  & 0.1362 &  0.1497 &  0.2322\\
		\hline
		\hline
		\multirow{3}{*}{AMIS}  &$M=10$&  0.9998 & 0.9997 & 0.9997 & 0.9996 & 0.9996 & 0.9995 \\
		%\hline 
		\multirow{3}{*}{$E=1000$} &$M=100$& 1.0000 & 1.0000 & 1.0000 & 0.9999 & 0.9997 &  0.9990 \\ %\hline
		&$M=200$& 1.0000 & 1.0000 & 1.0000 & 1.0000 & 0.9998 & 0.9994 \\
		%\hline
		& $M=500$& {\bf 1.0000} & 1.0000 & 1.0000 & 1.0000 & 0.9998 & {\bf 0.9989} \\
		\hline
		\hline
		\multirow{3}{*}{AMIS}  &$M=10$& 0.9155 & 0.9117 & 0.8981 & 0.8987 & 0.8891 & {\bf 0.8878} \\
		%\hline 
		\multirow{ 3}{*}{$E=2000$}&$M=100$& 0.9998 & 0.9986 &  0.9934 & 0.9784 & 0.9559 & 0.9072
		\\ %\hline
		&$M=200$& 1.0000 & 1.0000 & 0.9998 & 0.9981 & 0.9888 & 0.9712 \\
		%\hline
		& $M=500$& {\bf 1.0000} & 1.0000 & 1.0000 & 0.9998 & 0.9984 & 0.9953 \\
		\hline
		\hline
		\multirow{3}{*}{AMIS}  &$M=10$& 0.3293 & 0.3402 & {\bf 0.3051} & 0.3381 & 0.3540 & 0.3443 \\
		%\hline 
		\multirow{ 3}{*}{$E=3000$}	&$M=100$& 0.9725 & 0.9040 & 0.7963  & 0.6384 & 0.4964 & 0.3816 \\
		&$M=200$& 0.9998 & 0.9977 & 0.9884 & 0.9527 & 0.8308 & 0.7119 \\
		%\hline
		& $M=500$& {\bf 1.0000} & 1.0000 & 0.9998 & 0.9988  & 0.9859 & 0.9566
		\\ 
		\hline
		\hline
		\multirow{3}{*}{AMIS}&$M=10$& 0.0766 & 0.0768 & {\bf 0.0695} & 0.0722 & 0.0699 & 0.0725 \\
		%\hline 
		\multirow{ 3}{*}{$E=5000$} &$M=100$& 0.1626 & 0.1176 & 0.0957 & 0.0810 & 0.0737 &  0.0656 \\
		&$M=200$& 0.8771 & 0.6040 & 0.2824 & 0.1473 & 0.1163 & 0.0899 \\
		%\hline
		& $M=500$& {\bf 1.0000} & 0.9982 & 0.9904 & 0.9449 & 0.7944 &  0.4532 \\ 
		\hline
	\end{tabular}
	%	}	
	\label{table_amis}
\end{table*}

{
	\subsection{Inference in an Astronomical model}\label{sec_Astro}
	
	%%%%%%%%%%%%%%%%%%%%%%%%%%%%%%%%%%%%%%%%%%%%%%%%%%%%%%%%%
	%\subsection{Inference in Kepler's models}\label{KeplerSect}
	%%%%%%%%%%%%%%%%%%%%%%%%%%%%%%%%%%%%%%%%%%%%%%%%%%%%%%%%%
	
	In recent years, the problem of revealing objects orbiting other stars has acquired large attention in Astronomy. Different techniques have been proposed to discover exo-objects but, nowadays, the radial velocity technique is still the most used \cite{Gregory2011,Barros2016,Trifonov2019}.  The model is highly non-linear and it is costly in terms of computation time (specially, for certain sets of parameters). The evaluation of the posterior involves numerically integrating a differential equation in time or an iterative procedure for solving a non-linear equation. Typically, the iteration is performed until a threshold is reached, or a certain number of  iterations (e.g., typically $10^6$ iterations), are performed. For the radial velocity model, this is needed for solving  Eq. \eqref{eq:kepler} described below. 
	%The problem of radial velocity curve fitting is applied in several related applications. It is similar to the problem of determining the orbits of spectroscopic binary stars \cite{Strassmeier1993,Galvez2006} or the stars surrounding the galactic center \cite{Gillessen2017}. 
	In the following, we describe an orbital model, which is equivalent for any N-body system observed from Earth, i.e. exoplanetary systems, binary stellar system, double pulsars, etc. 
	
	\begin{table}[t]
		\centering
		\caption{Description of parameters in Eq.~\eqref{eq:rv}.}
		\small
		\begin{tabular}{lll} % Column formatting, @{} suppresses leading/trailing space
			\hline
			Parameter & Description & Units \\
			\hline
			\multicolumn{3}{l}{For each planet}\\
			\hline
			$\zeta_i$        & amplitude of the curve & m\,s$^{-1}$ \\
			${u}_{i,k}$      & true anomaly     & rad \\
			$\omega_{i}$      & longitude of periastron & rad \\ 
			$e_i$        & orbit's eccentricity    & \ldots \\
			$P_i$        & orbital period        & s \\
			$\tau_i$       & time of periastron passage & s \\
			\hline
			\multicolumn{3}{l}{\footnotesize Below: not depending on the number of objects/satellite }\\
			\hline
			$V_0$      & mean radial velocity   & m\,s$^{-1}$ \\
			%$\mathbf{s}$         & jitter               & m\,s$^{-1}$ \\
			\hline
		\end{tabular}
		\label{tab:rvpar}
	\end{table}

	%%%%%%%%%%%%%%%%%%%%%%%%%%%%%%%%%%%
	\subsubsection{Likelihood function and prior densities}
	%%%%%%%%%%%%%%%%%%%%%%%%%%%%%%%%%%%
	When analysing radial velocity data of an exoplanetary system, it is commonly accepted that the \emph{wobbling} of the star around the centre of mass is caused by the sum of the gravitational force of each planet independently and that they do not interact with each other. Each planet follows a Keplerian orbit and the radial velocity of the host star is given by
	\begin{equation}
		{y}_{k} = V_0 + \sum\limits_{i = 1}^{S} \zeta_i \left[ \cos \left( {u}_{i,k} + \omega_{i} \right) + e_i \cos \left( \omega_{i} \right) \right] +\xi_k,
		\label{eq:rv}
	\end{equation}
	with $k=1,\ldots,K$. The number of objects in the system is $S$, that is consider known in this experiment (for the sake of simplicity). Note that the iteration index $i=1,...,S$ denotes the $i$-th object/planet.
	Both ${y}_{k}$, ${u}_{i,k}$ depend on time $t$, and $\xi_k$ is a Gaussian noise perturbation with variance $\sigma_e^2$. For simplicity, we consider this value known, $\sigma_e^2=1$.   
	The meaning of each parameter in Eq.~\eqref{eq:rv} is given in Table~\ref{tab:rvpar}. The likelihood
	function is defined by \eqref{eq:rv} and some indicator variables described below. 
	The angle ${u}_{i,k}$ is 
	the true anomaly of the planet $i$ and it can be determined from
	\begin{equation}
		\frac{d{u}_{i,k}}{dt} = \frac{2\pi}{P_i} \frac{\left( 1 + e_i \cos {u_{i,k}} \right)^2}{\left( 1 - e_i \right)^\frac{3}{2}}
		\label{eq:trueanomaly}
	\end{equation}
	This equation has analytical solution. As a result, the true anomaly $u_{i,k}$ can be determined from the mean anomaly $M_{i,k}$. However, the analytical solution contains a non linear term that needs to be determined by iterating. First, we define the mean anomaly $M_{i,k}$ as
	\begin{equation}
		M_{i,k} = \frac{2\pi}{P_i} \left( t - \tau_i \right),
		\label{eq:meananomaly2}
	\end{equation}
	where $\tau_i$ is the time of periastron passage of the planet $i$ and $P_i$ is the period
	of the orbit (see Table~\ref{tab:rvpar}). Then, through the Kepler's equation, 
	\begin{equation}
		M_{i,k} = E_{i,k} - e_i \sin E_{i,k},
		\label{eq:kepler}
	\end{equation}
	we have to obtain $E_{i,k}$, which is the eccentric anomaly. Equation~\eqref{eq:kepler} has no analytic solution and it must be solved by an iterative procedure. A Newton-Raphson method is typically used to find the roots of this equation \cite{Press2002}. For certain sets of parameters this iterative procedure can be particularly slow.
	\newline 
	\newline 
	Finally, we can  also obtain $u_{i,k}$ from
	\begin{equation}
		\tan \frac{u_{i,k}}{2} = \sqrt{ \frac{1 + e_i}{1 - e_i}} \, \tan \frac{E_{i,k}}{2}, 
		\label{eq:eccentricanomaly}
	\end{equation}
	%
	%The result of the model for a given time $t$ is compared to the received data $y_{r,t}$.
	Hence, the vector of variables to infer, $\mathbf{x}$, is
	\begin{equation}
		\mathbf{x} = [V_0, \zeta_1, \omega_{1,t}, e_1, P_1, \tau_1, \ldots, \zeta_S, \omega_S, e_S, P_S, \tau_S],
	\end{equation}
	For a single object (e.g., a planet or a natural satellite), the dimension of $\mathbf{x}$ is $d_X = 5+1=6$, with two objects the dimension of $\mathbf{x}$ is is $d_x = 11$ etc. Generally, we have $d_x=1+5S$. Note that the observation model in Eq. \eqref{eq:rv} induces the likelihood function $p(\y|\x)$, where $\y=[y_1,...,y_K]$.
	\newline
	{\bf Priors}. As prior densities we consider uniform pdfs in the following intervals: $V_0\in [-20,20]$, $\zeta_{i} \in [0,50]$, $e_{i} \in [0,1]$, $P_i \in [0,365]$, $\omega_{i,k} \in [0,2\pi]$, $\tau_{i} \in [0,P_i]$  (i.e., the prior is zero outside these intervals), for all $i=1,\ldots,S$. This means that the likelihood function is zero when the particles fall out of these intervals. Note that the interval of $\tau_{i}$ is conditioned to the value $P_i$. This parameter is the time of periastron passage, i.e. the time passed since the object passed the closest point in its orbit. It has the same units of $P_i$ and can take values from 0 to $P_i$.
	\subsubsection{Experiment setting and results}
	%%%%%%%%%%%%%%%%%%%%%%%%%%%%%%%%%%%%%%
	
	We generate a set of data $\{y_{k}\}_{t=1}^K$ with $K=50$, and $S=2$ objects (so that $d_x=11$), according to the observation model above. We set $V=2$,  $\zeta_{1}=25$, $\omega_{1}=0.61$, $e_{1}=0.1$, $P_{1}=15$, $\tau_{1}=3$ (for the first object) and $\zeta_{2}=5$, $\omega_{2}=0.17$, $e_{2}=0.3$, $P_{2}=115$, $\tau_{2}=25$ (for the second object). We compare a standard IS scheme using the prior as proposal and  the NN-AIS+U  scheme (using again the prior as uniform proposal component) using the parsimonious scheme with acceptance function A3 in Eq. \eqref{eq_luca2}. In NN-AIS+U, we consider $N=10000$, $T=100$ and $L=10^6$. The total number of evaluations of the posterior is then $NT=10^6$ for NN-AIS+U. For the   standard IS scheme, we consider different number of samples $\{10^6,2\cdot 10^6,  3 \cdot 10^6, 4 \cdot 10^6\}$. We  compute the Relative MSE (RMSE) in estimation of the $11$ parameters in $\x$, averaged over all the components. The results are also averaged over $200$ independent runs. Table  \ref{table_MSE_astro} provides the RMSE and the computational time,  normalized with respect to the time spent by the  standard IS scheme with $10^6$ samples.
	We can observe that, in order to obtain the same performance of NN-AIS+U in terms of RMSE, the IS schemes require much more computational time than NN-AIS+U. Therefore, this is an example with a real-world model where the inequality \eqref{EqSuperCost} is fulfilled.

	%In all experiments, we set $R=5$ and $T=50$ and average the results over $500$ independent runs. 
	%We consider three different experiments: {\bf (E1)} $S=0$, i.e., no object, {\bf (E2)} $S=1$ (one object) and {\bf (E2)} the case of two objects $S=2$. We set $V_0=2$, in all cases. For the first object in {\bf E1} and {\bf E2}, we set $\zeta_{1,0}=25$, $\omega_{1,0}=0.61$, $e_{1,0}=0.1$, $P_{1,0}=15$, $\tau_{1,0}=3$. For {\bf E2}, we also consider a second object with $\zeta_{2,0}=5$, $\omega_{2,0}=0.17$, $e_{2,0}=0.3$, $P_{2,0}=115$, $\tau_{2,0}=25$ (in that case $S=2$). Note that the SNR associate to the second object is low (so that the detection of this planet is not straightforward). The rest of trajectories are generated according to the transition model (and the corresponding measurements $y_{r,t}$ according to the observation model). 
	% We consider $N=10^5$ total number of particles and just $M=100$ summary particles for CPF in Table \ref{GCPFtable} ($\frac{M}{N}=10^{-3}$). %%%%%Note that ABC cannot be applied since we are not able to generate arti 

	\begin{table*}[!ht]
		{
			\centering
			%\small
			\caption{Relative Mean Square Errors (MSE) and normalized computational time.}
			\vspace{-0.2cm}
			\begin{center}
				%	{\footnotesize
				\begin{tabular}{|c|c|c|c|c|c|}
					\hline	
					% & & & & & & &  \\
					{\bf Methods} &  {\bf NN-AIS+U}  &   {\bf IS}  &   {\bf IS} & {\bf IS} & {\bf IS} \\
					\hline
					\hline
					RMSE & 5.755  & 9.439 & 7.943 & 6.524  &  5.431\\
					normalized time & 1.53 &  1 & 1.91  & 3.20  & 4.17  \\
					posterior evaluations ($E$) & $10^6$ &  $10^6$ & $2\cdot 10^6$ &  $3\cdot 10^6$  & $4\cdot 10^6$   \\		
					\hline
				\end{tabular}
			\end{center}
			%	}	
			\label{table_MSE_astro}
		}
	\end{table*}

}

\subsection{Retrieval of biophysical parameters inverting an RTM model}\label{sec_rROSAIL}

\begin{figure}[h!]
	\centering
	\includegraphics[width=8cm]{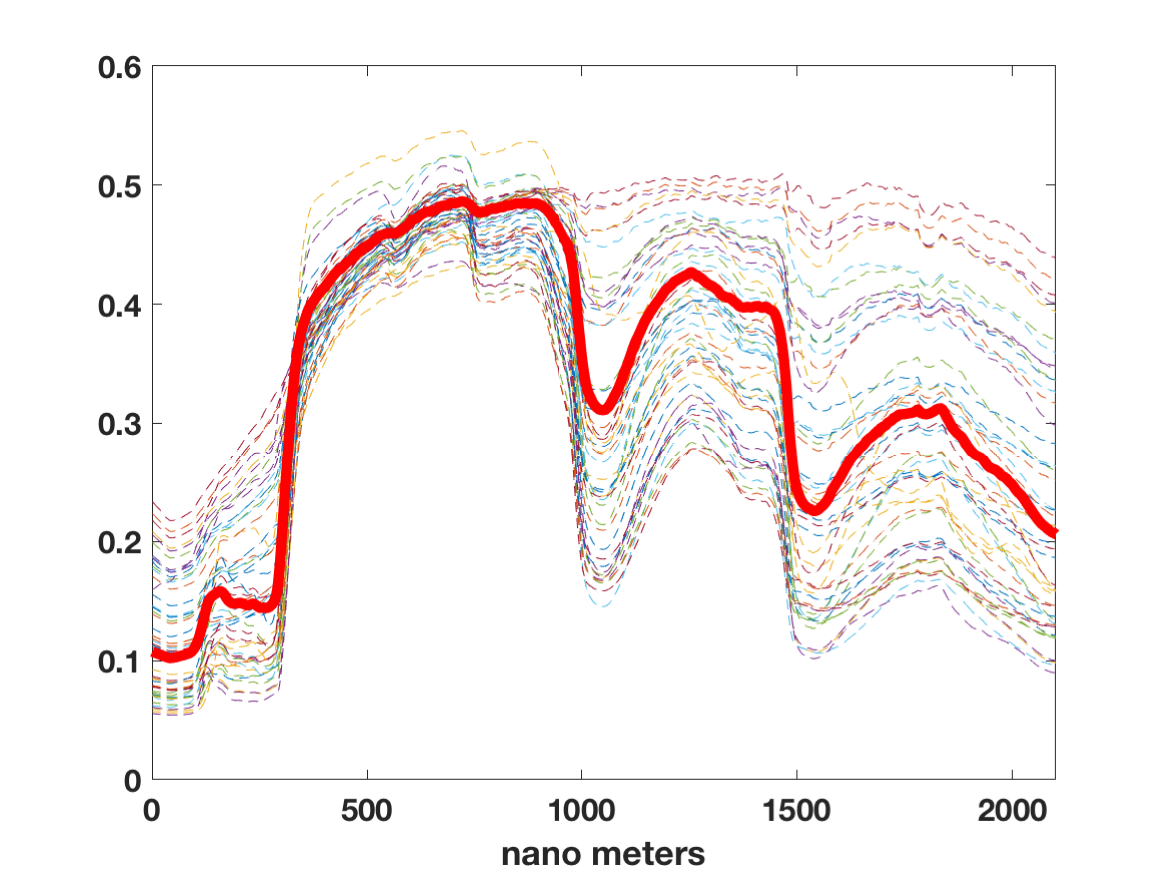}
	\caption{An example of vector of data ${\bf y}_r$ (hyperspectral reflectances, shown with solid line) and the model values corresponding to 50 different samples, ${\bf f}^{(i)}={\bf f}({\bf x}_r^{(i)})$ (dashed lines). Each component of the vector ${\bf y}_r$ corresponds to a different wavelength (nm).}
	\label{FigEX_rrosail}
\end{figure}

In this experiment, we apply NN-AIS to retrieve biophysical parameters of a sequence of problems involving the radiatrive transfer PROSAIL model. The purpose is to show the ability of NN-AIS to share information from related inverse problems easily.  The combined PROSPECT leaf optical properties model and SAIL canopy bidirectional reflectance model, also referred to as PROSAIL, have been used for almost two decades to study plant canopy spectral and directional reflectance in the solar domain \cite{jacquemoud2009prospect+}. PROSAIL has also been used to develop new methods for retrieval of vegetation biophysical properties. It links the spectral variation of canopy reflectance, which is mainly related to leaf biochemical contents, with its directional variation, which is primarily related to canopy architecture and soil/vegetation contrast. This link is key to simultaneous estimation of canopy biophysical/structural variables for applications in agriculture, plant physiology, and ecology at different scales. PROSAIL has become one of the most popular radiative transfer tools due to its ease of use, general robustness, and consistent validation by lab/field/space experiments over the years. 
\newline
{\bf Inversion of PROSAIL.} The context is Bayesian inversion of an observation model ${\bf h}(\x)$.\footnote{The MATLAB code of PROSAIL is available in \url{http://teledetection.ipgp.jussieu.fr/prosail/}.}
%\begin{align}
%	\y = {\bf F}(\x) + {\bf u},
%\end{align}
%where ${\bf F}: \mathbb{R}^{d_x} \to \mathbb{R}^{d_y}$ is the forward model that relates parameters $\x$ with observations $\y$, and ${\bf u}$ is some stochastic noise. 
In our setting, the observation model is PROSAIL, which models reflectance in terms of leaf optical properties and canopy level characteristics. We choose only leaf optical properties
as the set parameters of interest
\begin{align}\label{eq:PROSAILparam}
	\x = [S_{st}, C_{hl}, C_{ar}, C_{br}, C_w, C_m]\in \mathbb{R}^6,
\end{align} 
described in Table \ref{tab:PROSAILparam}. %Then, $d_x=6$ in this experiment.
In Table \ref{tab:canopy}, we show the fixed values of canopy level characteristics, which are determined by the leaf area index (LAI), the average leaf angle inclination (ALA), the hot-spot parameter (Hotspot), and the parameters of system geometry described by the solar zenith angle ($\theta_s$), view zenith angle ($\theta_\nu$), and the relative azimuth angle between both angles ($\Delta\Theta$).
The observation model is 
$\y={\bf h}(\x)+{\bf v}$,
where ${\bf v}\sim \mathcal{N}({\bf 0}, \sigma^2{\bf I}_{d_y})$ with $\sigma=1$.
The observed data, denoted $\y\in \mathbb{R}^{d_y}$ with $d_y=2101$, corresponds to the detected spectra.
We generated synthetic spectra and the goal is to infer $\x$ studying the corresponding posterior distribution. The Gaussian noise ${\bf v}\sim \mathcal{N}({\bf 0}, \sigma^2{\bf I})$ jointly with PROSAIL, ${\bf h}(\x)$, induces the following likelihood function 
\begin{align}
	\ell(\y|\x) = \mathcal{N}(\y|{\bf h}(\x),\sigma^2{\bf I}).
\end{align}
We set the prior $g(\x)$ as a product of indicator variables $S_{st} \in [1,3]$, $C_{hl} \in [0,100]$, $C_{ar} \in [0,25]$, $C_{br} \in [0,1]$, $C_w \in [0,0.05]$ and $C_b \in [0,0.02]$, i.e., the prior is zero outside these intervals.\footnote{We have employed the ranges suggested \url{http://opticleaf.ipgp.fr/index.php?page=prospect}.} The complete posterior is then $p(\x|\y) = \frac{1}{Z}\ell(\y|\x)g(\x)$. It is important to remark that PROSAIL is an {\it highly non-linear} model and its inversion is a very complicated problem, as shown the remote sensing literature \cite{CampsValls18sciasi,CampsValls19nsr}. 
%We are interested in inferring the parameters $\x$. Different possibilities exist: $\x_\text{MMSE}$, $\x_\text{MAP}$, etc. 
\newline
{\bf Sequential inversion for image recovery.} In remote sensing, the goal is usually to recover of an image formed by $R$ pixels. A set of physical parameters $\x_r$ is associated to the $r$-th pixel. Hence, the corresponding vector of observations $\y_r$ is also associate to each pixel. We have then  a collection of inverse problems, where we desire to retrieve $\x_r$ given $\y_r$, one for each pixel.
Mathematically, let consider $R$  measurements, $\{\y_r\}_{r=1}^R$, associated each to a different inverse problem, under the  PROSAIL model, i.e., a mapping ${\bf h}(\x):\mathbb{R}^{d_x}\rightarrow \mathbb{R}^{d_y}$,
\begin{align}
	\y_r = {\bf h}(\x_r) + {\bf v}_r, \quad r=1,\dots,R.
\end{align}
We assume ${\bf v}_r\sim \mathcal{N}({\bf 0}, \sigma^2{\bf I}_{d_y})$,  for all $r=1,\dots,R$, with $d_y=2101$, and $\sigma=1$, and thus we have a $R$ posterior distributions $p_r(\x_r|\y_r)$ for $r=1,\dots,R$ (we recall that $\x_r \in \mathbb{R}^{d_x}$,with $d_x=6$). We solve them sequentially while reusing information. Some examples of data $\y_r$ and model values are given in Figure \ref{FigEX_rrosail}.
%There exist different scenarios. We can treat each of these  {{blue}problems independently} and hence study the posterior of $\x_r$ given $\y_r$ for $r=1,\dots,R$. Alternatively, we can consider the information provided by one problem and use it to solve the next one (i.e. in a sequential manner). Our scheme easily allows for this recycle. For instance, the estimated MAP or MMSE in one problem can be used as initial nodes for building the interpolator in the following problem. Note that no extra model evaluations are required since we store all past model evaluations. 

\begin{table}[h]
	\centering
	\caption{Description of parameters in Eq.~\eqref{eq:PROSAILparam}.}
	\small
	\begin{tabular}{lll} % Column formatting, @{} suppresses leading/trailing space
		\hline
		Parameter & Description & Units \\
		%		\hline
		%		\multicolumn{3}{l}{For each planet}\\
		\hline
		$S_{st}$        & structure coefficient & --- \\
		$C_{hl}$      & chlorophyll content     & $\mu$g\,cm$^{-2}$ \\
		$C_{ar}$      & carotenoid content & $\mu$g\,cm$^{-2}$ \\ 
		$C_{br}$   & brown pigment content  & --- \\
		$C_w$       &  water content        & cm \\
		$C_m$       &  dry matter content & g\,cm$^{-2}$ \\
		\hline
	\end{tabular}
	\label{tab:PROSAILparam}
\end{table}

\begin{table}[h]
	\centering
	\caption{Characteristics of the simulation used in the PROSAIL model.}
	\vspace{0.1cm}
	\small
	\begin{tabular}{lcccccc} % Column formatting, @{} suppresses leading/trailing space
		\hline
		%		Parameter & Description & Units \\
		%		\hline
		%		\multicolumn{3}{l}{For each planet}\\
		%		\hline
		\multirow{2}{*}{Canopy level} & LAI & ALA &Hotspot & $\theta_s$ & $\theta_\nu$ & $\Delta \Theta$ \\
		& 5 & 30 & 0.01 & 30 & 10 & 90 \\
		\hline
	\end{tabular}
	\label{tab:canopy}
\end{table}

\begin{figure*}[!ht]
	\centering
	\centerline{
		\subfigure[]{\includegraphics[width=0.3\textwidth]{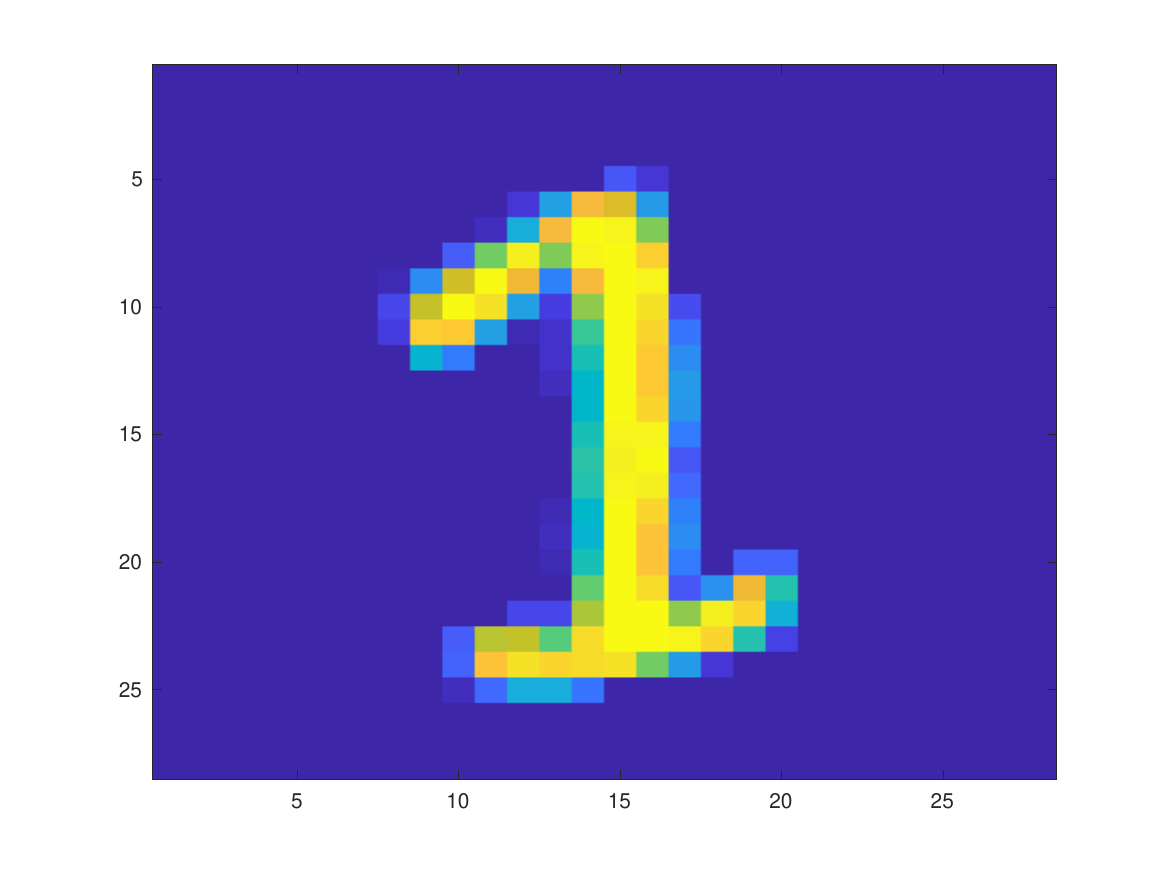}}
		\subfigure[]{\includegraphics[width=0.3\textwidth]{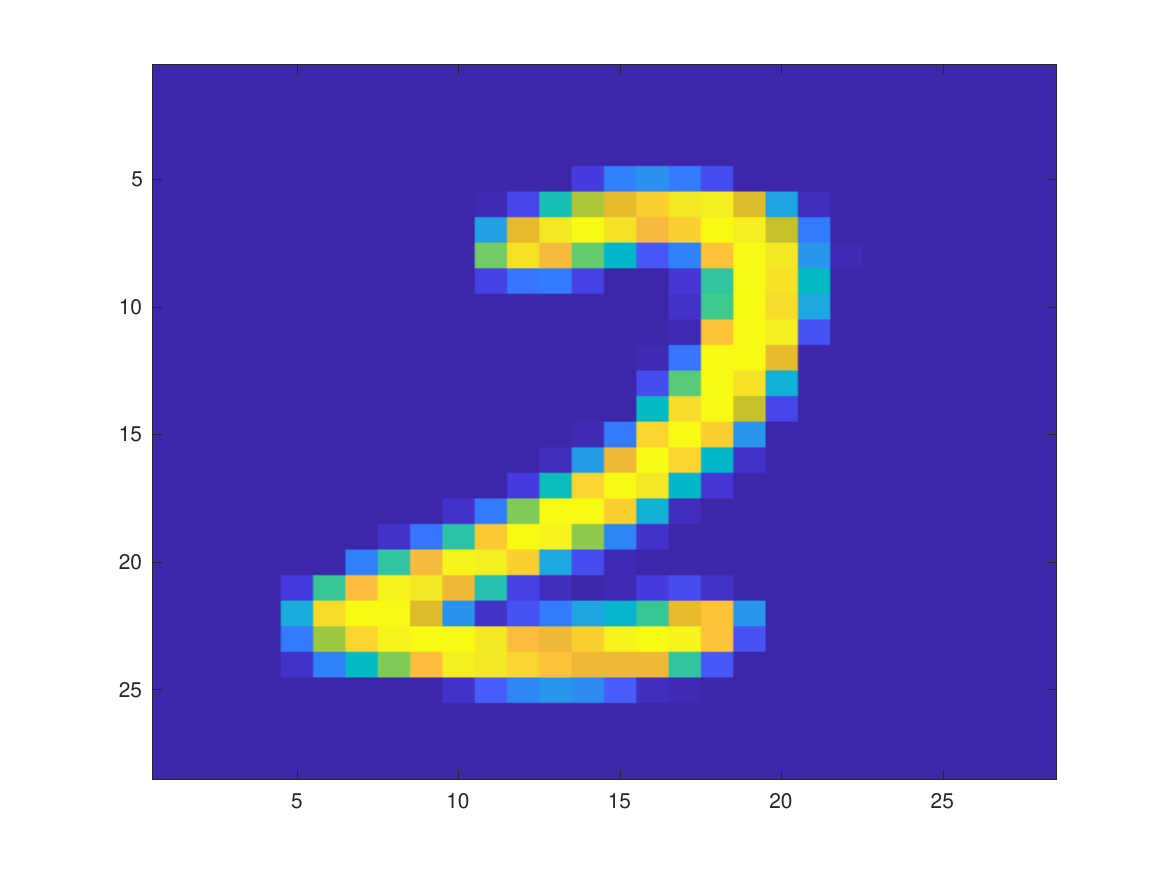}}
		\subfigure[]{ \includegraphics[width=0.3\textwidth]{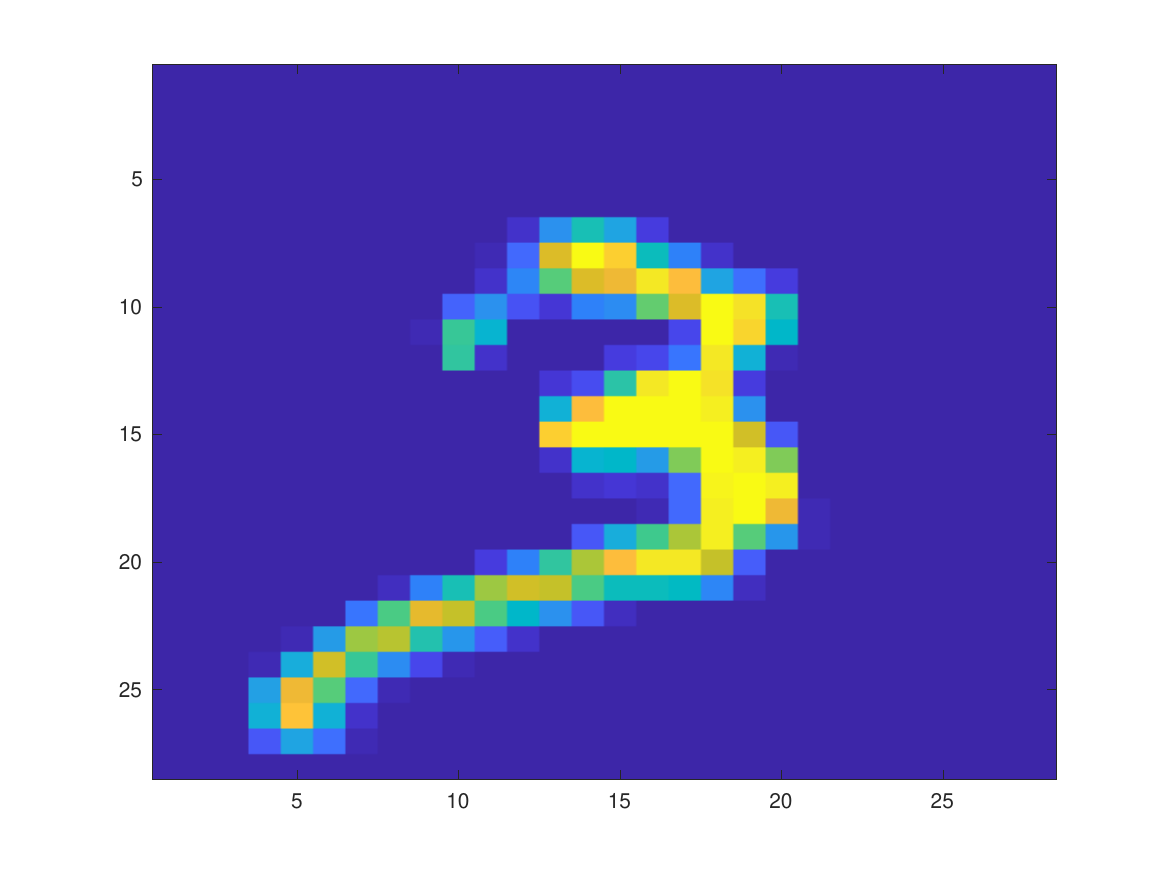}}
	}
	\centerline{
		\subfigure[]{ \includegraphics[width=0.3\textwidth]{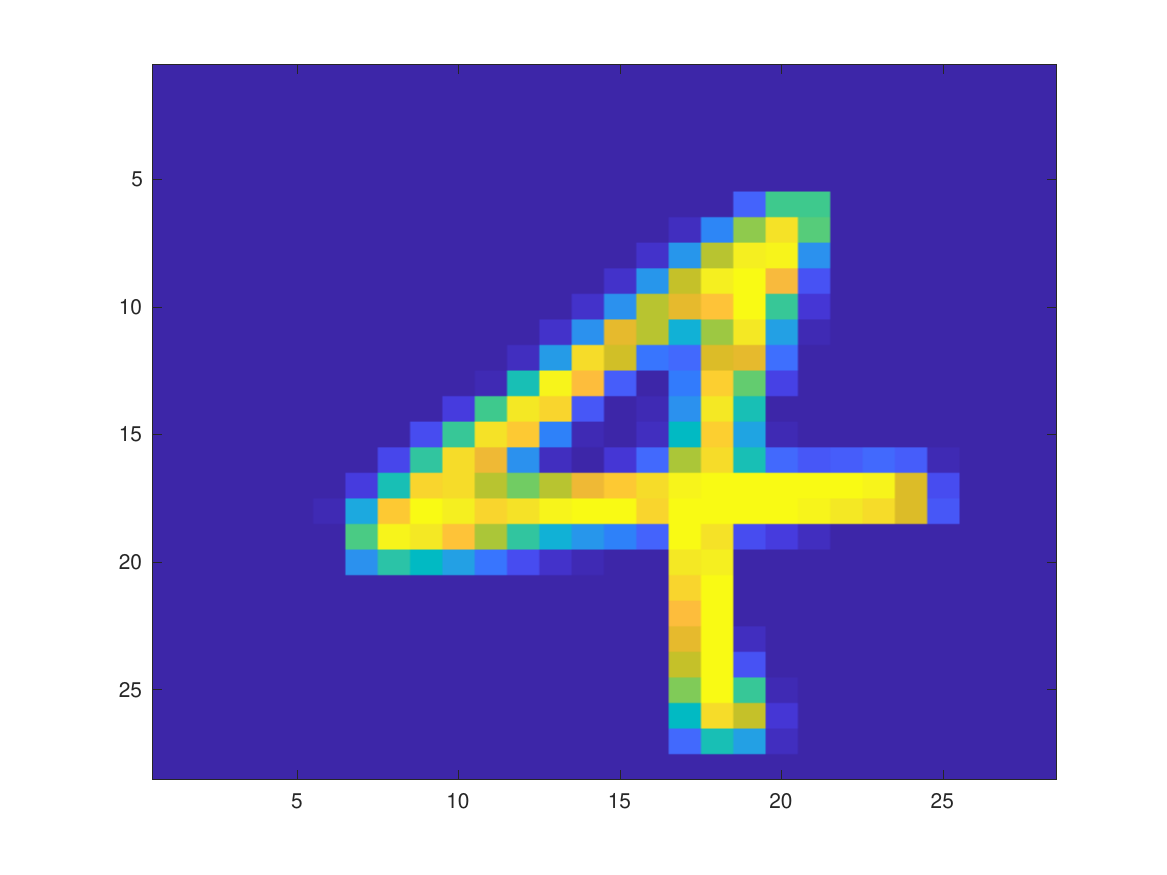}}
		\subfigure[]{	\includegraphics[width=0.3\textwidth]{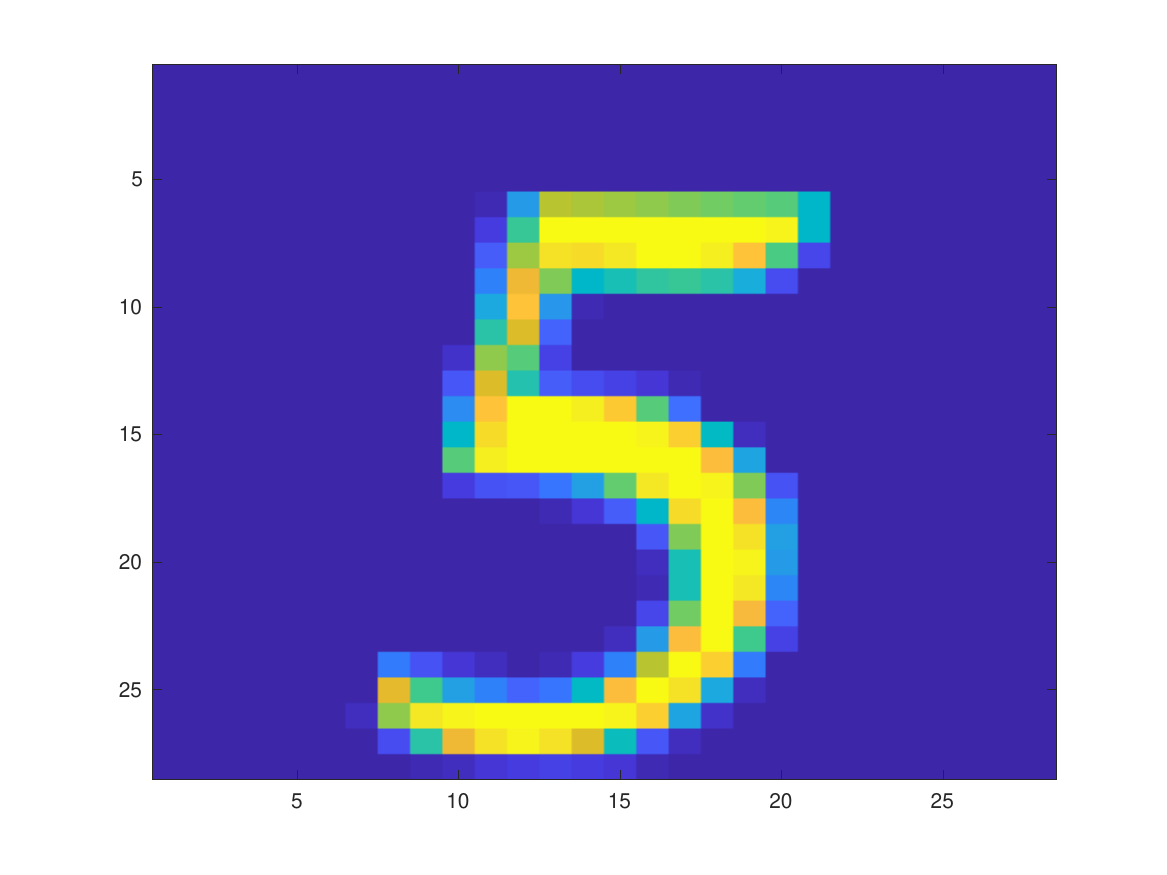}  }
		\subfigure[]{  \includegraphics[width=0.3\textwidth]{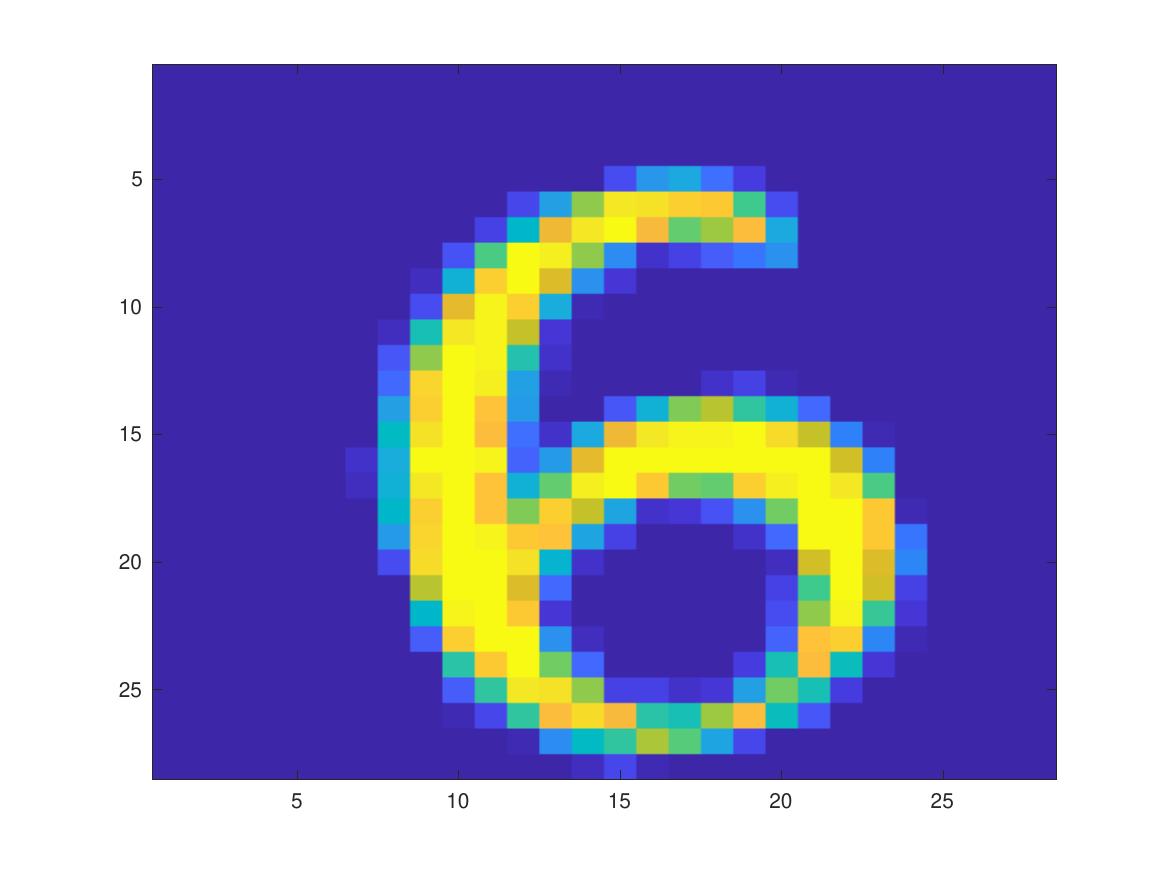}}
	}
	\vspace{-0.3cm}
	\caption{Patterns of the true parameter values (scaled according to range of each parameter), i.e., the ground-truths.}
	\label{fig:patternsTRUE}
\end{figure*}

\begin{figure*}[!ht]
	\centering
	\centerline{
		\subfigure[]{\includegraphics[width=0.3\textwidth]{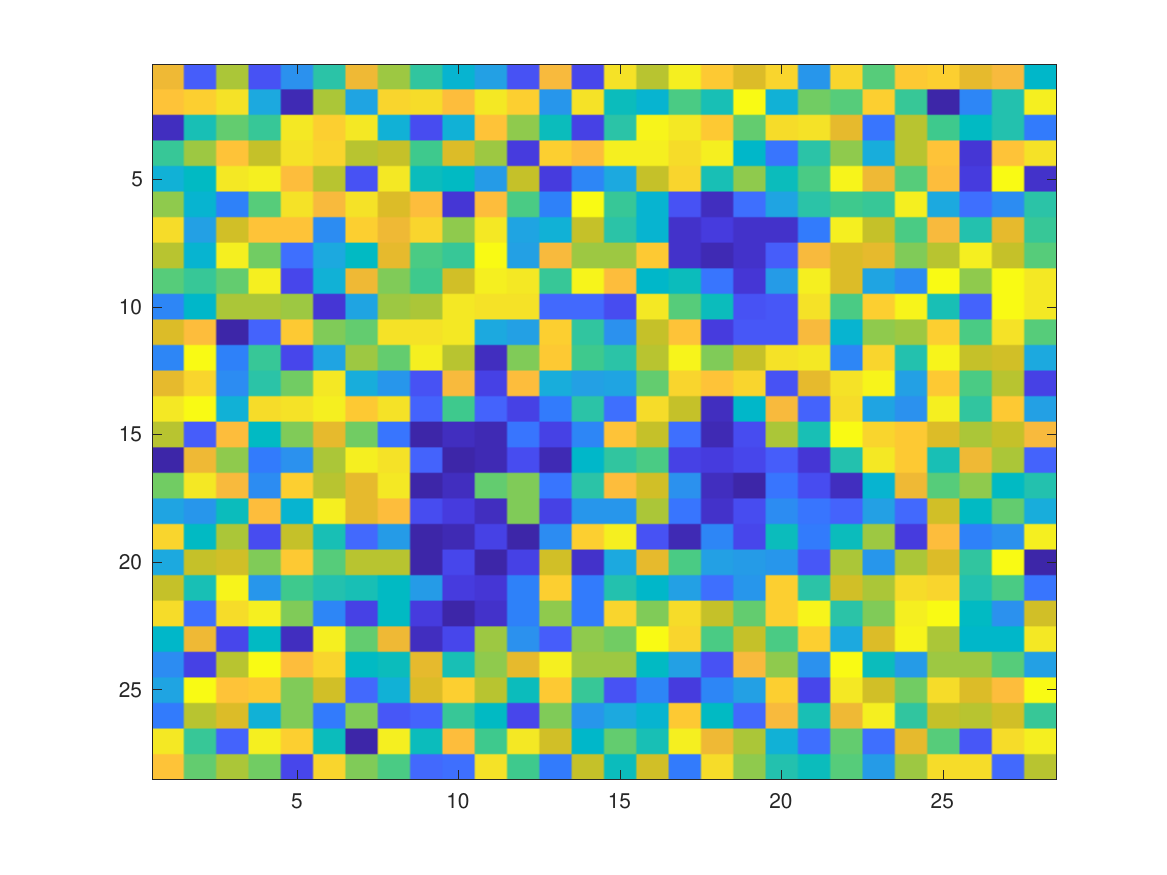}}
		\subfigure[]{\includegraphics[width=0.3\textwidth]{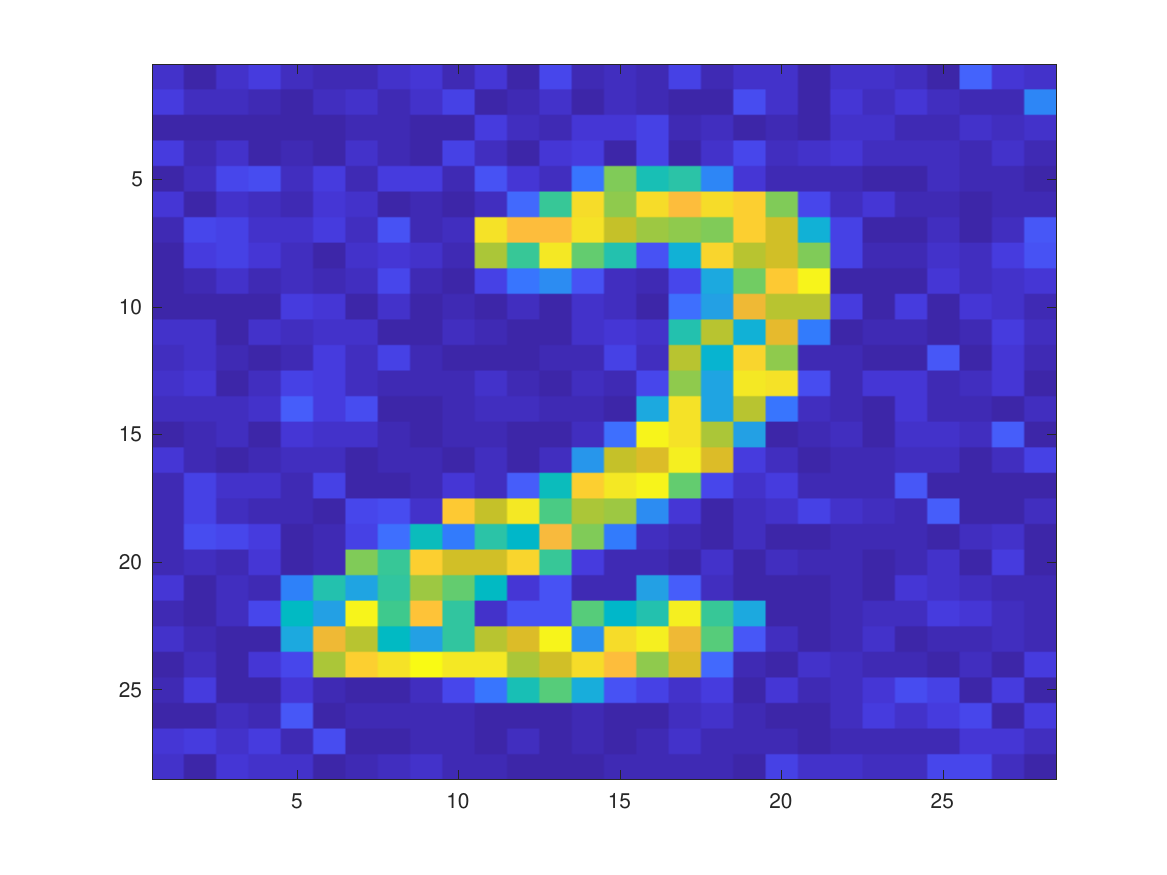}}
		\subfigure[]{ \includegraphics[width=0.3\textwidth]{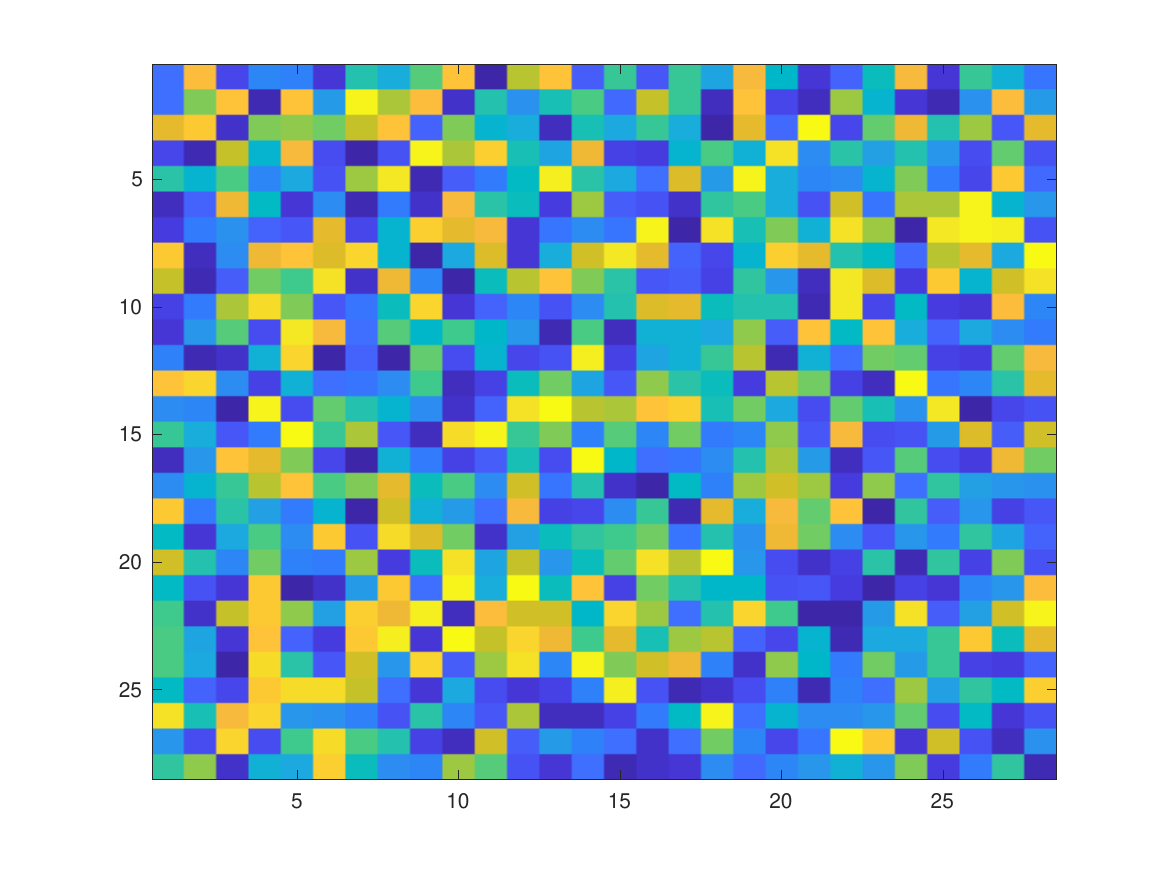}}
	}
	\centerline{
		\subfigure[]{ \includegraphics[width=0.3\textwidth]{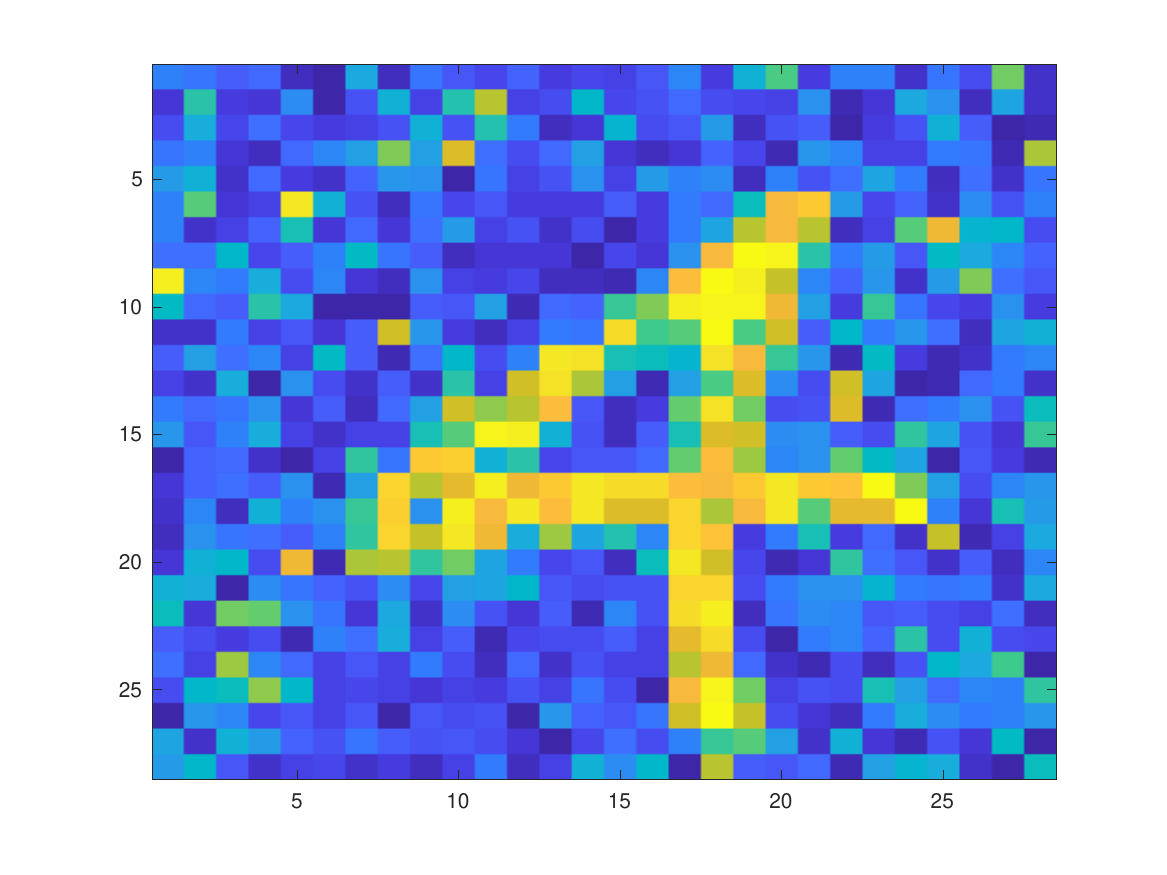}}
		\subfigure[]{\includegraphics[width=0.3\textwidth]{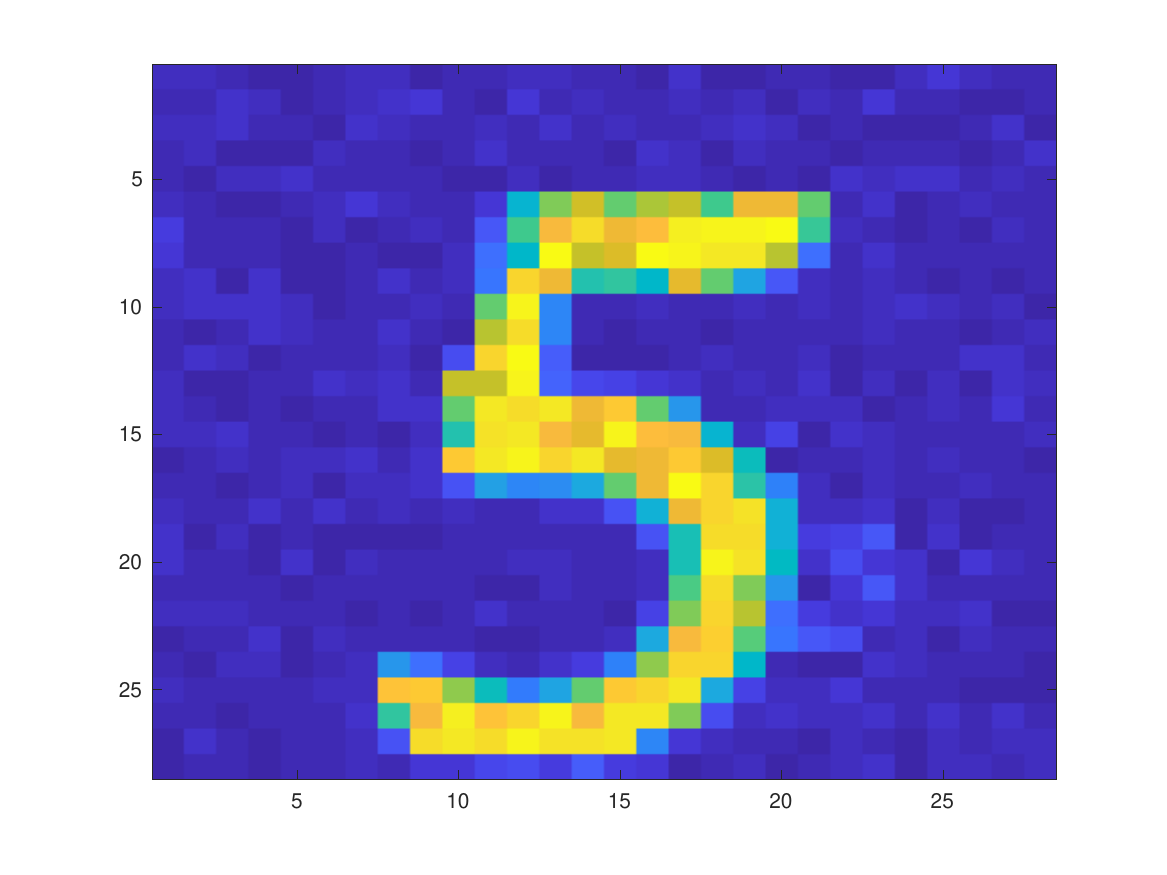} } 
		\subfigure[]{\includegraphics[width=0.3\textwidth]{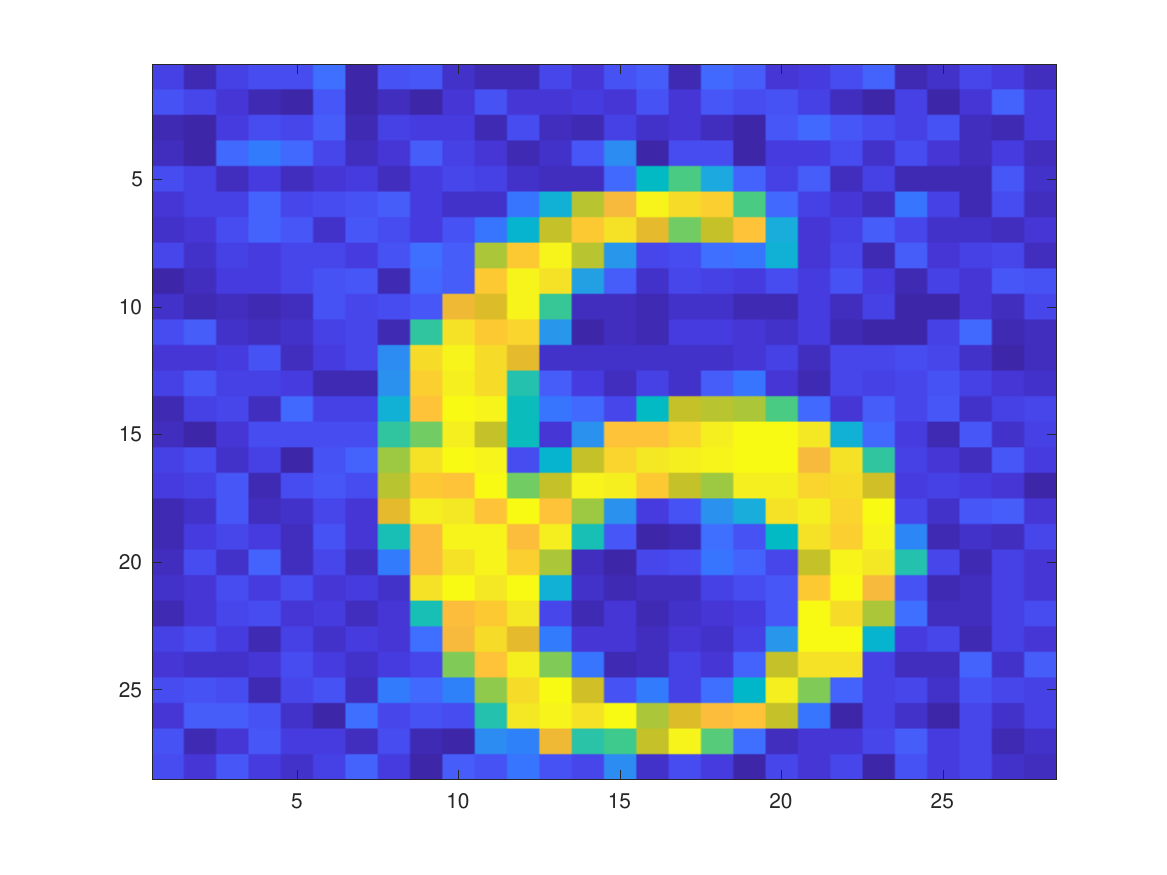}}
	}
	\vspace{-0.3cm}
	\caption{Recovered by standard IS. We can observe the difficulty in the retrieval of the first and third parameter.}
	\label{fig:patternsUNIF}
\end{figure*}

\begin{figure*}[!ht]
	\centering
	\centerline{
		\subfigure[]{\includegraphics[width=0.3\textwidth]{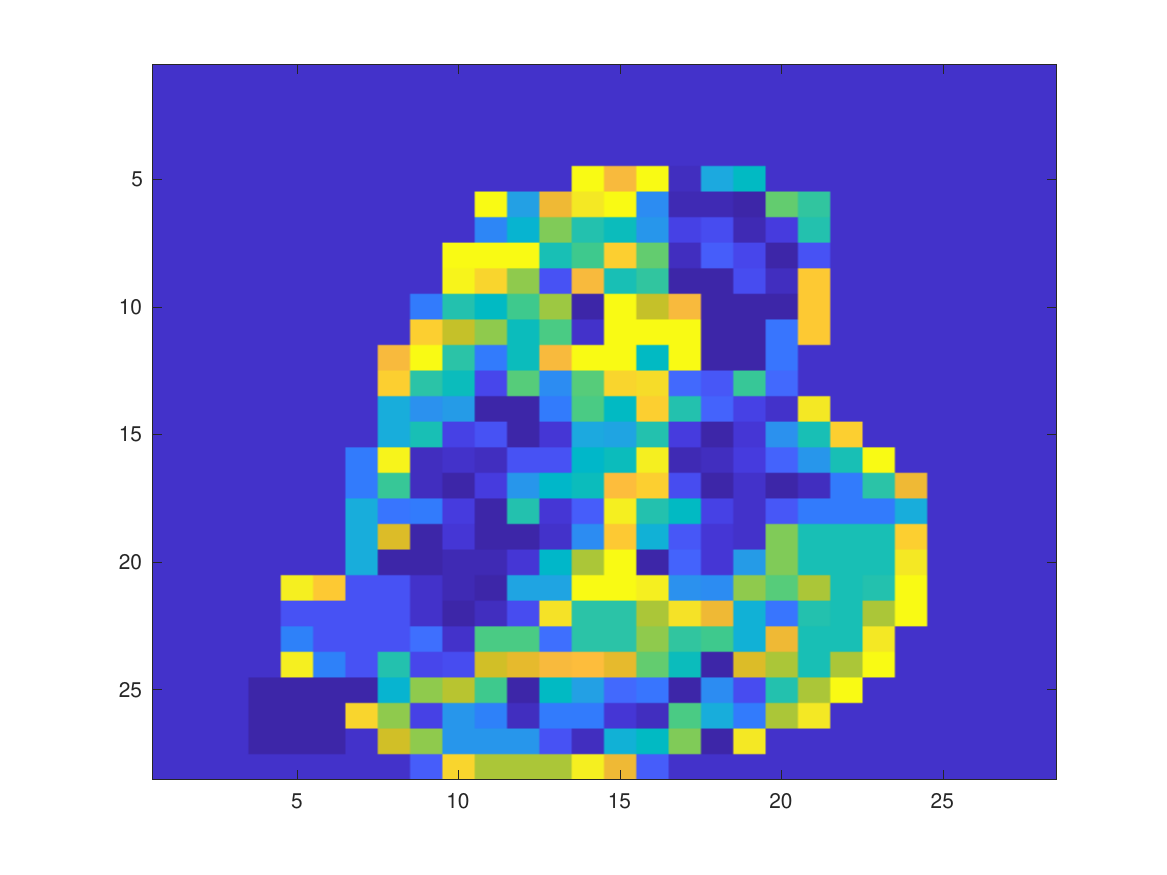}}
		\subfigure[]{\includegraphics[width=0.3\textwidth]{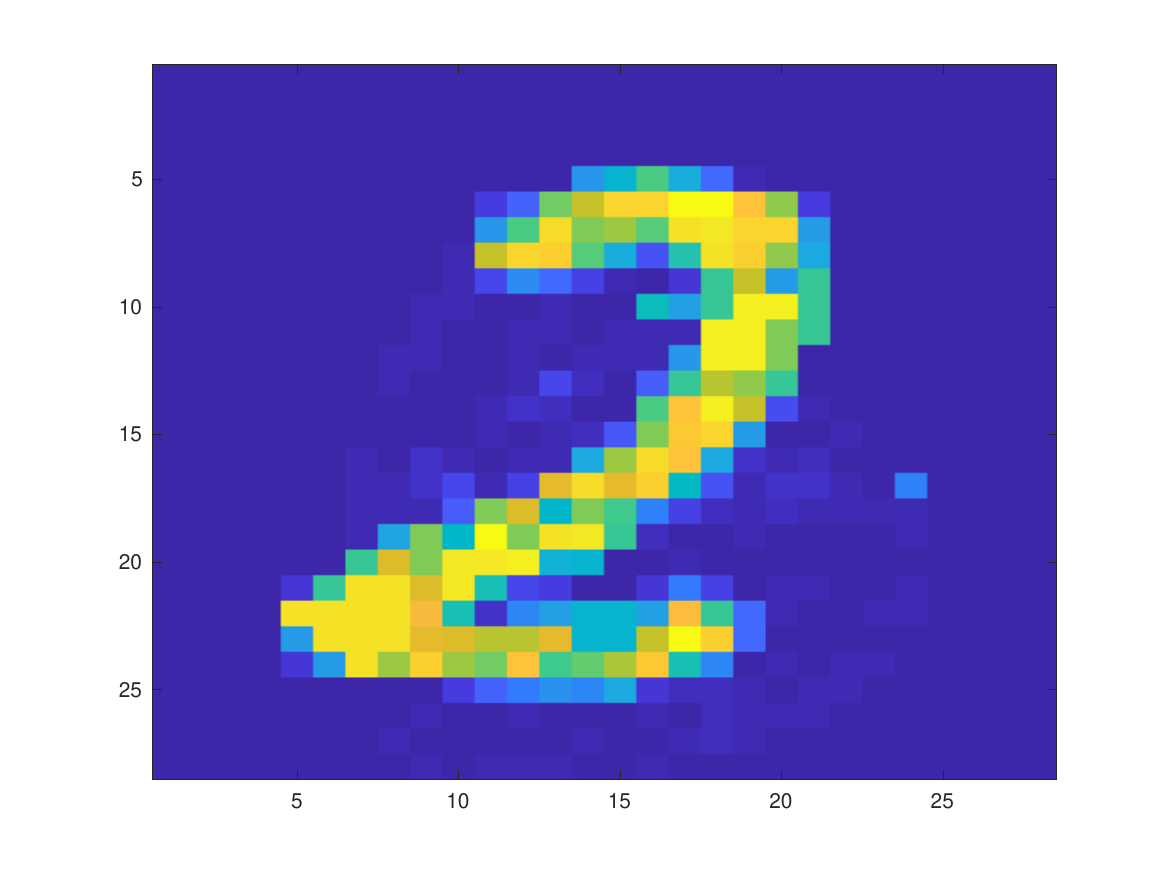}}
		\subfigure[]{ \includegraphics[width=0.3\textwidth]{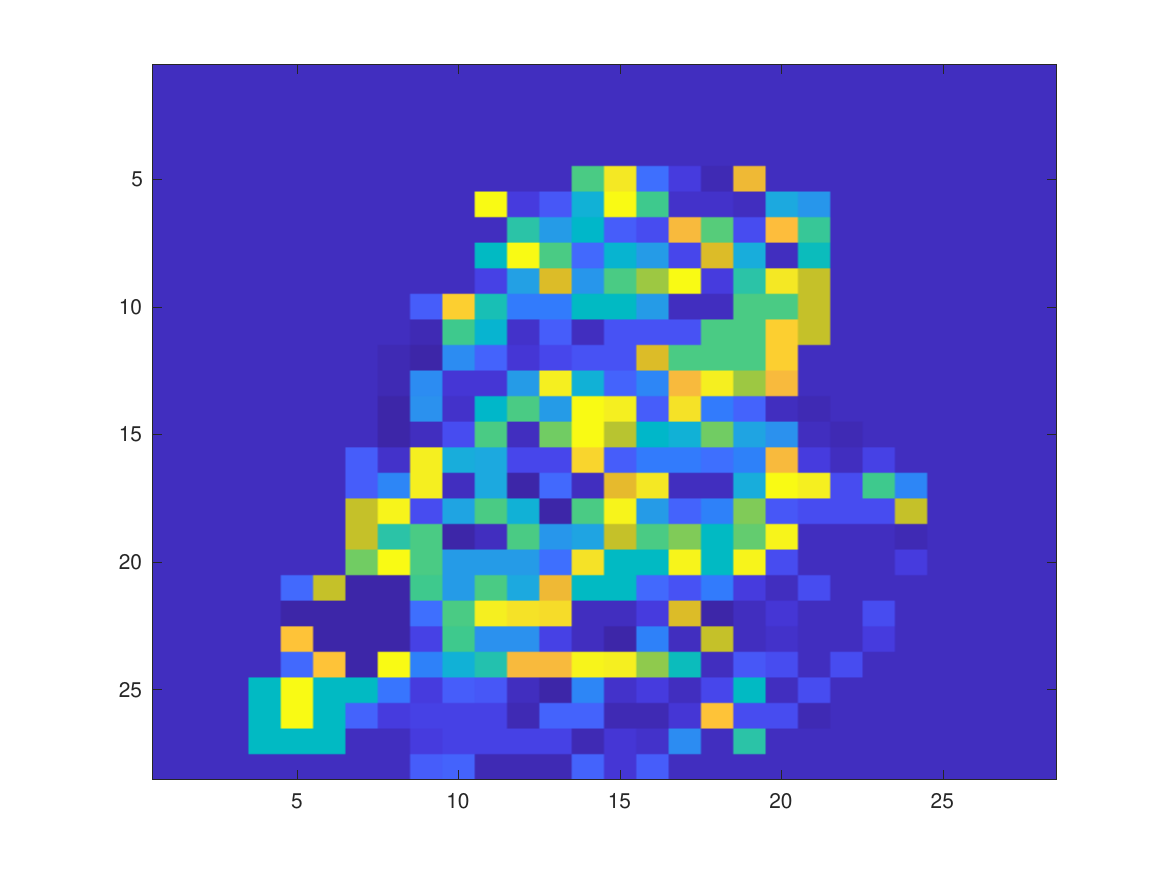}}
	}
	\centerline{
		\subfigure[]{  \includegraphics[width=0.3\textwidth]{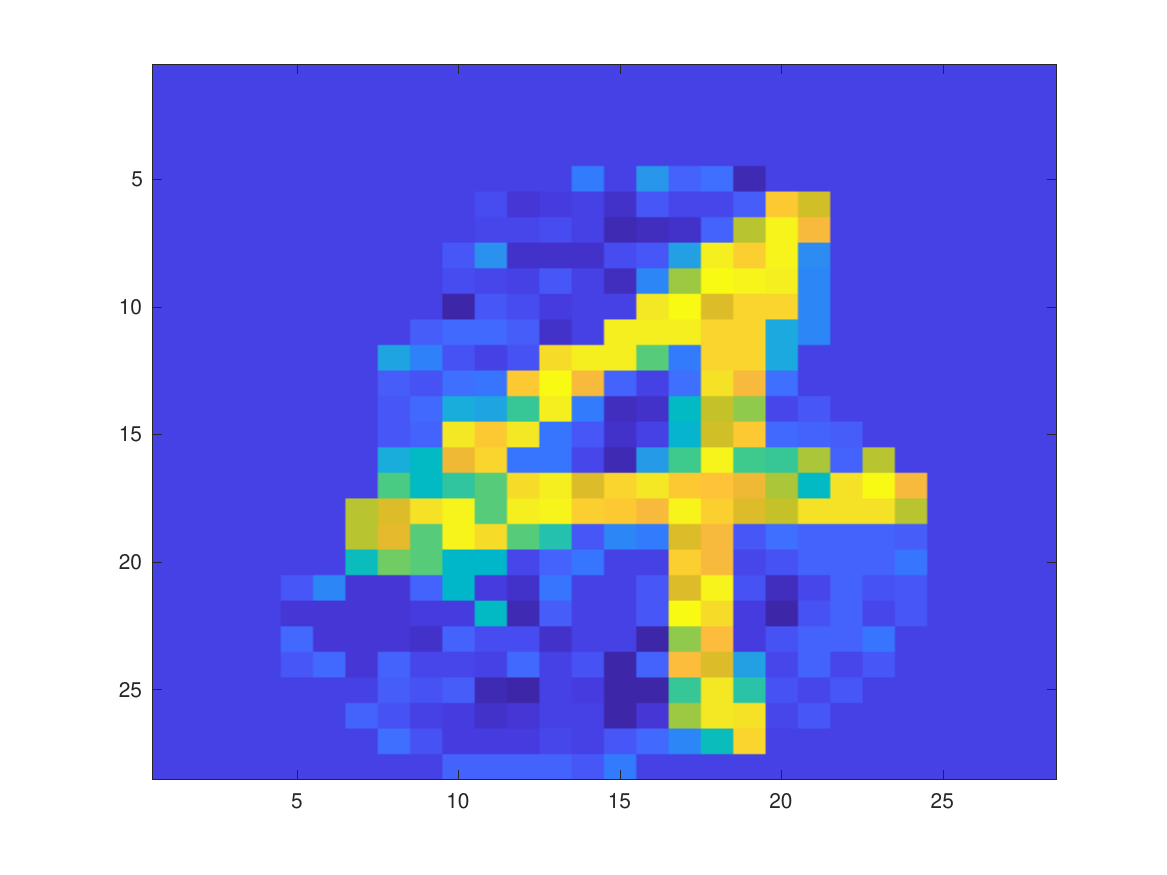}}
		\subfigure[]{  	\includegraphics[width=0.3\textwidth]{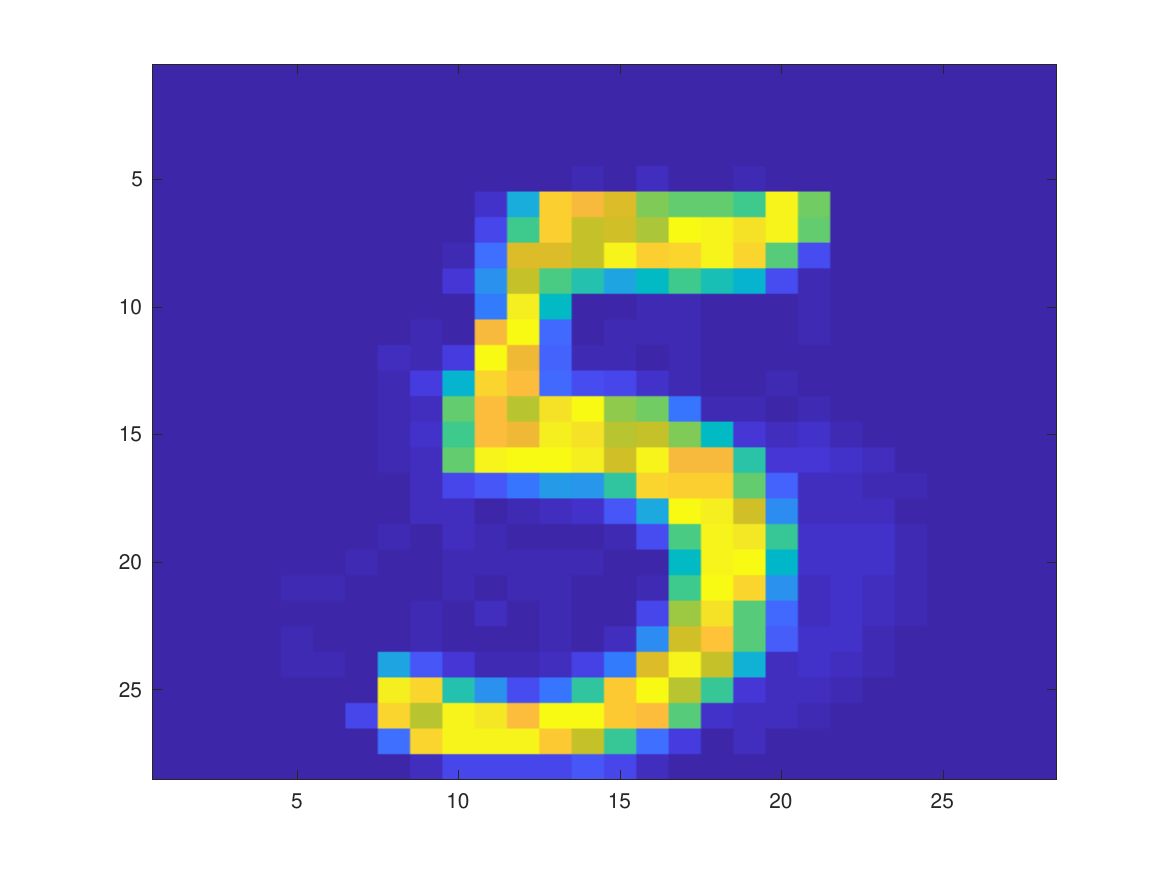} } 
		\subfigure[]{  \includegraphics[width=0.3\textwidth]{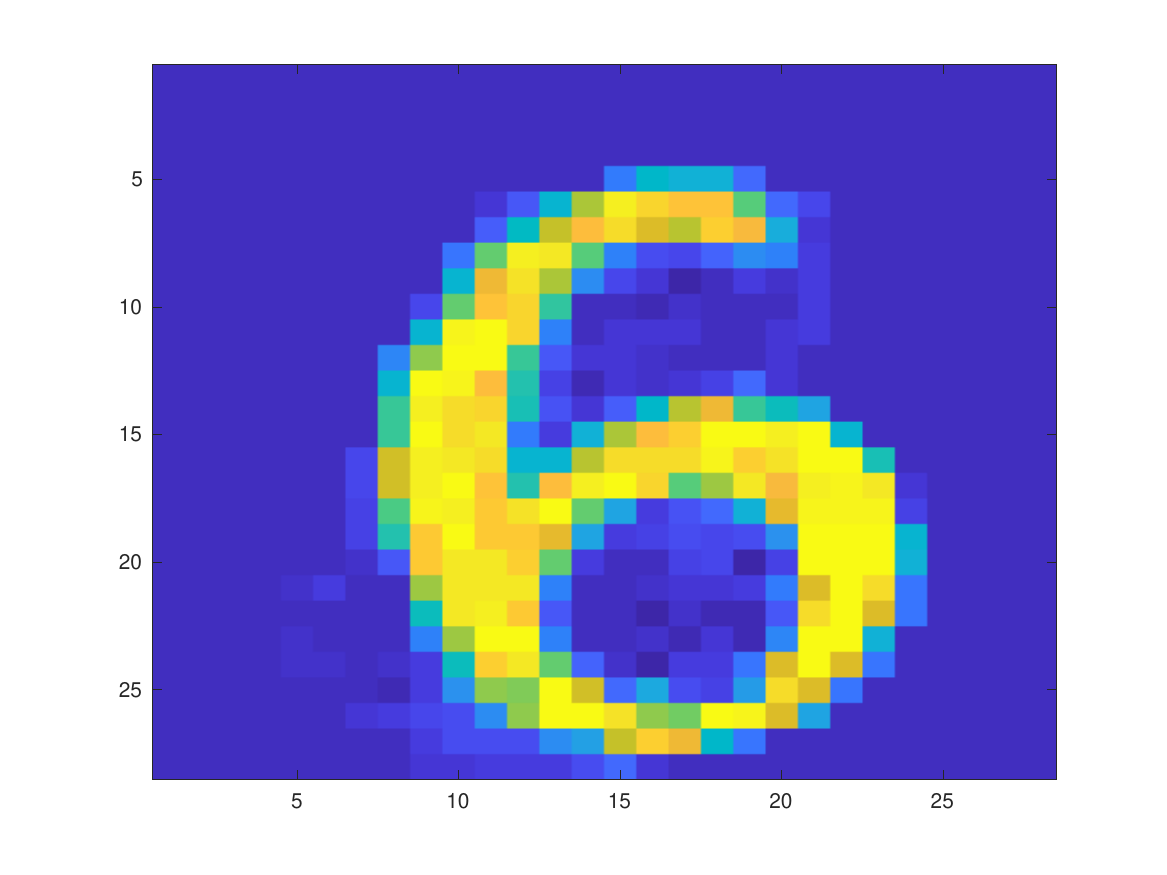}}
	}
	\vspace{-0.3cm}
	\caption{Recovered by  NN-AIS. We can observe the retrieval of first and third parameter is not completely successful.}
	\label{fig:patternsSticky}
\end{figure*}

\noindent {\bf Experiment.} 
In a real data settings, physical and geographical patterns are associated to the parameters $\x_r$ in the image. In order to check the performance of each algorithm, we consider synthetic data. Thus,  in this experiment, we havev also generated synthetic patterns in order to simulate a real scenario. In particular, we produce six patterns (recall $\x \in \mathbb{R}^6$) that represent handwritten digits (see Figure \ref{fig:patternsTRUE}).
Hence, in this setting, we have $R=784$ different observation vectors $\y_r$, $r=1,\dots,R$, for which we want to estimate the vectors of true values $\x_r$, $r=1,\dots,R$. Each observation corresponds to a single pixel of a 28$\times$28 image. 
%These observations were generated such the corresponding true values show different patterns when plotted. 
%Since $\x_r \in \mathbb{R}^6$, we have a total of six different patterns displayed in images of $28 \times 28$ pixels (see Figure \ref{fig:patternsTRUE}) that we want to recover.
We also compute the maximum a-posteriori (MAP) of $p_r(\x_r|\y_r)$, $\x_{r,\text{MAP}}$, as estimate of $\x_r$. 
\newline 
{\bf Methods.}
We use the NN-AIS scheme to estimate $\x_{r,\text{MAP}}$ for $r=1,\dots, 784$, and compare it against IS using the prior as proposal density, in terms of relative squared error and by looking at the recovered images. As parameters of our scheme we chose $N_\text{init}=1000$, $T=20$, $N=250$ and $L=10^5$. The $N_\text{init}$ initial points were taken at random in the domain except for $11$ points that were placed in the vertices of the domain. 
Our scheme allows for sharing information from problem to the next one, so
we also use the $\widehat{\x}_{s,\text{MAP}}$ for $s=1,\dots,r-1$ as initial nodes when estimating $\x_{r,\text{MAP}}$. Note that this is completely fair since the model has been already evaluated at those points. The comparison is fair in terms of model evaluations, with a total of $E=6000$ for each $r=1,\dots,784$. 
\newline 
{\bf Results.}
The results are shown in Figures 
\ref{fig:patternsUNIF} and \ref{fig:patternsSticky}. It can be seen that both standard IS and NN-AIS are able to correctly recover components 2, 4, 5 and 6 of $\x_r$ ($r=1,\dots,784$), i.e., the images of ``2'', ``4'', ``5'' and ``6'' in both Figure \ref{fig:patternsUNIF} and Figure \ref{fig:patternsSticky}  look  very close to the true ones (Figures \ref{fig:patternsTRUE}(b),(d),(e) and (f) respectively). The images recovered by NN-AIS have lower noise though. The components 1 and 3 of the $\x_r$'s are completely lost with standard IS (see Figure \ref{fig:patternsUNIF}), whereas NN-AIS is able at least to achieve to recover the boundaries of the corresponding patterns. Indeed, NN-AIS obtains a much lower error in estimation, as it is shown in Table \ref{table_RMAE} and Table \ref{table_MAE}. The difficulty in recovering the components 1 (i.e., $S_{st}$) and 3 (i.e., $C_{ar}$) deserves further studies. This issue could be related to some relevant features of PROSAIL (e.g., the average partial derivatives with respect to these two components). We leave the study of these specific issues for future work. %We leave it as future work.
In Table \ref{table_spectra}, we also show the averaged error in the spectra produced by both methods as compared to the true observations.

\begin{table*}[!ht]
	\centering
	%\small
	\caption{Relative Mean Absolute Errors (RMAE) for each component (averaged over all spectra).}
	\vspace{0.2cm}
	%	{\footnotesize
	\begin{tabular}{|c|c|c|c|c|c|c|c|}
		\hline	
		% & & & & & & &  \\
		Components & 1 & 2 & 3 & 4 & 5 & 6 & {\bf Mean}\\
		\hline
		\hline
		Stand. IS & 0.7556  & 0.4397 & 2.9431 & 0.6247 &  0.2096 & 0.2782  &  {\bf 2.8516} \\
		Sequential NN-AIS & 0.2045 &  0.2245 & 0.8891  & 0.1985 &   0.1425 &  0.1320 & {\bf 1.0715} \\
		\hline
	\end{tabular}
	%	}	
	\label{table_RMAE}
\end{table*}

\begin{table*}[!ht]
	\centering
	%\small
	\caption{ Mean Absolute  Errors (RMAE) for each component (averaged over all  spectra).}
	\vspace{0.2cm}
	%	{\footnotesize
	\begin{tabular}{|c|c|c|c|c|c|c|c|}
		\hline	
		% & & & & & & &  \\
		Components & 1 & 2 & 3 & 4 & 5 & 6 & {\bf Mean}\\
		\hline
		\hline
		Stand. IS &  0.9760 &  6.1754 &  9.8204 & 0.1348 &  0.0016 &  0.0012 & {\bf 0.8752} \\
		Sequential NN-AIS  &  0.2641  & 3.1535 & 2.9667 & 0.0428 &  0.0011 &  0.0006 &  {\bf 0.2985} \\
		\hline
	\end{tabular}
	%	}	
	\label{table_MAE}
\end{table*}

\begin{table*}[!ht]
	\centering
	%\small
	\caption{ Absolute and relative error (averaged over all the  pixels) in the transformed domain (``reconstruction of the spectra'')}
	\vspace{0.2cm}
	%	{\footnotesize
	\begin{tabular}{|c|c|c|}
		\hline	
		% & & & & & & &  \\
		& {\bf Absolute} & {\bf Relative}\\
		\hline
		\hline
		Stand. IS & 66.4395 & 0.0802\\
		Sequential NN-AIS & 11.0844 & 0.0198\\
		\hline
	\end{tabular}
	%	}	
	\label{table_spectra}
\end{table*}

%%%%%%%%%%%%%%%%%%%%%%%%%%%%%%%%
%%%%%%%%%%%%%%%%%%%%%%%%%%%%%%%%
\section{Conclusions and future lines}
%%%%%%%%%%%%%%%%%%%%%%%%%%%%%%%%
%%%%%%%%%%%%%%%%%%%%%%%%%%%%%%%%

In this work, we introduced a novel framework of  adaptive importance sampling algorithms. % based on a two-stage deep IS approach. 
The key idea is the use of a non-parametric proposal density built by a regression procedure (the emulator), that mimics the true shape of  posterior pdf. Hence, the proposal pdf represents also a surrogate model, that is in turn adapted through the iterations by adding new support points. The regression (e.g., obtained by nearest neighbors and Gaussian processes) can be applied directly on the posterior domain or, alternatively, in just one piece of the likelihood, such as an arbitrary physical model.
Drawing from the emulator is possible by a deep architecture of two nested IS layers. More sophisticated deep structures, employing a a chain of emulators, have been described. 

RADIS is an extremely efficient importance sampling scheme since  the emulator (used as proposal pdf)  becomes closer and closer to the true posterior, as new nodes are incorporated.   As  a  consequence,  RADIS  asymptotically  converges  to  an exact sampler under mild conditions.
Several numerical experiments and theoretical supports  confirm these statements.   
% When used as an importance sampler, it has higher efficiency than other IS algorithms with fixed proposals, since the non-parametric proposal gets closer and closer to the target density. When used as an optimization technique, it is a gradient-free algorithm that adapts the proposal in an efficient way using past information. In addition, at the end of the algorithm, the resulting proposal constitutes a cheap surrogate model of the true target function. This is particularly appealing in inverse problems where one is interested in obtaining an emulator of the costly physical model. 
%\newline
Robust accelerating versions of RADIS have been also presented, as well as combinations with other benchmark AIS algorithms. 
{ Cheap constructions of the emulator have been also discussed and tested.} 
%The combination of RADIS with a suitable extra parametric proposal reduces the dependence on the starting nodes and ensure the convergence of the non-parametric proposal to the posterior. 
%Finally, we have described the use of RADIS in sequential settings, where information is shared among different inversion problems.
The use of RADIS within a sequential Monte Carlo scheme will be considered in future works.
Furthermore, as future research lines, we also plan to analyze in depth the PROSAIL inversion problem, approximating the partial derivatives with respect some specific parameters by RADIS. Moreover, we also plan to consider the adaptation of the auxiliary proposal $\bar{q}_\text{aux}(\x)$, adding also  additional layers in the proposed deep architecture.

%%%%%%%%%%%%%%%%
\section{Acknowledgements}
%%%%%%%%%%%%%%%%%
This work has been supported by Spanish government via grant FPU19/00815.

\bibliographystyle{plain}

\bibliography{bibliografia,gus}

\begin{appendices}

%%%%%%%%%%%%%%%%%%%%%%%%%%%%%%%%%%%%%%%%%%%%%% 
\section{Theoretical support}\label{APP:theo}
%%%%%%%%%%%%%%%%%%%%%%%%%%%%%%%%%%%%%%%%%%%%%% 

%\begin{itemize}
%	\item la convergencia del metodo es la de IS convencional, pero tenemos que asegurarnos que sticky$+$param tenga colas mas pesadas que la target para evitar pesos infinitos...
%\end{itemize}

In this section, we discuss several theoretical aspects of RADIS.
First, we address the error in the approximate sampling and evaluation of the interpolating proposal.
Then, we show that the adaptive construction of the proposal decreases the distance with respect to the true target as the number of nodes $J_t$ grows. Finally, we show that this also minimizes the variance of the IS weights.

\subsection{Sampling Importance Resampling (SIR)}\label{App:SIRasymptotics}

Let $\widehat{\pi}_t(\x)$ the unnormalized interpolating proposal from which we aim to sample. Its normalizing  constant $c_t$ is not important in this first part. The SIR method allows to sample from the density $\widehat{\pi}_t$ by resampling a sample drawn from another auxiliary (importance) density \cite{Rubin88}\cite[Chapter 24]{gelman2004applied}. This method is also referred as the weighted bootstrap in \cite[Sect. 3.2]{smith1992bayesian}. The SIR algorithm is as follows:
\begin{enumerate}
	\item Draw $\{\x_1,\dots,\x_L\}$ i.i.d. from $\bar{q}_\text{aux}(\x)$, that is a density with fatter tails than $\widehat{\pi}_t(\x)$.
	\item Calculate the importance weights for each $\x_i$
	$$\gamma_i = \gamma(\x_i) = \frac{\widehat{\pi}_t(\x_i)}{\bar{q}_\text{aux}\x_i)}.$$
	\item Resample $N$ ($N\leq L$) values $\{\x^*_1,\dots,\x^*_N\}$ from $\{\x_1,\dots,\x_L\}$ with probabilities proportional to $\gamma_i$ assigned to $\x_i$. 
\end{enumerate}
If $L\rightarrow \infty$, or more precisely $\frac{L}{N}\rightarrow \infty$, then the set $\{\x^*_1,\dots,\x^*_N\}$  is asymptotically distributed  as $\widehat{\pi}_t(\x)$.  
Thus,  the choice of $L$ and $N$ is important for two factors: (i) to reduce the dependence of the $\x^*_i$'s, and (ii) to have the distribution of $\x^*_i$ as close to $\widehat{\pi}_t$ as possible. The relative magnitude between $N$ and $L$ controls this dependence, while only the magnitude of $L$ affects how close the distribution of the resampled particle is to the density  $\widehat{\pi}_t$. 
%The proposal $\bar{q}_\text{aux}(\x)$ must have the same  support of $\widehat{\pi}_t(\x)$, and it should have fatter tails than $\widehat{\pi}$, to avoid to have weights with  infinite variance \cite{Robert04}.
%Overall, the closer $\bar{q}_\text{aux}(\x)$ is to $\widehat{\pi}$ the greater the efficiency will be. 
\newline
\newline
{\bf Bias and correlation in SIR.} Under mild conditions, as $\frac{L}{N}\to \infty$, the density of resampled particle converges to
$ \widehat{\pi}_t(\x)$. 
For more details see \cite[Sect. 6.2.4]{givens2012computational}, \cite[Sect. 3.2]{smith1992bayesian} and \cite[page 6 ]{robert2020markov}. 
As SIR is an approximate sampling algorithm, it has some bias\footnote{Measured as the difference in the probability of some set between the target pdf and the ``equivalent'' pdf}. 
If the first and second moment of the IS weight $\gamma(\x) = \frac{\widehat{\pi}(\x_i)}{\bar{q}_\text{aux}(\x)}$ exists, it can be shown that this bias vanishes at $\mathcal{O}(L^{-1})$ rate \cite[Chapter 24]{gelman2004applied}. 
In \cite[Sect. 3.2]{smith1992bayesian}, they show the convergence of the cdf of the resampled particle as $L \to \infty$ in the the univariate case.  
\newline
Resampling $N$ times from a unique pool of $L$ samples from $\bar{q}_\text{aux}(\x)$ introduces correlation in the resampled sample. However, when $N \ll L$, this correlation is negligible. Some heuristics suggest $\frac{L}{N} = 20$ \cite{Rubin88}, or $\frac{L}{N}\geq 10$ \cite{smith1992bayesian}. For more details in the relation of the values of $L$ and $N$ see \cite{gelman2004applied}[Sect. 24.3].
%Resampling from a set of $L$ samples of $\bar{q}_\text{aux}(\x)$ , weighted according to $\widehat{\pi}$, accounts to sampling from a particle approximation of $\widehat{\pi}$, a pdf which is actually different from $\widehat{\pi}$. 
In \cite{martino2018group} (see Figure 5 and Appendix A therein), it is shown the ``equivalent'' density of a resampled particle for a fixed value of $L$, which converges to the target pdf as $L$ diverges.
%$$ \widetilde{q}(\x) = \widehat{\pi}_t(\x)\int_{\mathcal{X}^{L-1}}\frac{1}{\widehat{c}_t}\left[\prod_{i\neq j}^L\bar{q}_\text{aux}\x_i)\right]d\x_{-j}.
%$$
%Assuming that $\bar{q}_\text{aux}(\x)$ has fatter tails than $\widehat{\pi}$, it can be shown that the pdf $ \widetilde{q}(\x) $ becomes closer and closer to $\frac{1}{c_t}\widehat{\pi}_t(\x)$ as $L$ grows.
Furthermore, for computing the denominator of the outer weights we need the normalizing constant of $\widehat{\pi}_t$ (see Eq. \eqref{eq:OuterISweights}).
In this sense, the inner IS also provides with an approximation by using the $L$ samples from $\bar{q}_\text{aux}(\x)$,
\begin{align}
	\widehat{c}_t = \frac{1}{L}\sum_{\ell=1}^{L}\frac{\widehat{\pi}_t(\z_\ell)}{\bar{q}_\text{aux}(\z_\ell)}.
\end{align} 
This estimate converges as $L \to \infty$
\cite{Robert04}.

\subsection{Variance of the IS weights}\label{App:varISweights}

Let $w(\x)=\frac{\pi(\x)}{\bar{q}(\x)}$ be the weight function evaluated at samples $\x \sim \bar{q}(\x)$. First of all, note that $E[w(\x)]=Z$. Below, we show that the variance of $w(\x)$ is proportional to the Pearson divergence between the posterior $\post$ and proposal $q$, i.e.,
\begin{align}
	\texttt{var}[w(\x)] &= \int_{\mathcal{X}} (w(\x) - Z)^2\bar{q}(\x)d\x \\
	&= \int_{\mathcal{X}} \left(\frac{\pi(\x)- Z\bar{q}(\x)}{\bar{q}(\x)}\right)^2\bar{q}(\x)d\x \\
	&=Z^2\int_{\mathcal{X}} \frac{(\post(\x)-\bar{q}(\x))^2}{\bar{q}(\x)}d\x = Z^2\chi^2(\post\|q),
\end{align}
where  $\chi^2(\post\|q)= \int_{\mathcal{X}} \frac{(\post(\x)-\bar{q}(\x))^2}{\bar{q}(\x)}d\x$,
is the Pearson divergence and we have used $\bar{\pi}(\x)=\frac{1}{Z} \pi(\x)$. Hence, if we construct a proposal such $\chi^2(\post\|q) \to 0$, we would obtain $\texttt{var}[\widehat{Z}]=0$. Moreover, the mean square error (MSE) of $\widehat{I}$ can also be shown to be bounded by this divergence (see e.g. \cite{akyildiz2019convergence})
\begin{align}
	\mathbb{E}[|I - \widehat{I}|^2] \leq \frac{C_f (\chi^2(\post\|\bar{q}) + 1)}{N}.
\end{align}
Thus, it is beneficial to reduce the $\chi^2(\post\|\bar{q})$ in order to obtain accurate IS estimators.
%%%%%%%%%%%%%%%%%%%%%%%%%%%%%%%
\subsection{Pearson divergence and $L_p$ distances}
%%%%%%%%%%%%%%%%%%%%%%%%%%%%

Now, we aim to show that $\chi^2(\post\|\bar{q})$ can be bounded in terms of the $L_2$ and $L_\infty$ distances, between $\bar{\pi}(\x)$ and $\bar{q}(\x)$. 
Using Holder's inequality and the fact that pdfs are always positive, we can write
\begin{align}
	\chi^2(\post\|\bar{q}) = \int_{\mathcal{X}} |\post(\x)-\bar{q}(\x)|\frac{|\post(\x)-\bar{q}(\x)|}{|\bar{q}(\x)|}d\x 
	&= \norm{(\post-q)\left(\frac{\post-\bar{q}}{\bar{q}}\right)}_{L_1} \nonumber\\
	&\leq \norm{\post-\bar{q}}_{L_2}\norm{\frac{\post-\bar{q}}{\bar{q}}}_{L_2}.
\end{align}
The $L_2$ distance can be easily shown to be bounded by $L_\infty$ distance (considering a bounded domain $\mathcal{X}$), i.e.,
\begin{align}\label{IneAqui0}
	\|\post - \bar{q}\|_{L_2} = 
	\left( \int_\mathcal{X} |\post(\x) - \bar{q}(\x)|^2d\x \right)^\frac{1}{2} &\leq \left(|\mathcal{X}|\max |\post(\x)-\bar{q}(\x)|^2  \right)^\frac{1}{2} \nonumber \\
	&= |\mathcal{X}|^\frac{1}{2} \|\post - \bar{q}\|_{L_\infty}.
\end{align}
Similarly, we have
\begin{align}\label{IneAqui1}
	\norm{\frac{\post - \bar{q}}{\bar{q}}}_{L_2} \leq |\mathcal{X}|^\frac{1}{2}\norm{\frac{\post - \bar{q}}{\bar{q}}}_{L_\infty}.
\end{align}
Thus, we can obtain the following result regarding the $L_\infty$ distance,
\begin{align}\label{IneAqui2}
	\chi^2(\post\|\bar{q}) \leq |\mathcal{X}| \norm{\frac{\post - \bar{q}}{\bar{q}}}_{L_\infty}
	\|\post - \bar{q}\|_{L_\infty}.
\end{align}
Since we choose $\bar{q}$ in order to have fatter tails than $\bar{\pi}$ and since $\bar{q}$, $\bar{\pi}$ are bounded, then the factor
$\|\frac{\post - \bar{q}}{\bar{q}}\|_{L_\infty}$ in  \eqref{IneAqui2} vanishes to zero if $\|{\post - \bar{q}}\|_{L_\infty} \to 0$.
Therefore, if $\|{\post - \bar{q}}\|_{L_\infty} \to 0$, we have 
$\chi^2(\post\|\bar{q}) \to 0$. Due to \eqref{IneAqui0}-\eqref{IneAqui1}, this result is also valid for the $L_2$ distance. 
In this work, we consider $\bar{q} = \bar{q}_t = \frac{1}{c_t}\widehat{\pi}_t$ such $\|{\pi - \widehat{\pi}_t}\|_{L_\infty} \to 0$ as $t \to \infty$ (see section below), and thus $\|{\post - \frac{1}{c_t}\widehat{\pi}_t}\|_{L_\infty} \to 0$, that  implies $\chi^2(\post\|\frac{1}{c_t}\widehat{\pi}_t) \to 0$.

\subsection{Convergence of the emulator to target function}\label{App:surrConv}

For simplicity, let us focus on the interpolation case and a bounded $\mathcal{X}$.
Here, we show that the interpolating constructions of Sect. \ref{sec:buildProp}, jointly with the adaptation process, lead to a proposal $\frac{1}{\widehat{c}_t}\widehat{\pi}_t(\x)$ that converges to $\post(\x)$. Since $\widehat{c}_t$ is an unbiased estimation of the area below $\widehat{\pi}_t(\x)$, we focus on  the convergence of $\widehat{\pi}_t(\x)$ to the unnormalized posterior $\pi(\x)$. As $\widehat{\pi}_t(\x)\rightarrow \pi(\x)$, then $\widehat{c}_t\rightarrow Z$.
In Sect. \ref{sec:robustSchemes}, we have introduced an extra parametric density $\bar{q}_\texttt{par}(\x)$ to  also ensure that new points can be added in any region of the domain $\mathcal{X}$  during the adaptation.
We show below that, when using the NN or GP constructions, the approximation error of $\widehat{\pi}_t$ depends on a quantity called {\it fill distance}, 
\begin{align}
	r_t = \max_x \min_{i=1,\dots,J_t} \|\x- \x_i\|_2,
\end{align}
%where $J_t$ denotes the total number of nodes at time $t$,
which measures the filling of the space. In other words, the greater the fill distance, the less covered the space is. For both constructions, decreasing the fill distance ensures that $\widehat{\pi}_t(\x)$ converges in $L_\infty$ norm to $\pi(\x)$. 
Using a $\bar{q}_\texttt{par}(\x)$ that is not negative in $\mathcal{X}$ ensures every region will be covered eventually, i.e., $r_t\to 0$ as $t\to\infty$.
\newline
\newline
{\bf NN construction.} If $\pi$ is Lipschitz continuous, we have that
\begin{align}
	\norm{\pi - \widehat{\pi}_t}_\infty \leq L_0 r_t,
\end{align}
where $L_0$ is the Lipschitz constant and $r_t$ denotes the fill distance \cite{llorente2020adaptive}[App. D.4]. Equivalently, we have \cite{butler2017measure}
\begin{align}
	\norm{\pi - \widehat{\pi}_t}_\infty \leq L_0 \max_{i=1,\dots,J_t} \text{diam}(\mathcal{R}_i),
\end{align}
that is, the approximation error is bounded by the biggest Voronoi cell. Covering the space (not necessarily with uniform points) ensure that $\max_i \text{diam}(\mathcal{R}_i) \to 0$ \cite{devroye2017measure} (equivalently $r_t \to 0$), and thus $\widehat{\pi}_t \to \pi$ as $t\to\infty$. 
%Note that convergence in $\norm{\cdot}_\infty$ implies convergence in $\norm{\cdot}_2$ (see, e.g., \cite{llorente2020adaptive}[App. D.1]).
\newline
\newline	
{\bf GP construction.} 
First, we recall a result valid when the GP regression is applied on $\pi$, not a transformation. It can be shown that the approximation error $\|\pi - \widehat{\pi}_t\|_\infty$ is bounded in terms of the fill distance (e.g. see \cite{llorente2020adaptive}[Sect. 7] and references therein)
\begin{align}
	\|\pi - \widehat{\pi}_t\|_\infty = \mathcal{O}(\lambda(r_t)).
\end{align}
The speed of convergence, i.e., the functional form of $\lambda(r_t)$, depends on the choice of kernel (e.g. under some circumstances and with Gaussian kernel, $\lambda(r_t)$ decays exponentially when $r_t \to 0$). 
\newline
In case we do not approximate $\pi(\x)$ directly, but we build an emulator of $\log \pi(\x)$ or just on the physical model ${\bf h}(\x)$, it is also possible to show the convergence of the posterior approximation. See, for instance, the error bounds in \cite[Theorem 4.2]{stuart2018posterior}. %That is, an approximation of the posterior based on the mentioned emulators converges to the true function if the emulators also converge. 

%{
%\begin{enumerate}
%	\item mostrar que $\frac{1}{c_t}\widehat{\pi}_t \to \pi$ as $t \to \infty$... y que por tanto muestreamos approx de la proposal ``optima''...
%	
%	\item Given normalized $p$ and $q$, the variance of the importance weight corresponds to the $\chi^2$ divergence between said distributions
%	\begin{align}
%	\chi^2(p\|q)= \int \frac{(p(\x)-\bar{q}(\x))^2}{\bar{q}(\x)}d\x.
%	\end{align}
%	Using Holder's inequality, we can bound the above divergence in terms of the $L_2$ distance 
%	\begin{align}
%	\int \frac{(p(\x)-\bar{q}(\x))^2}{\bar{q}(\x)}d\x = \int |p(\x)-\bar{q}(\x)|\frac{|p(\x)-\bar{q}(\x)|}{|\bar{q}(\x)|}d\x 
%	&= \norm{(p-q)\left(\frac{p-q}{q}\right)}_{L_1} \\
%	&\leq \norm{p-q}_{L_2}\norm{\frac{p-q}{q}}_{L_2}.
%	\end{align}
%	Thus, if we make $\norm{p-q}_{L_2}\to 0$ then $\chi^2(p\|q)\to0$.
%\end{enumerate}
%}

%{ ver si hay mas resultados para poner...  }

%%%%%%%%%%%%%%%%%%%%%%%%%%%%%%%%%%%%%%%%%%%%%% 
%%%%%%%%%%%%%%%%%%%%%%%%%%%%%%%%%%%%%%%%%%%%%%
\section{A special interesting case for NN-AIS}
\label{AppQueMeGusta}
%%%%%%%%%%%%%%%%%%%%%%%%%%%%%%%%%%%%%%%%%%%%%% 
%%%%%%%%%%%%%%%%%%%%%%%%%%%%%%%%%%%%%%%%%%%%%%
%{ $J_t$ aqui?}

Here, We focus on NN-AIS. We consider a bounded $\mathcal{X}$ and building $\widehat{\pi}_t$ with a nearest neighbor (NN) approach.  In  Sect. \ref{sec:buildProp}, we show that the NN emulator at iteration $t$ is given by
\begin{align}
	\widehat{\pi}_t(\x)&= \sum_{i=1}^{J_t} \pi(\x_i)\mathbb{I}_{\mathcal{R}_i}(\x) = \sum_{i=1}^{J_t} \pi(\x_i)|\mathcal{R}_i|\left[\frac{1}{|\mathcal{R}_i|}\mathbb{I}_{\mathcal{R}_i}(\x)\right], \\
	&=\sum_{i=1}^{J_t} \nu_i \ p_i(\x), 
\end{align}		
where $|\mathcal{R}_i|$ is the measure of $i$-th Voronoi region (see Eq. \eqref{eq:DefVoronoiCells} for the definition of $\mathcal{R}_i$), 
$\nu_i=\pi(\x_i)|\mathcal{R}_i|$, and $p_i(\x)=\frac{1}{|\mathcal{R}_i|}\mathbb{I}_{\mathcal{R}_i}(\x)$ are uniform densities over $\mathcal{R}_i$. Hence, $\widehat{\pi}_t(\x)$ is a mixture of ${J_t}$ uniform densities where the mixture weight is proportional to $\nu_i$. The normalizing constant of $\widehat{\pi}_t(\x)$ is given by
\begin{align}
	c_t=\sum_{i=1}^{J_t} \nu_i= \sum_{i=1}^{J_t} \pi(\x_i)|\mathcal{R}_i|,
\end{align}
so that the normalized proposal based on the NN emulator is 
\begin{align*}
	\frac{1}{c_t}\widehat{\pi}_t(\x)
	&=\frac{1}{c_t}\sum_{i=1}^{J_t} \nu_i \ p_i(\x)=\sum_{i=1}^{J_t} \bar{\nu}_i \ p_i(\x),
\end{align*}
where 
$$
\bar{\nu}_i=\frac{\nu_i}{c_t}=\frac{\pi(\x_i)|\mathcal{R}_i|}{\sum_{j=1}^{J_t} \pi(\x_j)|\mathcal{R}_j|}, \quad i=1,...,N,
$$
are also normalized. In order to sample $\frac{1}{c_t}\widehat{\pi}_t(\x)$, we would first (i) draw an index $i^*$ from the set $\{1,\dots,{J_t}\}$ with probabilities $\bar{\nu}_i = \frac{1}{c_t}\nu_i$ ($i=1,\dots,{J_t}$), and then (ii) sample from $p_{i^*}(\x)$. 
In practice, we do not know  the measures $|\mathcal{R}_i|$ and we are not able to draw samples uniformly in $\mathcal{R}_i$. Hence,  we use  SIR method to solve the problem  drawing from an auxiliary pdf $\bar{q}_\text{aux}(\x)$ (see \ref{APP:theo}), as we have proposed in RADIS. Namely, we resample from the set $\{\z_{t,\ell}\}_{\ell=1}^L \sim \bar{q}_\text{aux}(\x)$ with probabilities proportional to $\gamma_{t,\ell} = \frac{\widehat{\pi}_t(\z_{t,\ell})}{\bar{q}_\text{aux}(\z_{t,\ell})}$.  Below, we consider the special case that  $\bar{q}_\text{aux}(\x)$ is uniform.
\newline
\newline
{\bf Approximating $\bar{\nu}_i$'s.} Let choose an uniform auxiliary density $\bar{q}_\text{aux}(\x)$, i.e., $\bar{q}_\text{aux}(\x) = \frac{1}{|\mathcal{X}|}$ for all $\x\in\mathcal{X}$.
We draw  $\{\z_{t,\ell}\}_{\ell=1}^L$ from the uniform $\bar{q}_\text{aux}(\x)$. Then, the IS weight associated with the $\ell$-th sample is
$$\gamma_{t,\ell} \propto \widehat{\pi}_t(\z_{t,\ell}) = \pi(\x_{k_\ell}),$$
where 
$$\x_{k_\ell} = \arg \min_{\x_k\in \mathcal{S}_{t}} \|\x_k - \z_\ell\|,$$ 
i.e., $\x_{k_\ell}$ represents the NN of $\z_\ell$ within the set of ${J_t}$ nodes. 
Consider now the $i$-th node $\x_i$. All samples whose NN is $\x_i$ have weight proportional to $\pi(\x_i)$. We denote those samples as the set
\begin{align}
	\mathcal{U}_i = \{\z_{t,\ell}:\ \x_i = \arg \min_{\x_k} \|\x_k - \z_{t,\ell}\|\}.
\end{align}
The number of samples within $\mathcal{U}_i$ can be written as $|\mathcal{U}_i| = {\sum_{\ell=1}^{L}\mathbb{I}(\x_{k_\ell}=\x_i)}$. The probability of resampling a $\z_{t,\ell}$ that comes from $\mathcal{U}_i$ is proportional to $|\mathcal{U}_i|\pi(\x_i)$ (since there are $|\mathcal{U}_i|$ samples with weight $\pi(\x_i)$). As $L \to \infty$, by the law of large numbers, we have these probabilities converge to the true ones
\begin{align}
	\frac{|\mathcal{U}_i|\pi(\x_i)}{\sum_{k=1}^ {J_t}|\mathcal{U}_k|\pi(\x_k)} \to \frac{|\mathcal{R}_i|\pi(\x_i)}{\sum_{k=1}^{J_t}|\mathcal{R}_k|\pi(\x_k)} = \bar{\nu}_i.
\end{align}
%\begin{align}
%	|\mathcal{U}_i| = \frac{\sum_{\ell=1}^{L}\mathbb{I}(\x_{k_\ell}=\x_i)}{L}
%\end{align}
%Resampling $\z_\ell$ with probability proportional to $\gamma_\ell$  means we pull out one sample from $\mathcal{U}_i$ with probability proportional to 
%	{$$ \mbox{NO SE PONEN NUNCA, ECUACIONES CON UNA SOLA "SIDE", DEFINELO BIEN !!!}$$
%	$$
%	\frac{\sum_{\ell=1}^{L}\mathbb{I}(\x_{k_\ell}=\x_i)}{L}\pi(\x_i),$$}
%where the term $\sum_{\ell=1}^{L}\mathbb{I}(\x_{k_\ell}=\x_i)$ accounts for the number of samples whose NN is $\x_i$.
%Note that, by the law of large numbers, when $L \to \infty $
%	$$ \frac{\sum_{\ell=1}^{L}\mathbb{I}(\x_{k_\ell}=\x_i)}{L}\pi(\x_i) \to
%	\frac{|\mathcal{R}_{i}|}{|\mathcal{X}|}\pi(\x_i) \propto |\mathcal{R}_i|\pi(\x_i),
%	$$
%and we will sample uniformly from $\mathcal{R}_i$ with the correct probability.
{\bf Rejection sampling.}
Note also that the samples within $\mathcal{U}_i$ form a particle approximation of the uniform density over $\mathcal{R}_i$. Indeed, taking one sample at random from $\mathcal{U}_i$ corresponds to applying rejection sampling on $p_i(\x)$. In order to see this, consider the rejection sampling setting where $p_i(\x)$ is the target probability and $\bar{q}_\text{aux}(\x) = \frac{1}{|\mathcal{X}|}$ is the proposal. Note that $\frac{p_i(\x)}{\bar{q}_\text{aux}(\x)} = \frac{|\mathcal{X}|}{|\mathcal{R}_i|}$ for all $\x \in \mathcal{R}_i$, and $\frac{p_i(\x)}{\bar{q}_\text{aux}(\x)} = 0$ for all $\x \not\in \mathcal{R}_i$, so $\bar{q}_\text{aux}(\x)$ is a valid proposal for rejection sampling with rejection constant $M= \frac{|\mathcal{X}|}{|\mathcal{R}_i|}$ \cite[Chapter 3]{MARTINO_book}. In rejection sampling, we draw $\z \sim \bar{q}_\text{aux}(\x)$,  $u\sim \mathcal{U}[0,1]$ and accept $\z$ if 
\begin{align}
	u\frac{|\mathcal{X}|}{|\mathcal{R}_i|}\bar{q}_\text{aux}(\z) \leq p_i(\z).
\end{align}
If the condition holds,  $\z$ is an independent sample from $p_i(\x)$. Otherwise we reject $\z$, draw another candidate $\z$ and so on.
Note that, when $\z \in \mathcal{R}_i$, we have
\begin{align}
	u \frac{|\mathcal{X}|}{|\mathcal{R}_i|}\frac{1}{|\mathcal{X}|}\leq \frac{1}{|\mathcal{R}_i|} \Longleftrightarrow u \leq 1,
\end{align}
so we always accept all $\z$'s that are closest to node $\x_i$, becoming i.i.d. samples from $p_i(\x)$. Conversely, when $\z \not\in \mathcal{R}_i$, we have the condition $u \leq 0$ that never holds, so that  we always reject them. Namely, the set $\mathcal{U}_i$ contains  i.i.d. samples from $p_i(\x)$, that have been obtained by rejection sampling.
\newline
{\bf Summary.} With the particular choice  $\bar{q}_\text{aux}(\x) = \frac{1}{|\mathcal{X}|}$ for all $\x \in \mathcal{X}$, the SIR approach in NN-AIS is equivalent to (i) estimating by Monte Carlo the mixture probabilities $\bar{\nu}_i$, and (ii) applying rejection sampling to sample uniformly within each Voronoi region $\mathcal{R}_i$.

%%%%%%%%%%%%%%%%%%%%%%%%%%%%%%%%%%%%%%%%%%%%%% 
\section{NN-AIS in unbonded domains}\label{AdaptiveSupportSect}
%%%%%%%%%%%%%%%%%%%%%%%%%%%%%%%%%%%%%%%%%%%%%% 

%{{red} utilizar q param para ayudar a cambiar el soporte... contextualizar...}
%\newline
%{{red}If NN interpolant...} 

In this section, we recall how to extend the applicability of the nearest neighbor (NN) construction (see Sect. \ref{sec:buildProp}) when the domain $\mathcal{X}$ is unbounded and show how to adapt the support of NN approximation.

%%%%%%%%%%%%%%%%%%%%%%%%%%%%%%%%%%%%
\subsection{NN-AIS with a fixed support in an unbounded domain}
%%%%%%%%%%%%%%%%%%%%%%%%%%%%%%%%%%%%

%If the support $\mathcal{X}$ is unbounded, the NN-based surrogate is not a valid proposal on its own since it is only defined for bounded domains. 

Consider again the following mixture proposal,
\begin{align}
	\varphi_t(\x) = \alpha_t \bar{q}_\texttt{par}(\x) + (1-\alpha_t)\frac{1}{c_t}\widehat{\pi}_t(\x),
\end{align}
where $\alpha_t\in [0,1]$ for all $t$, and $\bar{q}_\texttt{par}(\x)$ is parametric pdf that covers properly the tails of the posterior $\pi$. Namely, $\bar{q}_\texttt{par}(\x)$ is defined in the unbounded domain $\mathcal{X}$ of $\pi$, whereas $\widehat{\pi}_t(\x)$ is built considering a bounded support $\mathcal{D}\subset \mathcal{X}$, decided in advance by the user. Hence, $\varphi_t$ is a valid proposal when using NN-AIS with unbounded $\mathcal{X}$. 
In this simple scenario, $\mathcal{D}$ is fixed and does not vary with the iteration $t$.
However, the information provided by the samples from $\bar{q}_\texttt{par}(\x)$ can be used to expand the support of $\widehat{\pi}_t$, i.e., such it has an adaptive support, as described below.

%%%%%%%%%%%%%%%%%%%%%%%
\subsection{Adapting support in NN-AIS}
%%%%%%%%%%%%%%%%%%%%%%
Let $\mathcal{X}$ be unbounded and  $\widehat{\pi}_t$ be the surrogate model built with NN. Let $\mathcal{D}_t \subset \mathcal{X}$ denote the compact subset of $\mathcal{X}$ where $\widehat{\pi}_t$ is defined, i.e., $\widehat{\pi}_t$ is zero outside $\mathcal{D}_t$. Note that $\mathcal{D}_t$ depends on $t$.
The set of current nodes $\mathcal{S}_t$ is used to define the boundaries of $\mathcal{D}_t$.
One possible way is as follows:
Take $\mathcal{D}_t$ as the hyperrectangle whose edges are defined by the maximum and minimum value, in each dimension, of the set $\mathcal{S}_t$, i.e., 
\begin{align}
	\mathcal{D}_t = \{\x\in\mathcal{X}: \ \min_{\s_{t-1} \in \mathcal{S}_{t-1}} s_{d,t-1} \leq x_{d} \leq \max_{\s_{t-1} \in \mathcal{S}_{t-1}} s_{d,t-1}, \enskip d=1,\dots,d_x \},
\end{align}
where $x_d$ denotes the $d$-th element of $\x$,  $\s_{t-1} = [s_{1,t-1}, \dots, s_{d_x,t-1}]\in\mathcal{S}_{t-1}$ and $\mathcal{S}_{t-1}$ denotes the set of nodes at iteration $t$. 
After adding new nodes, we update the bounds of the hyperrectangle. Note that only samples from $\bar{q}_\texttt{par}(\x)$ that fall outside $\mathcal{D}_t$ will expand it.  Note that, the size of $\mathcal{D}_t$ is always increasing but controlled by the tail of $\bar{\pi}$. Indeed, the candidate samples drawn in the tails of $\bar{\pi}$ will have very low values of $\pi$, so that the probability of sampling those regions will be negligible (then these regions will be never used). In the case we use a uniform $\bar{q}_\text{aux}(\x)$ in $\mathcal{D}_t$ to sample $\widehat{\pi}_t$, note that $\bar{q}_\text{aux}(\x)$ actually depends on $t$ and is changing at every iteration whenever $\mathcal{D}_t$ changes.
\end{appendices}

\end{document}